\documentclass[apj,numberedappendix,twocolappendix]{emulateapj}
\usepackage{natbib}
\usepackage{amsmath}
\usepackage{epstopdf}
\usepackage{xcolor}
\newcommand{\myemail}{nwt2de@virginia.edu}
\shorttitle{Companions to APOGEE Stars I}
\shortauthors{Troup et al.}

\begin{document}

\title{Companions to APOGEE Stars I: A Milky Way-Spanning Catalog of Stellar and Substellar Companion Candidates and their Diverse Hosts}

\author{Nicholas W. Troup\altaffilmark{1a}, 
		David L. Nidever\altaffilmark{2,3,24},		
		Nathan De Lee\altaffilmark{4,5}, 
		Joleen Carlberg\altaffilmark{6},
		Steven R. Majewski\altaffilmark{1},
		Martin Fernandez\altaffilmark{7},
		Kevin Covey\altaffilmark{7},
		S. Drew Chojnowski\altaffilmark{8},
	    Joshua Pepper\altaffilmark{9}, 
	    Duy T. Nguyen\altaffilmark{1},
		Keivan Stassun\altaffilmark{4},    
		Duy Cuong Nguyen\altaffilmark{10},
		John P. Wisniewski\altaffilmark{11},
		Scott W. Fleming\altaffilmark{12,13},	
		Dmitry Bizyaev\altaffilmark{14,15},
		Peter M. Frinchaboy\altaffilmark{23},
		D. A. Garc\'{i}a-Hern\'{a}ndez\altaffilmark{20,21},
		Jian Ge\altaffilmark{16},
		Fred Hearty\altaffilmark{17,18}, 
		Szabolcs Meszaros\altaffilmark{19},
		Kaike Pan\altaffilmark{14},
		Carlos Allende Prieto\altaffilmark{20,21},
		Donald P. Schneider\altaffilmark{17,18},
		Matthew D. Shetrone\altaffilmark{22},
		Michael F. Skrutskie\altaffilmark{1},
		John Wilson\altaffilmark{1},
		Olga Zamora\altaffilmark{20,21}}

\altaffiltext{1}{Department of Astronomy, University of Virginia, Charlottesville, VA 22904-4325, USA $^{\mathrm{A}}$\myemail}
\altaffiltext{2}{University of Michigan, 1085 S University Ave, Ann Arbor, MI 48109, USA}
\altaffiltext{3}{Large Synoptic Survey Telescope, 950 North Cherry Ave, Tuscon, AZ 85719, USA}
\altaffiltext{4}{Department of Physics, Geology, and Engineering Tech, Northern Kentucky University, Highland Heights, KY 41099, USA}
\altaffiltext{5}{Department of Physics and Astronomy, Vanderbilt University, Nashville, TN, USA}
\altaffiltext{6}{NASA Goddard Spaceflight Center, Greenbelt, MD, USA}
\altaffiltext{7}{Western Washington University, Bellingham, WA 98225, USA}
\altaffiltext{8}{New Mexico State University, Las Cruces, NM, USA}
\altaffiltext{9}{Lehigh University, Bethlehem, PA, USA}
\altaffiltext{10}{University of Toronto, Toronto, Ontario, Canada}
\altaffiltext{11}{University of Oklahoma, Norman, OK, USA}
\altaffiltext{12}{Space Telescope Science Institute, Baltimore, MD, USA}
\altaffiltext{13}{Computer Sciences Corporation, Baltimore, MD, USA}
\altaffiltext{14}{Apache Point Observatory and New Mexico State University, P.O. Box 59, Sunspot, NM, 88349-0059, USA}
\altaffiltext{15}{Sternberg Astronomical Institute, Moscow State University, Moscow, Russia}
\altaffiltext{16}{Department of Astronomy, University of Florida, Gainesville, FL 32611, USA}
\altaffiltext{17}{Department of Astronomy \& Astrophysics, The Pennsylvania State University, University Park, PA 16802, USA}
\altaffiltext{18}{Center for Exoplanets and Habitable Worlds, The Pennsylvania State University, University Park, PA 16802, USA}
\altaffiltext{19}{ELTE Gothard Astrophysical Observatory, H-9704 Szombathely, Szent Imre Herceg st. 112, Hungary}
\altaffiltext{20}{Instituto de Astrof\'{\i}sica de Canarias, Via L\'actea s/n, 38205 La Laguna, Tenerife, Spain}
\altaffiltext{21}{Departamento de Astrof\'{\i}sica, Universidad de La Laguna, 38206 La Laguna, Tenerife, Spain}
\altaffiltext{22}{University of Texas, Austin, TX, USA}
\altaffiltext{23}{Department of Physics \& Astronomy, Texas Christian University, TCU Box 298840, Fort Worth, TX 76129 (p.frinchaboy@tcu.edu)}	
\altaffiltext{24}{Steward Observatory 933 North Cherry Ave, Tuscon, AZ 85719, USA}

\begin{abstract}
In its three years of operation, the Sloan Digital Sky Survey (SDSS-III) Apache Point Observatory Galactic Evolution Experiment (APOGEE-1) observed $>$14,000 stars with enough epochs over a sufficient temporal baseline for the fitting of Keplerian orbits. \textcolor{black}{We present the custom orbit-fitting pipeline used to create this catalog, which includes novel quality metrics that account for the phase and velocity coverage of a fitted Keplerian orbit.} With a typical RV precision of $\sim100-200$ m s$^{-1}$, APOGEE can probe systems with small separation companions down to a few Jupiter masses. Here we present initial results from a catalog of 382 of the most compelling stellar and substellar companion candidates detected by APOGEE, which orbit a variety of host stars in diverse Galactic environments. Of these, 376 have no previously known small separation companion. The distribution of companion candidates in this catalog shows evidence for an extremely truncated brown dwarf (BD) desert with a paucity of BD companions only for systems with $a < 0.1-0.2$ AU, with no indication of a desert at larger orbital separation. We propose a few potential explanations of this result, some which invoke this catalog's many small separation companion candidates found orbiting evolved stars. Furthermore, 16 BD and planet candidates have been identified around metal-poor ([Fe/H] $< -0.5$) stars in this catalog, which may challenge the core accretion model for companions $>10 M_{Jup}$. Finally, we find all types of companions are ubiquitous throughout the Galactic disk with candidate planetary-mass and BD companions to distances of $\sim6$ and $\sim16$ kpc, respectively.
\end{abstract}
\keywords{binaries: close --- binaries: spectroscopic ---  brown dwarfs --- Galaxy: stellar content --- planetary systems}

\section{Introduction}
Over the past few decades, it has been established that solitary Milky Way stars are the exception rather than the rule. Previous studies of stellar multiplicity have shown that more than half of stellar systems contain two or more bound stars, and that stars in these systems span a wide range of separations and mass ratios \citep[e.g.,][]{Raghavan2010, Duchene2013}. With the advent of the enormous database of confirmed and candidate systems generated by the large-scale planet-hunting mission \textit{Kepler} \citep{Borucki2010}, planetary companions are also thought to be quite commonplace, including an unexpected class of short-period Jupiter-mass planet, the first discovered by \cite{Mayor1995}. These ``hot Jupiters,'' have been explained by inward orbital migration during their formation \citep{Masset2003}. Interestingly, while both exoplanets and stellar-mass companions have been found in extremely short-period orbits, there has been a paucity of brown dwarf (BD\footnote{For this paper we define a brown dwarf companion as a companion with a mass between the Deuterium-burning (0.013 $M_{\odot}$) and Hydrogen-burning (0.080 $M_{\odot}$) limits}) companions orbiting Sun-like stars, a phenomenon known as the ``brown dwarf desert'' \citep{Marcy2000}. However, more recent work has shown that this desert might be limited in extent, with no desert for wide ($a < 1000$ AU) companions \citep{Gizis2001}, and may not be as ``dry'' as initially thought when considering stars more massive than the Sun \citep{Guillot2014}.

Traditionally, solar-like dwarf stars have been the primary targets for exoplanet searches and stellar multiplicity studies. However, recently some work has been done with evolved stars \citep[e.g.,][]{Reffert2006, Lovis2007, Johnson2007,Wittenmyer2011,Zielinski2012}. Currently, there are only approximately 50 known planet-hosting giant stars, compared to the $> 1000$ known dwarf-star planet hosts \citep{Jones2014b}, but even this small sample of giant star hosts has produced some interesting results. As a star like the Sun expands into a red giant, its atmosphere will engulf the innermost planets \citep[e.g.,][]{Villaver2009,Villaver2014a}. Stronger tidal dissipation from the expanding star may also lead to more distant companions also being consumed. Possible observational signatures of planetary engulfment have been identified in the chemical abundances and peculiarly high rotational velocities seen in some giant stars  \citep[e.g.,][]{Massarotti2008, Adamow2012, Carlberg2012}. However, \cite{Silvotti2014} have found hot Jupiters orbiting subdwarf B stars, which suggests that some Jovian planets may survive within the extended envelope of their host star during its red giant phase.

It is becoming clear that the properties of the host star plays an important role in the types of companions that can form with it. It has been established that metal-rich host stars are more likely to host Jovian planets than their metal-poor counterparts \citep{Fischer2005}. This relation is believed to be a consequence of the core accretion model of planet formation, which requires a potential Jovian planet to acquire $\sim 5-10 M_{\oplus}$ worth of solid material before the central star expels the hydrogen and helium gas from the protoplanetary disk  \citep{Matsuo2007}. Similar trends relating individual elemental abundances to planet occurrence rate have also been found \citep[e.g.,][]{Bodaghee2003, Robinson2006,Adibekyan2012a}. Stellar binaries are formed via a separate mechanism, and it is disputed whether or not metallicity plays a role in binary fraction \citep{Abt2008}. Binarity has generally been found to be higher in lower metallicity populations \citep[e.g.,][]{Carney2003}. However, a higher fraction of stellar binaries has been found among metal-rich F-type dwarfs in the field compared to their metal-poor counterparts \citep{Hettinger2015}. It is not clear whether brown dwarf formation follows star or planet formation trends more closely. Planet occurrence rate has also been shown to depend on the mass of the host star, with higher-mass hosts being less likely to host a planet than lower-mass hosts \citep[e.g.,][]{Reffert2015}.  

Most exoplanet and multiplicity surveys have also focused on targeting stars within the solar neighborhood because of the aforementioned concentration on solar-like dwarf stars, and the greater difficulty in measuring transit signals and RVs for these types of stars at great distances. Because of these limitations, there is a limited understanding of the Galactic distribution of companions. Microlensing surveys such as The Optical Gravitational Lensing Experiment \citep[OGLE;][] {Udalski2003} have discovered potential planetary-mass candidates in the Galactic Bulge \citep{Shvartzvald2014}, but few other planets have been found farther than $\sim 1$ kpc from the Sun. Furthermore, the vast majority of planets have been identified among Galactic field stars, while only a few planets have been discovered in open clusters \citep[e.g.,][]{Lovis2007,Brucalassi2014}.

\subsection{The Role of APOGEE}
Many of the aforementioned discoveries came through small and large-scale stellar transit monitoring, the use of single-object spectroscopy, or the combination thereof.  A logical step forward in this field is the use of large-scale multi-object spectroscopy to complement current and future large photometric surveys such as those by \textit{Kepler}\citep{Borucki2010} and TESS\citep{Ricker2014}. The Sloan Digital Sky Survey III \citep[SDSS-III;][]{Eisenstein2011} Multi-object APO Radial Velocity Exoplanet Large-area Survey \citep[MARVELS;][]{Ge2008} used this approach to observe $\sim 10,000$ stars and discovered several BD and low-mass stellar companions \citep{Lee2011,Wisniewski2012,Fleming2012,Ma2013,Mack2013c,DeLee2013,Wright2013,Jiang2013}.

The SDSS-III Apache Point Observatory Galactic Evolution Experiment \citep[APOGEE][]{Majewski2015} is a large-scale, systematic, high-resolution ($R = 22,500$), $H$-band ($1.51 \mathrm{\mu m} < \lambda < 1.69 \mathrm{\mu m}$), spectroscopic survey of the chemical and kinematical distribution of Milky Way stars. APOGEE acquired high $S/N$ ($>100$) spectra of over 146,000 stars distributed across the Galactic bulge, disk, and halo. To achieve this $S/N$, many of the stars had to be observed for long net integration times -- up to 24 hours. To accomplish this goal, and to gain sensitivity to temporal variations in radial velocity (RV) indicative of stellar companions, the APOGEE survey observed most stars over multiple epochs. In three years of operations, APOGEE observed over 14,000 stars enough times ($\ge 8$) and over a sufficient temporal baseline to collect spectra yielding high quality RV measurements suitable to not only reliably detect RV variability, but also to construct reliable Keplerian orbital fits to search for companions of a wide range of masses. With a typical radial velocity precision of $\sim$ 100-200 m s$^{-1}$, APOGEE can detect RV oscillations typical of those expected from relatively short-period companions down to a few Jupiter-masses ($10^{-3} M_{\odot}$). And because of APOGEE's design as a systematic probe of Galactic structure, this sample probes stellar populations not traditionally sought in exoplanet and stellar multiplicity studies in regions of the Milky Way well beyond the solar neighborhood.

\subsection{Paper Overview}
In this paper, we present the first catalog of 382 candidate companions detected by APOGEE. In \S \ref{sec:Obs}, we give a brief description of the nature of the APOGEE observations, with a general description of the APOGEE data reduction in \S \ref{sec:Reduction}. Section \ref{sec:Reduction} also introduces the \texttt{apOrbit} pipeline, describing how the radial velocities and orbital parameters are derived, and introduces novel quality criteria which quantifies and accounts for both the phase and velocity space coverage of the fitted Keplerian model. Section \ref{sec:Catalog} presents APOGEE's first catalog of candidate companions to stars observed by APOGEE, and in particular, describes how we select the statistically significant RV variable sample, and the final ``gold sample'' of candidate companions. In \S \ref{sec:Results}, we discuss global analysis of this gold sample. Finally, in \S \ref{sec:Future} we describe planned future efforts with this and future, expanded catalogs, and we summarize conclusions drawn from the gold sample in \S \ref{sec:Conclusion}. Verification efforts of the \texttt{apOrbit} pipeline are described in Appendix \ref{sec:Verification}, and instruction on how to access and use the catalog are presented in Appendix \ref{sec:data}.

\section{APOGEE Radial Velocity Observations} \label{sec:Obs}
All APOGEE-1 observations were taken using fibers connected to either the Sloan 2.5 m telescope \citep{Gunn2006} or the NMSU 1-m telescope at Apache Point Observatory \citep[APO;][]{Majewski2015}. In normal use on the Sloan 2.5 m telescope, APOGEE employs a massively multiplexed, fiber-fed spectrograph capable of recording 300 spectra at a time. For full details on the APOGEE instrument see \cite{Wilson2015}. 

\begin{figure}[h!]
\center
\includegraphics[width=1\columnwidth]{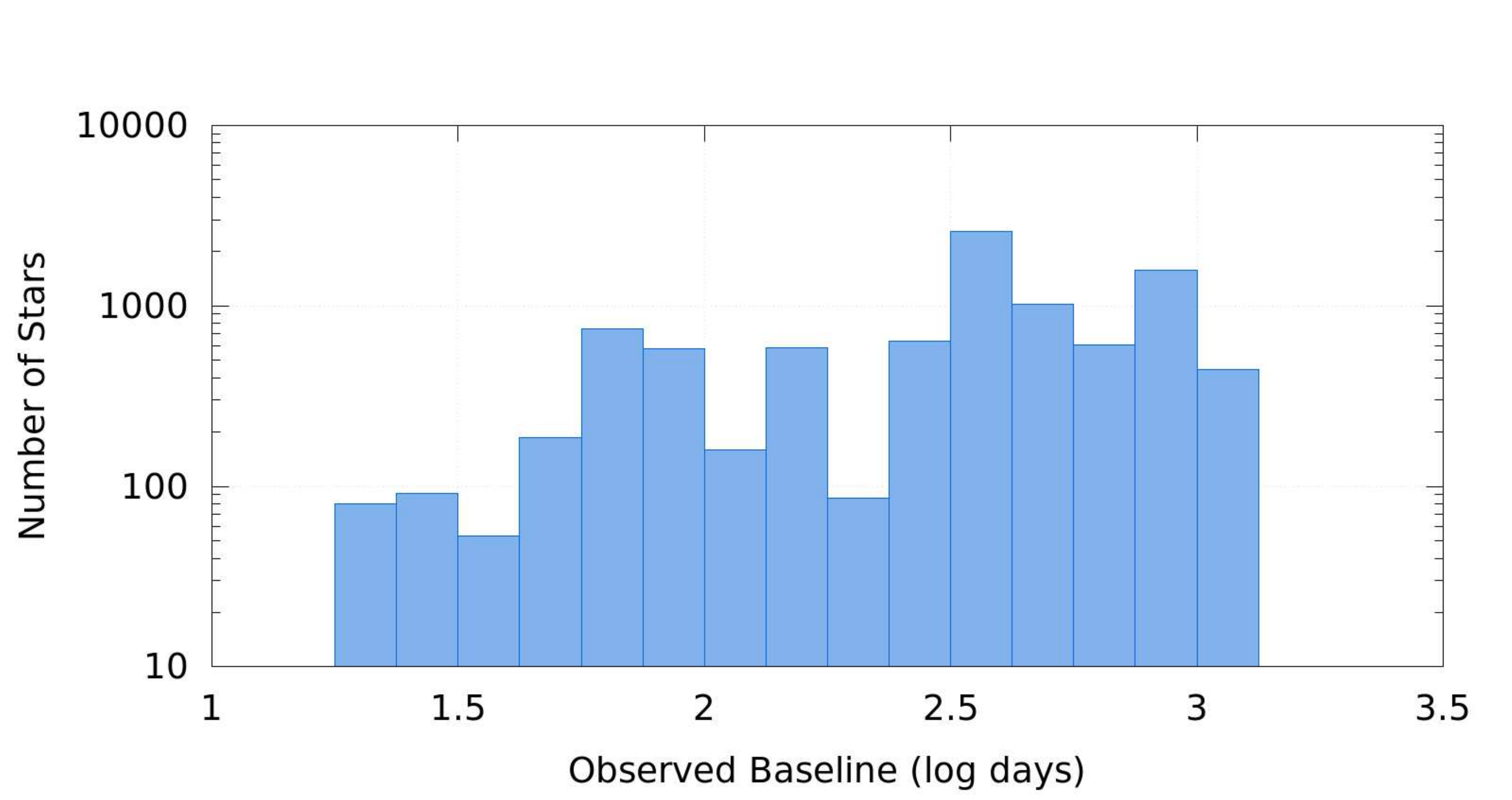}
\includegraphics[width=1\columnwidth]{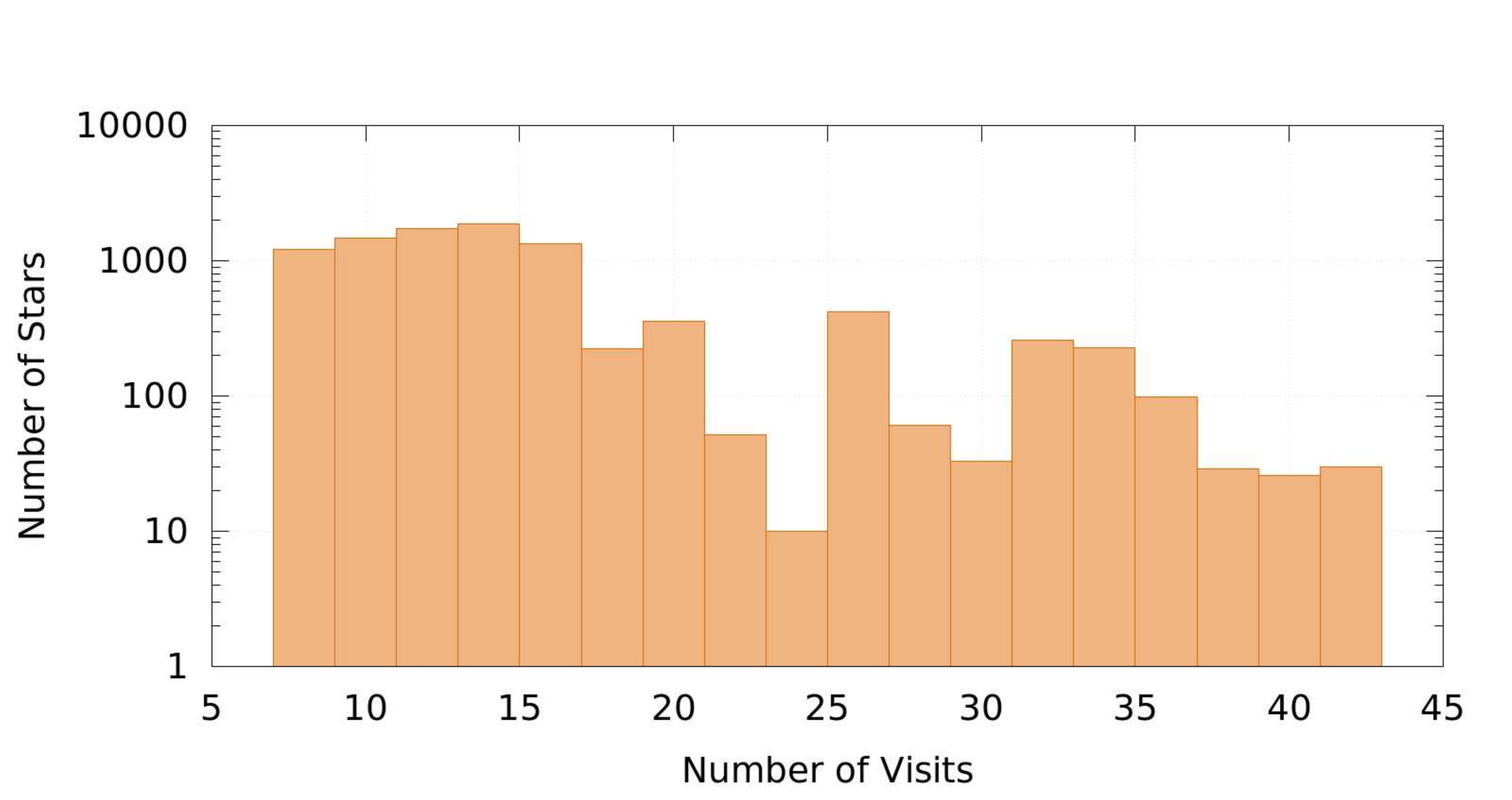}
\includegraphics[width=1\columnwidth]{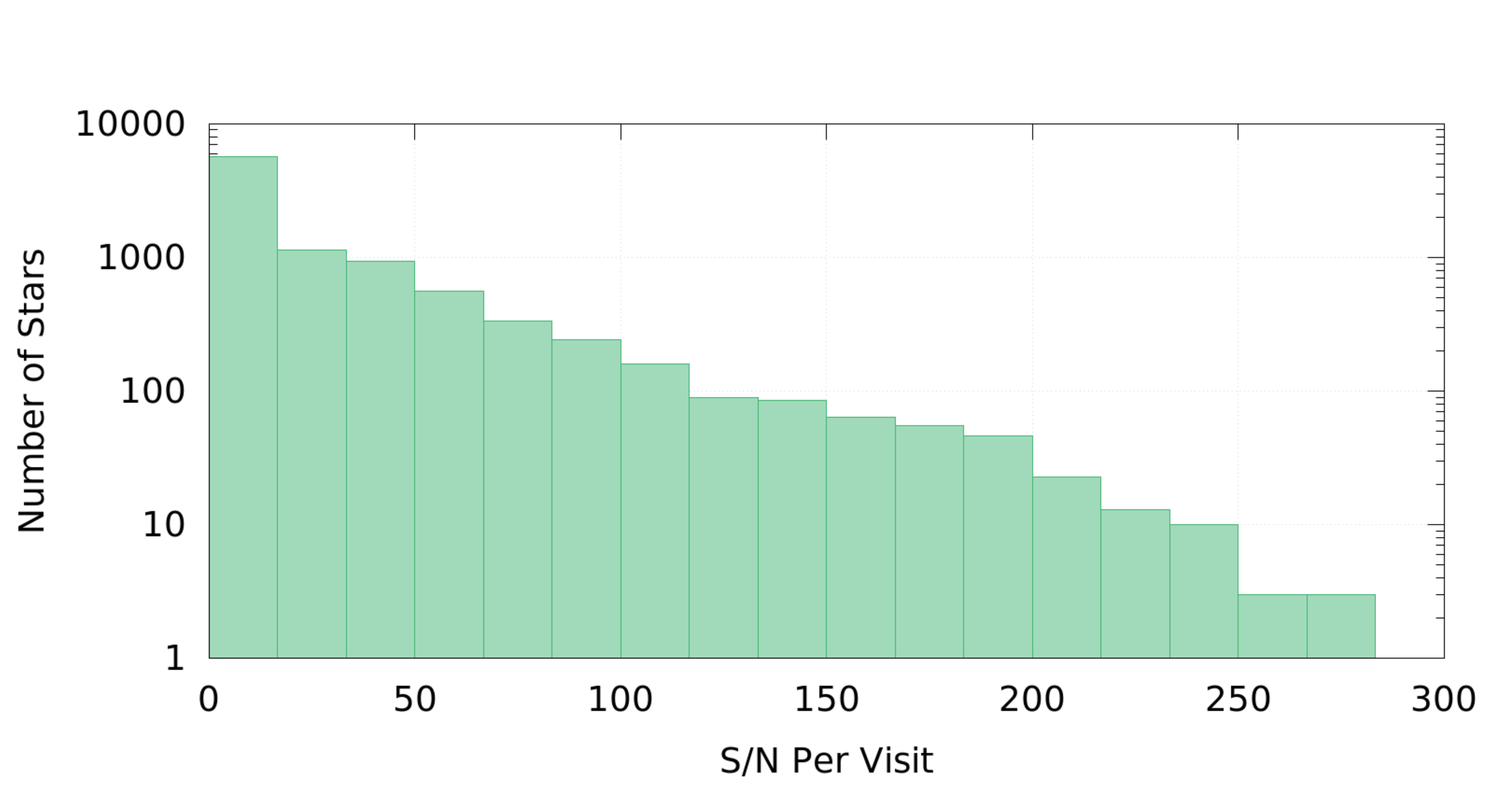}
\caption{\textit{Top Panel:} Distribution of the observed baseline for the 14,840 stars with at least eight visits. The median baseline for this set of stars is slightly over a year at 384 days. \textit{Middle Panel:} Distribution of the number of visits to the same set of stars, with 13 being the median number of visits. \textit{Bottom Panel:} Distribution of the average $S/N$ per visit for the same set of stars, with a median $S/N$ per visit of 12.2.} \label{fig:BaseVisitHist}
\end{figure}

Of the 146,000 stars observed in APOGEE-1, 14,840 had at least eight visits; these stars were selected for analysis here. APOGEE first light observations were obtained in May of 2011 and APOGEE-1 observations concluded at the end of SDSS-III in July of 2014, providing a maximum temporal baseline of slightly more than three years ($\sim$1000 days). Figure \ref{fig:BaseVisitHist} shows the distribution of temporal baselines for stars submitted for Keplerian orbit fitting, as well as the distribution of the number of visits to each of these stars.  An APOGEE ``visit'' is defined as the combined spectrum of a source from a single night's observations, typically $\sim 1$ hour of exposure. For main survey targets, the number of visits scheduled for a star depends on its $H$ magnitude, with fainter targets needing more visits to acquire the APOGEE target accumulated $S/N$ of 100 per half-resolution element. For stars with at least eight visits, individual visit spectra obtained a median $S/N$ of 12.2. Visits are required to be separated by $\ge$3 days, and must span $\ge$30 days at minimum to gauge the potential binarity of the source. Special targets such as stars used for calibration or ancillary science programs often have additional visits and employ a non-standard cadence. For example, some stars observed during commissioning were re-observed at the end of the survey as a consistency check (see Appendix \ref{sec:caveats}), so these stars may have visits separated by over two years. For a more detailed description of APOGEE targeting and observing strategy see \cite{Zasowski2013} and \cite{Majewski2015}.

\section{Data Reduction and the \texttt{apOrbit} Pipeline} \label{sec:Reduction}
Because the results of the present work depend critically on an understanding of the RVs and their uncertainties, we first review those aspects of the data reduction process most relevant to the derivation of the RVs. For more information on processing steps that lead to the creation of the individual visit spectra, as well as more information regarding the main APOGEE data reduction pipeline (\texttt{apogeereduce}) see \cite{Nidever2015}. 

After producing the individual visit spectra, \texttt{apogeereduce} performs initial radial velocity corrections on the visit spectra (described briefly in \S \ref{sec:RVder}), and combines them into a single spectrum for each star. The APOGEE Stellar Parameters and Chemical Abundances pipeline \citep[ASPCAP;][]{GarciaPerez2015} then matches this combined spectrum to a library of synthetic spectra \citep{Zamora2015}, constructed by using extensive atomic/molecular linelists\citep{Shetrone2015}, automatically delivering accurate stellar atmospheric parameters ($T_{\rm eff}$ within $\sim$100 K, $\log g$ and [Fe/H]  within $\sim 0.1$ dex) and the abundances of up to 15 chemical elements (Fe, C, N, O, Na, Mg, Al, Si, S, K, Ca, Ti, V, Mn, Ni). Both the model synthetic spectrum and stellar parameters derived for the star are used in the production of the final RVs used in orbit fitting as described in \S \ref{sec:RVder} and to derive the properties for the primary star as described in \S \ref{sec:spDer}.

\subsection{Derivation of Radial Velocities} \label{sec:RVder}
The main APOGEE pipeline retains RVs from two methods: 1) The APOGEE reduction pipeline initially selects, through $\chi^2$ minimization, an RV template from a coarse grid of synthetic spectra (the ``RV mini-grid''). This template is cross-correlated against the spectrum to produce absolute RVs.
2) The pipeline cross-correlates the visit spectra with a combined spectrum of all visits and applies a barycentric correction to acquire heliocentric RVs. These RVs are stored as APOGEE data products.

To ensure the highest precision RVs, we preformed the additional step of using the best-fit synthetic spectrum chosen by ASPCAP as the RV template. The grid of synthetic spectra used by ASPCAP is much finer than the RV mini-grid with additional dimensions to account for [$\alpha$/M], [C/M], and [N/M]. In addition, the final model spectrum is achieved through cubic B\'ezier interpolation in the grid of spectra. Therefore, the ASPCAP best-fit template is a significant improvement over the RV mini-grid template and provides a high-quality match to the observed combined spectrum. This approach combines the advantages of using a noiseless synthetic spectrum as a template and using the combined observed spectrum to mitigate the chances of template mismatch. In the cases when mismatch did occur (e.g., due to a poor or failed ASPCAP solution), we deferred to the RVs derived from the combined observed spectrum template. In either case, the RVs we used for orbit fitting were heliocentric RVs.

\subsubsection{Analysis of RV Precision}
\begin{figure*}
\centering
\includegraphics[height=0.49\textwidth, angle=-90]{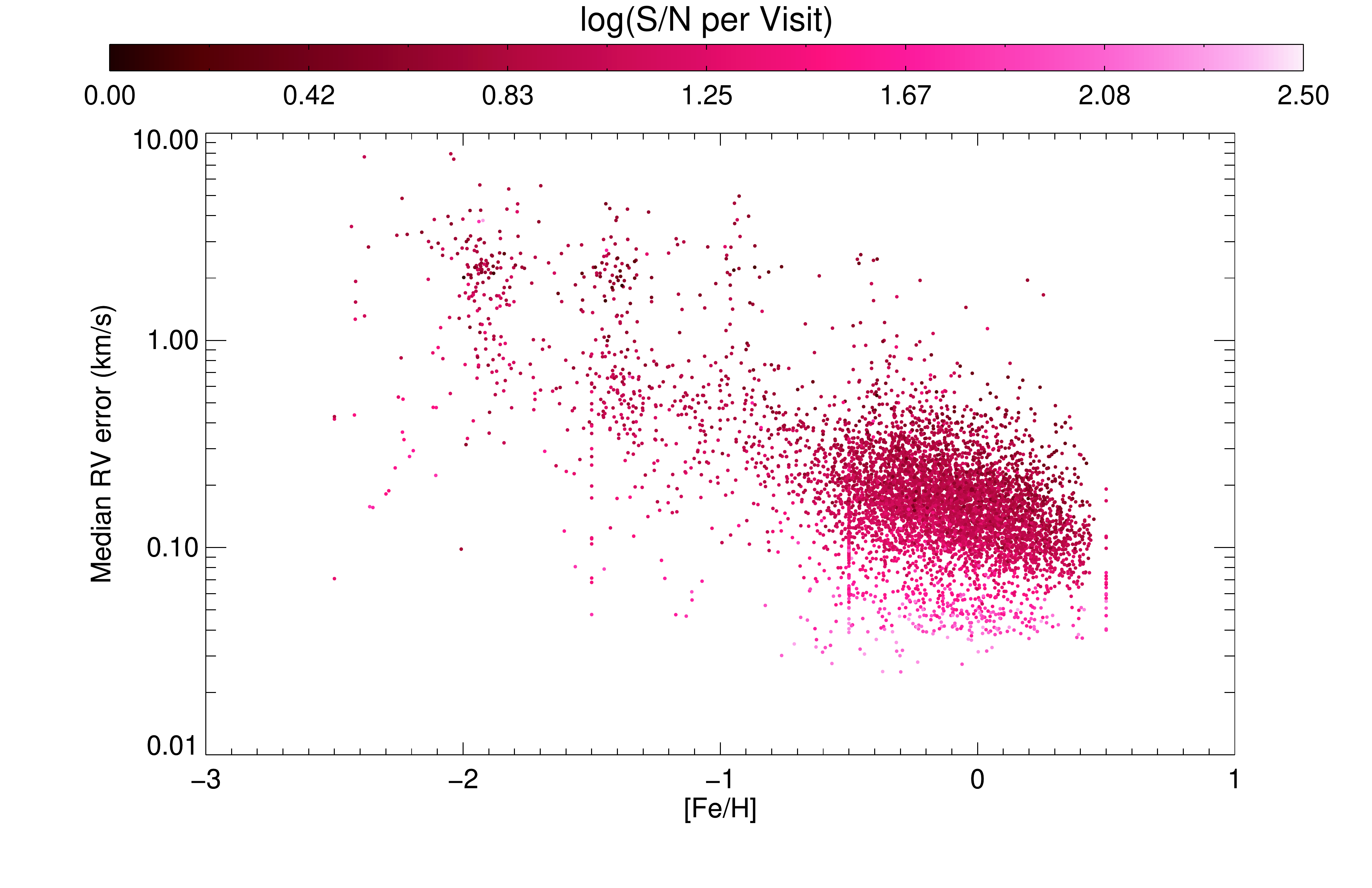}
\includegraphics[height=0.49\textwidth, angle=-90]{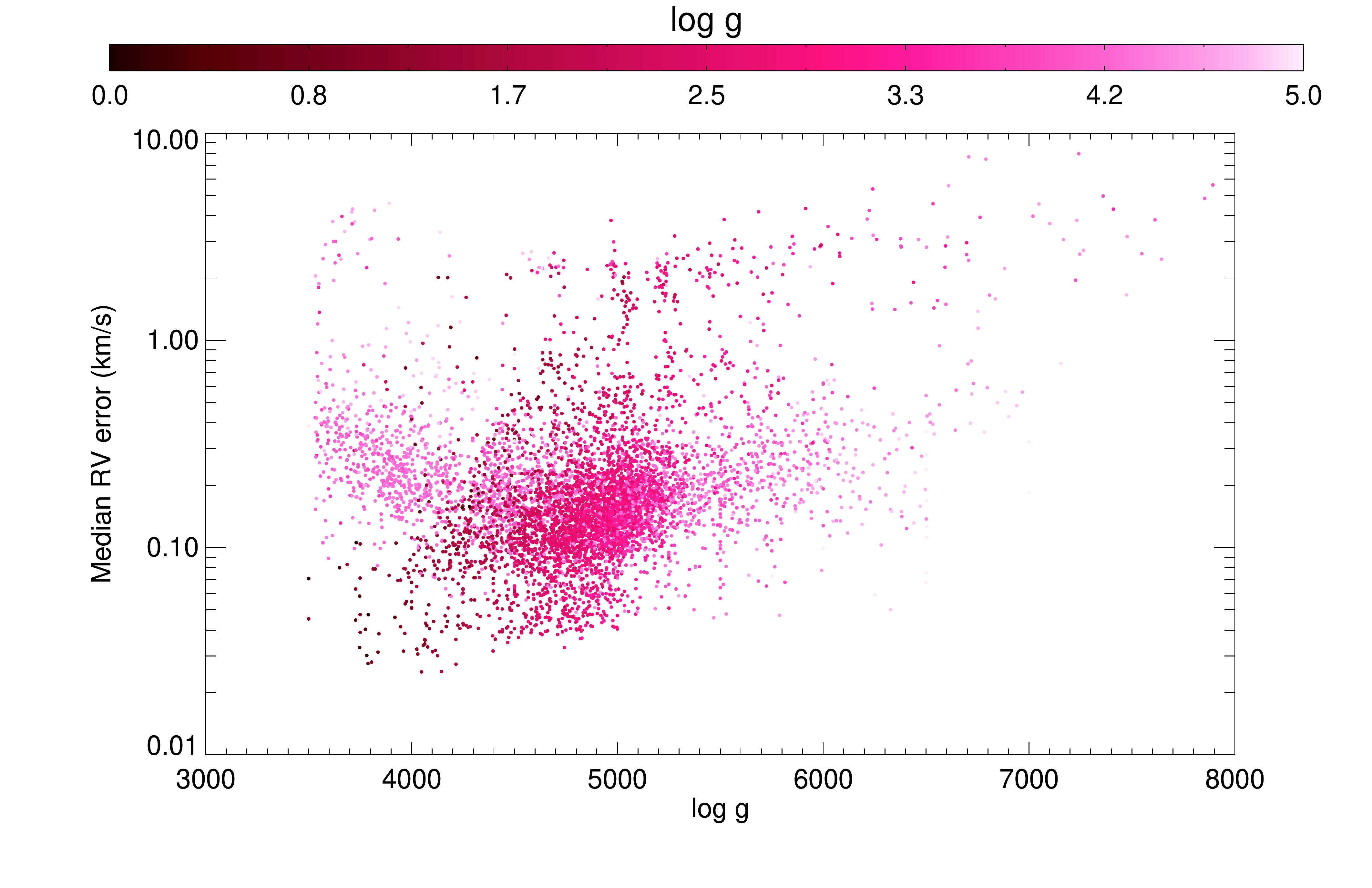}
\caption{\textit{Left Panel:} Precision of individual APOGEE visit RVs as a function of the metallicity ([Fe/H]) of the star with the color scale indicating the logarithm of the $S/N$ per visit. \textit{Right Panel:} Precision of individual APOGEE visit RVs as a function of the effective temperature ($T_{\rm eff}$) of the star with the color scale indicating the surface gravity ($\log g$) of the star.} \label{fig:RVerrDist}
\end{figure*}

To fully understand the types of companions to which we are sensitive, we need a clear understanding of dependencies of the RV precision on stellar parameters. Therefore, we created an empirical model of the RV precision based on the primary derived stellar parameters ($T_{\rm eff}$, $\log g$, [Fe/H]) and the $S/N$ for each visit of the star:
\begin{eqnarray}
\begin{split}
\log \sigma_v = 1.56 + (4.87\times 10^{-5}) T_{\rm eff} + 0.135 \log g \\
- 0.518 [Fe/H] - (5.55\times 10^{-3}) S/N,
\end{split}
\label{eqn:RVprec}
\end{eqnarray}
where $S/N$ is the signal-to-noise ratio of the visit spectrum from which the RV measurement was derived, and $\sigma_v$ is the RV measurement error in m s$^{-1}$. This model was determined by fitting a linear function of each parameter of interest using all APOGEE stars with at least 8 visits, excluding stars used as telluric standards and stars that have unreliable stellar parameters. The left panel of Figure \ref{fig:RVerrDist} displays two of the stronger effects on RV error: [Fe/H] and $S/N$ per visit. The effects of $\log g$ and $T_{\rm eff}$ are illustrated in the right panel. These effects are closely related to the strength and number of absorption lines in the spectra. For a typical solar metallicity ([Fe/H]$ = 0$) giant ($T_{\rm eff} = 4000 \mathrm{K}$, $\log g = 3$) and typical solar metallicity dwarf ($T_{\rm eff} = 5000 \mathrm{K}$, $\log g = 4.5$) stars with $S/N = 10$, we derive a typical RV precision of $\sim$130 m s$^{-1}$ and $\sim$230 m s$^{-1}$, respectively per visit. These are the random RV uncertainties reported by the APOGEE pipeline, and are likely to be underestimates of the true uncertainty (see Appendix \ref{sec:caveats}).

\subsubsection{Selection of Usable RVs and RV Variable Stars} \label{sec:selRVs}
RV measurements from observations with $S/N < 5$, as well visits that produced failure conditions in the RV pipeline, were not included in the final RV curves submitted to the orbit fitter. This reduced the number of stars for which Keplerian orbits could be attempted from 14,840 to 9454 stars.

Likely RV variable stars were selected using the following statistic:
\begin{eqnarray}
\Sigma_{RV} = \mathrm{stddev}\left(\frac{\mathbf{v} - \tilde{v}}{\mathbf{\sigma_v}} \right) \ge 2.5, \label{eq:rvvar}
\end{eqnarray}
where $\mathbf{v}$ and $\mathbf{\sigma_v}$ are the RV measurements and their uncertainties, and $\tilde{v}$ is the median RV measurement for the star. The criterion was motivated by the false positive analysis presented in Appendix \ref{sec:simRes}. 
There are also several additional pieces of information that we used to pre-reject stars that would have resulted in poor or erroneous Keplerian orbit fits. Therefore we also removed stars with the following criteria: 
\begin{itemize}
\item The system's primary must be characterized with reliable stellar parameters ($T_{\rm eff}, \log g$, [Fe/H]), so the ASPCAP $\texttt{STAR\_BAD}$ flag must not be set for the star. Derivations of the RVs and the physical parameters of the system both rely on reasonable estimates of the stellar parameters of the host star. 
\item The star cannot have been used as a telluric standard. These stars are selected for APOGEE observation for their nearly featureless spectra, so it is likely that RVs derived for these stars are unreliable and would lead to false positive signals.
\item The combined spectrum from which the stellar parameters and RVs were derived cannot be contaminated with spurious signals due to poor combination of the visit spectra, so the \texttt{SUSPECT\_RV\_COMBINATION} flag must not be set for the star. This criterion also catches the double-lined spectroscopic binaries (SB2s) that would have resulted in poor stellar parameters, RVs, or orbital parameters from our current pipelines. 
\end{itemize}
This preselection reduced the number of stars for which Keplerian orbit fits were attempted from 9454 to 907. This is not to say the stars excluded do not have any sort of RV variation, but the false positive interpretation cannot be ruled out for these stars, so we elected not to include them. 

\subsection{Derivation of Primary Stellar Parameters} \label{sec:spDer}
To determine masses of potential companions, a reasonable estimate of the primary star's mass is required. The measurement of masses for the primary stars in this sample is based on the spectroscopic stellar parameters ($T_{\rm eff}$, $\log g$, [Fe/H]) derived for each star. Between $\texttt{apogeereduce}$ and ASPCAP, stellar parameters are derived up to three times for each source. The first approach uses the stellar parameters from the RV template selected for determining initial visit-level RVs. These parameters are available for every star, but are also the least precise of the three methods, so they should only be used as a last resort. The next set of stellar parameters made available are from the raw ASPCAP output. Except in the rare cases where ASPCAP fails to converge (which are removed from the final sample), these are available for all stars. Finally, calibrations are applied to the raw ASPCAP results based on comparisons with manual analysis of cluster stars \citep{Meszaros2013,Holtzman2015}. These parameters are only available for giant stars in a specific temperature range ($3500 < T_{\rm eff} < 6000$ K), but are the most reliable in absolute terms. To summarize, in order of preference, we adopted: (1) stellar parameters from the calibrated ASPCAP parameters, (2) uncalibrated ASPCAP parameters, (3) parameters used by the much coarser RV mini-grid. 

\begin{figure*}
\includegraphics[height = 0.5\textwidth, angle=-90]{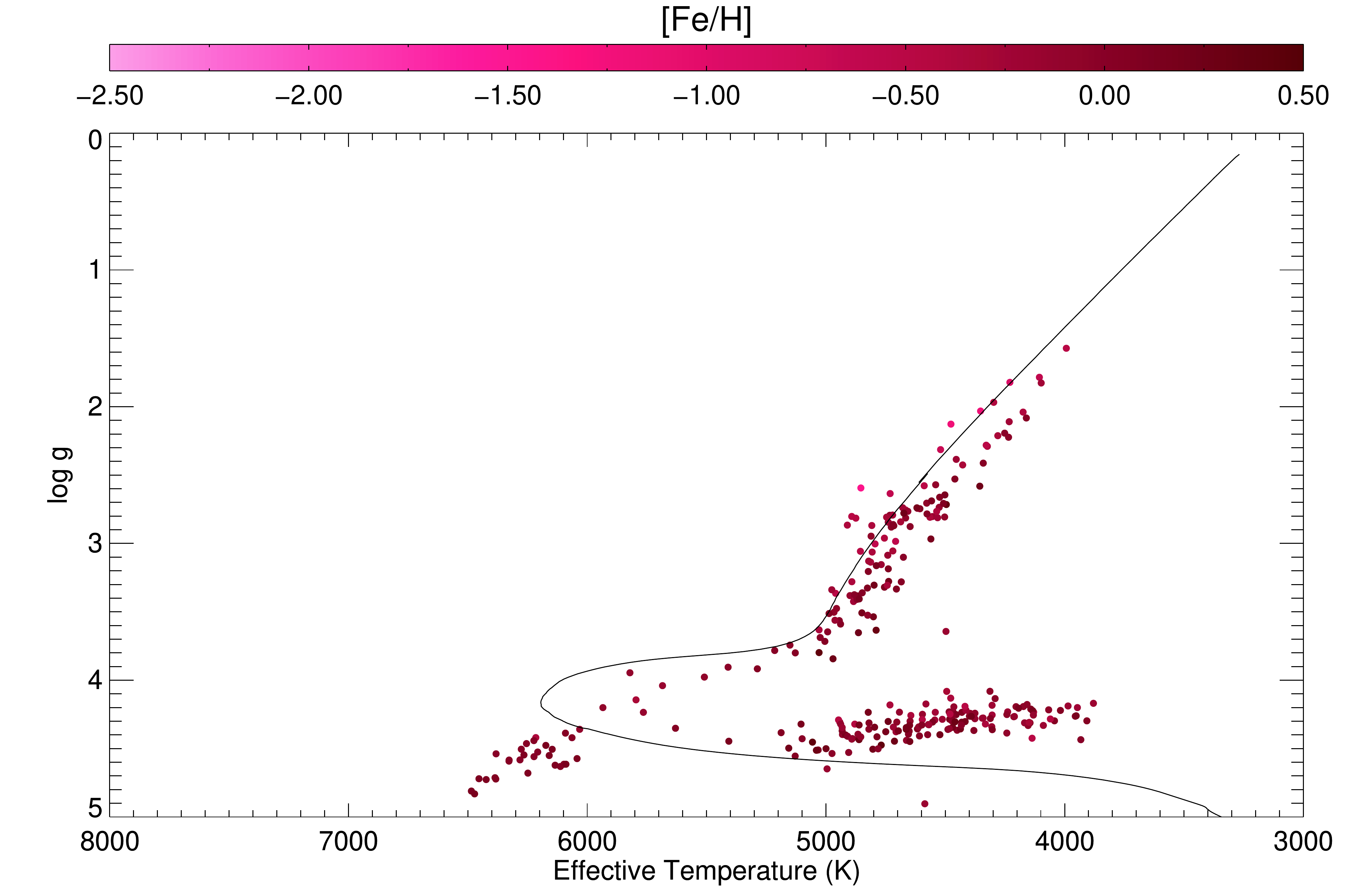}
\includegraphics[height = 0.5\textwidth, angle=-90]{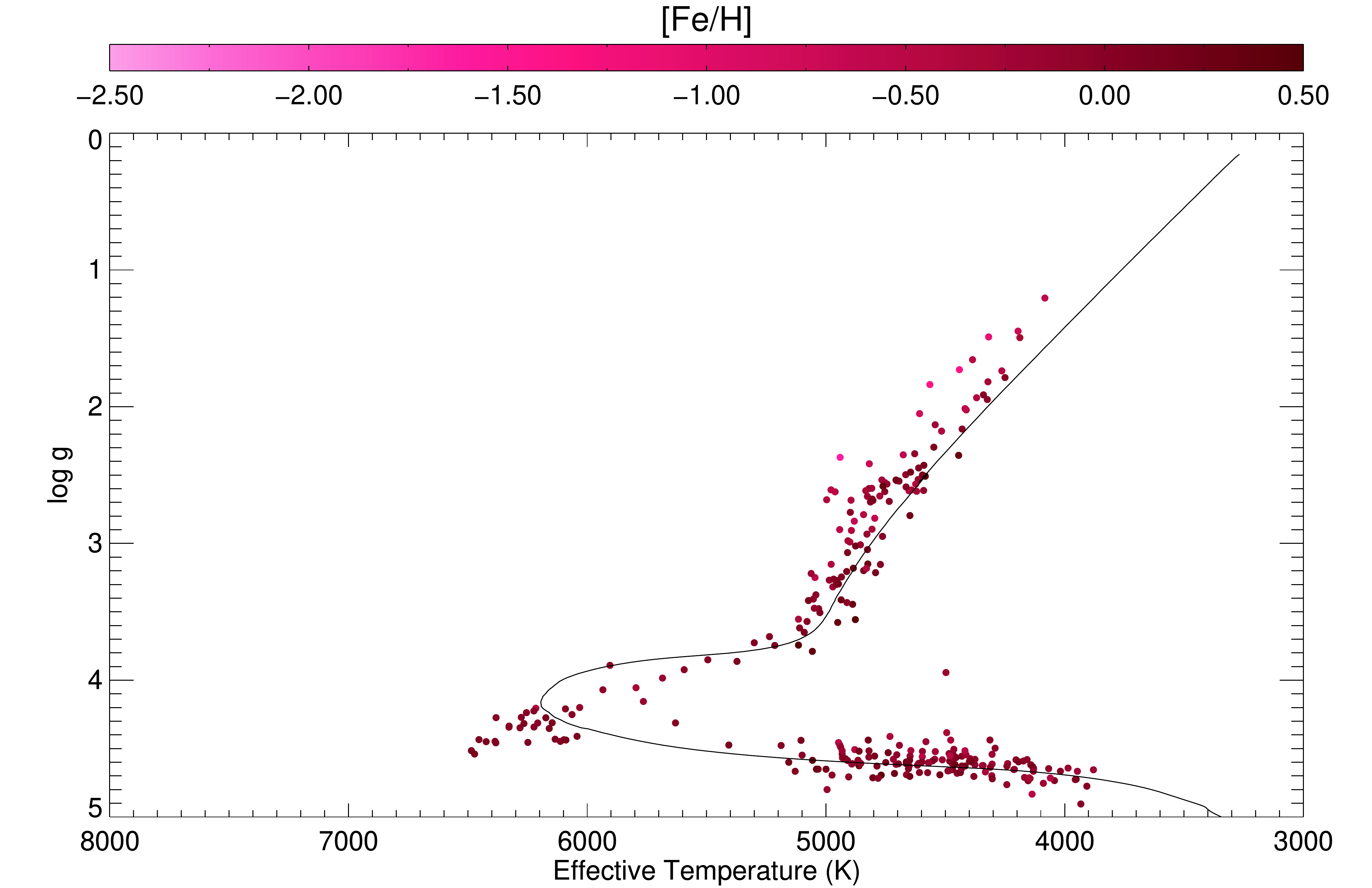}
\caption{Spectroscopic HR diagrams of stars in the field of M67 observed by APOGEE with the stars' $T_{\rm eff}$ and $\log g$ as the abscissa and ordinate. The points are color-coded by host star metallicity. A 5 Gyr solar-metallicity isochrone is also included for comparison. \textit{Left Panel:} Uncalibrated parameters (for both giants and dwarfs). Note the $\log g$ is underestimated by $\sim 0.5$ for stars at $T_{\rm eff} \sim 4000$ K. \textit{Right Panel:} Calibrated parameters, with giants using the ASPCAP calibrated parameters and dwarfs adopting the $\log g$ correction from Equation \ref{eq:logg_cor}. }
\label{fig:dwarfCal}
\end{figure*}

All of the dwarfs in this catalog rely on uncalibrated parameters. Unfortunately this leads to systematically overestimated $\log g$ values for cool dwarfs when compared to Dartmouth isochrones (Figure \ref{fig:dwarfCal}). We apply a simple linear correction to calibrate dwarf $\log g$ values:
\begin{eqnarray}
(\log g)_{cal} = \log g - (3\times 10^{-4})(T_{\rm eff} - 5500 \, \mathrm{K}),
\label{eq:logg_cor}
\end{eqnarray}
where $\log g$ and $T_{\rm eff}$ are the uncalibrated suface gravity and effective temperature. The results of this calibration can be seen in Figure \ref{fig:dwarfCal}.

\subsubsection{Primary Star Classification} \label{sec:hostStarSort}
\begin{figure}[h!]
\includegraphics[width=\columnwidth]{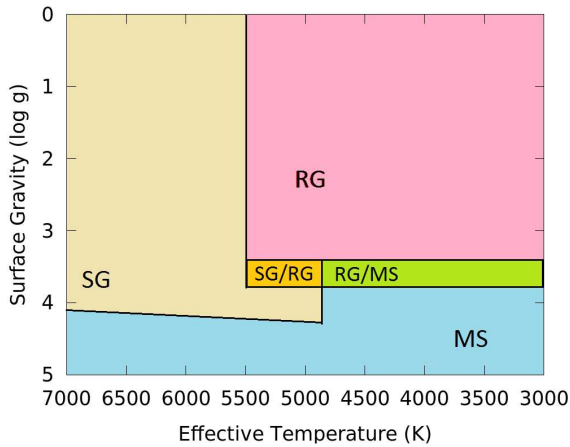}
\caption{Classification scheme of Red Giant (RG), Subgiant (SG), and main-sequence dwarf stars (MS) in $\log g$ -- $T_{\rm eff}$ space. Red Clump (RC) and pre-main sequence stars (PMS) transcend these boundries as they selected through alternate means. The areas labeled with SG/RG or MS/RG are regions where the star can be either classification depending on its metallicity. The upper left corner of this plot does not contain any stars in this sample, so the SG classification there is simply in place to cover the phase space.} \label{fig:primClass}
\end{figure}

Before any further stellar properties are estimated, we divide the stars in this sample into 5 classes defined by the following crteria:
\begin{enumerate}
\item \textbf{Pre-Main Sequence (PMS):} Stars flagged in APOGEE as young stellar cluster members (IC348 and Orion).
\item \textbf{Red Clump (RC):} Stars in the APOGEE RC Catalog \citep{Bovy2014}.
\item \textbf{Red Giant (RG):} Stars not selected as RC or PMS stars with
\begin{eqnarray*}
T_{\rm eff} &<& 5500 \, \mathrm{K},\\
\log g &<& 3.7 +0.1\mathrm{[Fe/H]}.
\end{eqnarray*} 
The second relation was derived by mapping the $\log g$ of the base of the giant branch as a function of [Fe/H] from Dartmouth isochrones \citep{Dotter2008} for typical ages expected of APOGEE giants.
\item \textbf{Subgiant (SG):} Stars not selected as RC or PMS stars with 
\begin{eqnarray*}
T_{\rm eff} &>& 4800  \, \mathrm{K}, \\
\log g &\ge& 3.7 +0.1\mathrm{[Fe/H]}, \\
\log g &\le& 4 - (7 \times 10^{-5})(T_{\rm eff}-8000\,\mathrm{K}). 
\end{eqnarray*}
The second relation only applies for $T_{\rm eff} < 5500 \, \mathrm{K}$. The third relation was determined by the $\log g$ at the highest $T_{\rm eff}$ of Dartmouth isochrones at a variety of ages and [Fe/H], roughly mapping the main-sequence turnoff (MSTO), and fitting a liner function to these points.
\item \textbf{Dwarf (MS):} Any star that does not fit into any of the above categories are classified as MS stars. 
\end{enumerate}
These classifications are saved for the catalog, and illustrated in Figure \ref{fig:primClass}.

\subsubsection{Derivation of Bolometric Magnitudes}
In addition to stellar parameters, we need an estimate of the stars' bolometric magnitudes to compare to the bolometric luminosities we calculate and use in the following derivations of the masses and radii of the primary stars. We adopt the extinction coefficient, $A_K$ from the APOGEE targeting data \citep{Zasowski2013}. If the APOGEE targeting $A_K$ is not populated or is less than zero, then we adopt the WISE all-sky K-band extinction. In the rare case ($<1\%$ of stars run through the $\texttt{apOrbit}$ pipeline) that neither quantity is available, we assume $A_K = 0$, and flag the star. The extinction-corrected $K_s$ magnitude is then $K_0 = K_s - A_K$. We derived the bolometric correction to the 2MASS $K_s$ band from Dartmouth isochrones: 
\begin{equation}
\begin{split}
BC_K =& (2.7+0.15\mathrm{[Fe/H]}) \\
-& (25+0.5\mathrm{[Fe/H]})X^{2-0.1\mathrm{[Fe/H]}}e^{-X} \\
\end{split}
\end{equation}
for PMS, dwarf and SG stars, where $X = \log T_{\rm eff} -3.5$, and
\begin{eqnarray}
BC_K = (6.8 - 0.2\mathrm{[Fe/H]})(3.96 - \log T_{\rm eff})
\end{eqnarray}
for RG and RC stars. This correction yields the bolometric magnitude of the star: $m_{bol} = K_0 + BC_K$.

\subsubsection{Derivation of Dwarf and Subgiant Primary Mass, Radius, and Distance}
For stars selected as dwarf and subgiant stars, we adopted the \cite{Torres2009} relations to estimate the mass and radius of the primary star:
\begin{equation}
\begin{split}
\log M_{\star} = a_1 + a_2X + a_3X^2 + a_4X^3 \\
 + a_5(\log g)^2 + a_6(\log g)^3 + a_7\mathrm{[Fe/H]},
\end{split}
\end{equation}
\begin{equation}
\begin{split}
\log R_{\star} = b_1 + b_2X + b_3X^2 + b_4X^3 \\
 + b_5(\log g)^2 + b_6(\log g)^3 + b_7\mathrm{[Fe/H]},
\end{split}
\end{equation}
where $X = \log T_{\rm eff} - 4.1$ and the coefficients, $a_{i}$ and $b_{i}$ are given in Table 4 of \cite{Torres2009}. This empirical relationship has a scatter of $6.4\%$ in mass and $3.2\%$ in radius, so for dwarfs and subgiants, we adopt $\sigma_{M_{\star}} = 0.064M_{\star}$ as the uncertainty in the mass, and $\sigma_{R_{\star}} = 0.032R_{\star}$. This information allows one to estimate the luminosity, $L_{\star}$, as well as the distance,$d$, to these stars: 
\begin{eqnarray}
L_{\star} = 4\pi R_{\star}^2 \sigma_{SB} T_{\rm eff}^4 \\
M_{bol} = 4.77-2.5 \log \left(\frac{L_{\star}}{L_{\odot}} \right) \\
d = 10^{1+0.2(m_{bol}-M_{bol})},
\end{eqnarray}
where $M_{bol}$ is the star's absolute bolometric magnitude. Uncertainty for these parameters are also derived through normal propagation of uncertainties, which yields a $13.5\%$ typical distance uncertainty for dwarfs and subgiants. A total of 340 of the 907 stars for which fitting was attempted used this prescription.

Unfortunately, the \cite{Torres2009} relations are not applicable to giant and pre-main sequence (PMS) stars. For example, using the \cite{Torres2009} relations to derive the mass of Arcturus ($T_{\rm eff} = 4286$ K, $\log g = 1.66$, [Fe/H] = -0.52) yields a mass of $3.5 M_{\odot}$ compared to the accepted mass of $1.08 M_{\odot}$ \citep{Ramirez2011}. Therefore, we must resort to alternate methods for estimating the mass of the primary.

\subsubsection{Derivation of Giant and Pre-Main Sequence Primary Mass, Radius, and Distance} \label{sec:giantPrim}
Efforts are currently underway to compile all published (or soon-to-be published) distance measurements to APOGEE stars. For stars selected as RG and RC stars, we employ a preliminary version of this distance catalog as the basis for our mass derivation. The most accurate distances for APOGEE stars are those derived from asteroseismic parameters from the APOGEE-Kepler catalog \citep[APOKASC;][]{Pinsonneault2014}.  These distances were given first priority because they only have $\sim 2\%$ random errors \citep{Rodrigues2014}. Unfortunately, no stars in this sample matched APOKASC stars with distance measurements, but we include it in the pipeline in hopes that future versions of the APOKASC catalog will overlap with future versions of this catalog. Our second choice, if the star is a RC star, is to use distances derived from the APOGEE RC catalog. These distances are cited to have $5-10\%$ random errors, and 71 stars of the 907 run through the $\texttt{apOrbit}$ pipeline are RC stars. If the star has neither of the above distances available, we adopt the spectrophotometic distance estimates derived by \cite{Santiago2015}, \cite{Hayden2015}, or \cite{Schultheis2014}, based on which estimate has the lowest error. These distances generally have $<15-20\%$ uncertainties, and for most of the RG stars run through the \texttt{apOrbit} pipeline (489 stars), we adopt these distances. The six PMS stars in this sample are located in the young cluster IC348 \citep[$d = 316 \pm 22$ pc;][]{Herbig1998}, so we adopt the distance to this cluster as the approximate distance to these stars. From the adopted distance, $d$, we estimate the luminosity of the star, and thus its radius and mass:
\begin{eqnarray}
M_{bol} &=& m_{bol} - 5\log(d) + 5 \\
L_{\star} &=& 10^{-0.4(M_{bol}-4.77)} L_{\odot} \\
R_{\star} &=& \sqrt{\frac{L_{\star}}{4\pi\sigma_{SB} T_{\rm eff}^4}} \\
M_{\star} &=& \frac{10^{\log g}R_{\star}^2}{G}
\end{eqnarray}
Following typical propagation of uncertainties, these techniques produce a mass uncertainty floor of $26\%$ due to the uncertainty in $\log g$. The median of mass uncertainties for these techniques is around $28\%$.

If a giant star has no distance measurement available, we adopt a characteristic mass from a \texttt{TRILEGAL} \citep{Girardi2005} simulation using parameters typical of APOGEE giants. The median mass for all stars in this simulation with $\log g <3.8$ and $3500$ K$ < T_{\rm eff} < 5000$ K in the direction of Galactic Coordinates ($\ell, b)$ =(0,40) is $M_{\star} = 1.6 \pm 0.6 M_{\odot}$ ($\sim 40\%$ mass uncertainty), which we adopt as the typical mass for all giant stars without a distance measurement. From this we derive $R_{\star} = (GM_{\star}/10^{\log g})^{1/2}$ and $d$, as for the dwarfs, both with typical estimated uncertainties of $25\%$. Fortunately, we only need to adopt this type of mass estimate for one star run through the $\texttt{apOrbit}$ pipeline.

\subsection{Keplerian Orbit Fitting} \label{sec:orbFit}

Once a star has mass and radius estimates, we can attempt to search for periodic signals and derive Keplerian orbits from its RV measurements. Only stars with at least eight ``good'' visits have enough degrees of freedom to attempt the six and seven parameter Keplerian orbit fits. For each star meeting this criterion, we attempt orbital fits with and without a long-term underlying linear trend. The linear fit accounts for additional long-term RV variability that may be indicative of an additional companion with a period longer than we can detect reliably, or long-term instrumental effects. 

\subsubsection{Period-Finding and Selection of Initial Conditions}
We employ the Fast $\chi^2$ Period Search (F$\chi^2$) algorithm \citep{Palmer2009} to search for periodic signals. This algorithm chooses the period based on the largest reduction in $\chi^2$ between a sinusoidal fit employing the first $n_h$ harmonics of a fundamental period, $p_i$, compared to a global $n_d$-degree polynomal fit.  The F$\chi^2$ algorithm uses harmonics of the fundamental period in its fits, which produces improved performance with non-circular orbits compared to the traditional Lomb-Scargle algorithm \citep{Scargle1982}. Another advantage of the F$\chi^2$ algorithm is a built-in avoidance of periodic signals introduced by the cadence of the data, i.e., inputting data taken every $n$ days will not return a $n$-day period as the best fit.

For our purposes, we employ three harmonics ($n_H = 3$), execute a search in four (logarithmic) period bins (0.3 to 3 Days, 3 to 30 days, 30 to 300 days, and 300 to 3000 days), and oversample ten times the default frequency sampling such that the frequency step is $\Delta f = 1/(10n_h\Delta T)$, where $\Delta T$ is the longest temporal baseline of the observations. The search is executed once with a constant ($n_d = 0$) fit and once with a linear fit ($n_d=1$). The periods in each bin, $p_j$ that produce the greatest reduction in $\chi^2$, $\Delta \chi^2_{max}$, are then assessed for their significance using the following criterion:
\begin{eqnarray}
P_{n-2n_h}(\Delta \chi^2_{max}) \ge 0.997,
\end{eqnarray}
where $P_{n-2n_h}(\Delta \chi^2_{max})$ is the probability for a $\chi^2$ distribution with $n-2n_h$ degrees of freedom, and $n$ is the number of RV epochs. The above limit is the equivalent of a $3\sigma$ detection. Periods that are not deemed significant by this metric are not used for full Keplerian orbit fitting. The significant periods ($p_j$) and their harmonics (1/3, 1/2, 2, and 3 times each value of $p_j$) are then each used for Keplerian orbit fitting.

\subsubsection{Derivation of Keplerian Orbits} \label{sec:KepOrbits}
Once the best periods are identified, Keplerian models with those periods are fit to the RV measurements using the $\texttt{MPFIT}$ algorithm \citep{Markwardt2009}. $\texttt{MPFIT}$ is a Levenberg-Marquardt non-linear least squares fitter implemented in IDL. This code is wrapped in an IDL code $\texttt{MP\_RVFIT}$ used in the MARVELS survey \citep{DeLee2013}. $\texttt{MP\_RVFIT}$ takes the input period and searches parameter space of the other Keplerian orbital parameters ($K,e,\Omega, T_p,$, and global velocity trends) and returns the Keplerian model that satisfies the period with the lowest $\chi^2$.

\begin{figure*}
\includegraphics[height=0.33\textwidth, angle=-90]{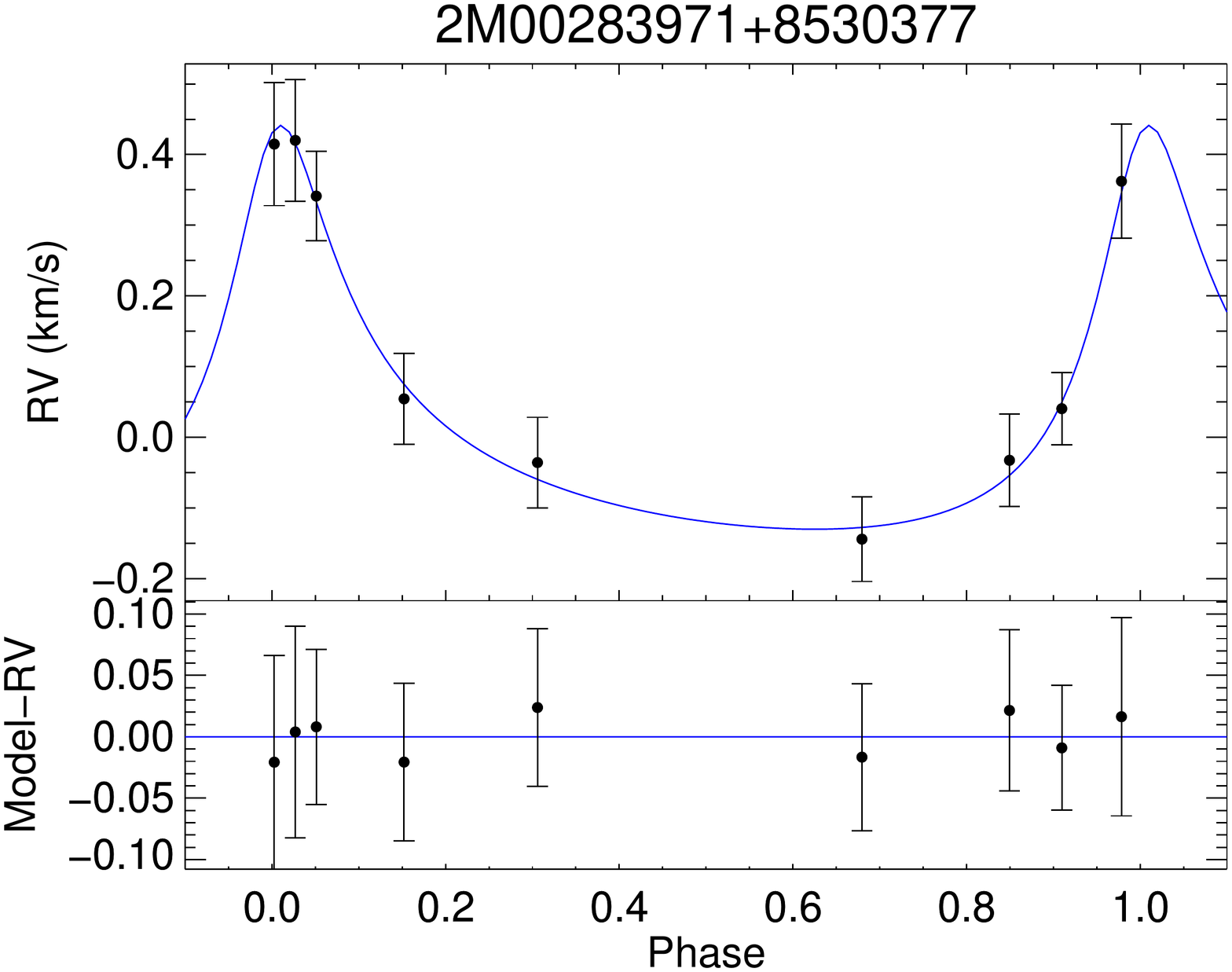}
\includegraphics[height=0.33\textwidth, angle=-90]{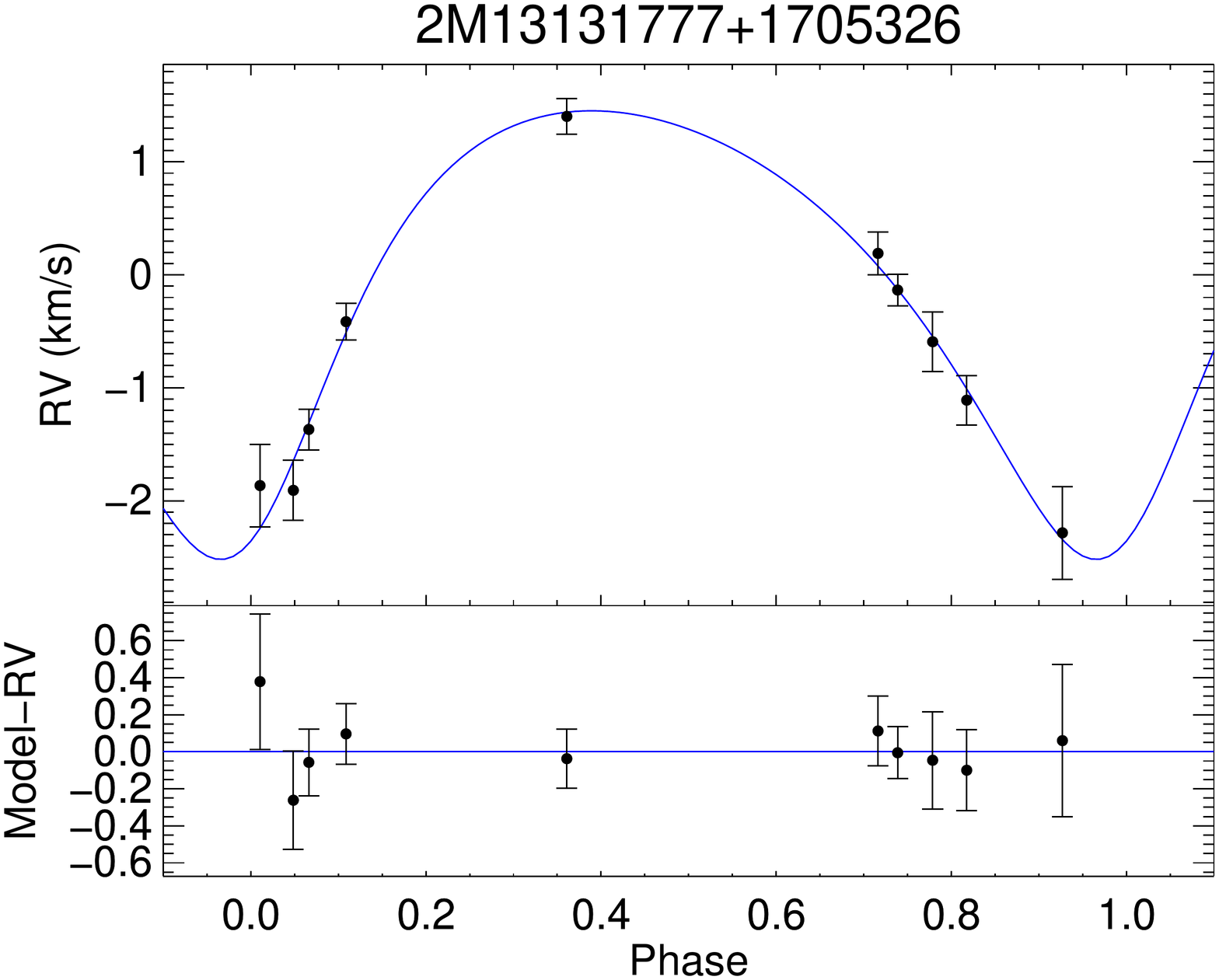}
\includegraphics[height=0.33\textwidth, angle=-90]{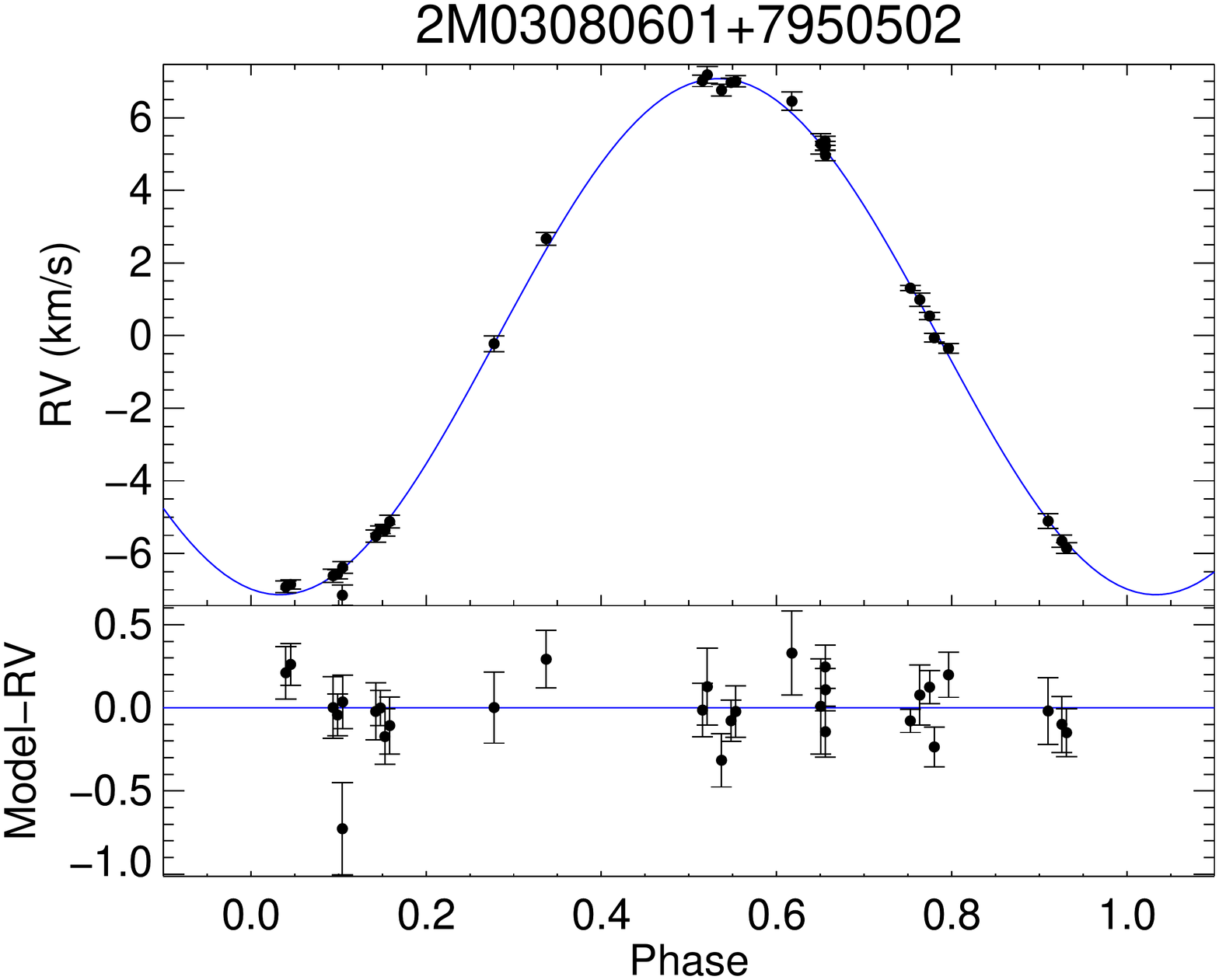}
\caption{RV curves for a few example systems. In each plot, the top panel presents the phased RV measurements with a line showing the best fit model and the bottom panel shows the residuals of the fit. Similar figures are available online for every star in the gold sample (see Appendix \ref{sec:data}). \textit{Left Panel:} A planetary-mass ($m\sin i = 4.60 M_{Jup}$) companion in a $P=41.3$ day, $a = 0.25$ AU orbit with $e = 0.566$, and $K = 0.29$ km s$^{-1}$. This orbit has uniformity index (See \S \ref{sec:fitCrit}) values of $U_N = 0.886$ and $V_N = 0.737$. \textit{Middle Panel:} A BD-mass companion ($m\sin i = 22.6 M_{Jup}$) companion in a $P=24.3$ day, $a = 0.15$ AU orbit with $e = 0.293$, $K = 1.99$ km s$^{-1}$. This orbit has uniformity index values of $U_N = 0.871$ and $V_N = 0.935$. \textit{Right Panel:} Binary System with a $m\sin i \approx 0.304 M_{\odot}$ secondary in a $P=184$ day, $a = 0.68$ AU orbit with $e = 0.004$, $K = 7.11$ km s$^{-1}$. This orbit has uniformity index values of $U_N = 0.937$ and $V_N = 0.869$.} 
\label{fig:exRVcurves}
\end{figure*}

Having a precise period is extremely important for acquiring an accurate Keplerian model, and simply submitting the periods from the period-finding algorithm to $\texttt{MP\_RVFIT}$ often leads to unsatisfactory results. Here we describe the bisector method implemented to achieve the best possible period. We initially submit the periods described above to $\texttt{MP\_RVFIT}$, and keep the three periods ($p_{k,0}$) that produce the best fits based on the modified reduced-chi-squared goodness of fit statistic, $\chi_{mod}^2$, described in \S \ref{sec:fitCrit}. For each of these periods we implement a bisector method to narrow in on the exact period. For each $p_k$ we run $\texttt{MP\_RVFIT}$ with three periods: $p_{k,0}$ and $p_{k,0} \pm \Delta p_{0}$, where $\Delta p_{0} = 0.5 p_{k,0}$. We then compare the $\chi^2_{mod}$ for the best fits for the three periods, and update $p_{k}$ and $\Delta p$ accordingly:
\begin{eqnarray}
\begin{split}
& \mathrm{If} \: \chi_{p_{k,i}}^2 \le \chi_{p_{k,i}\pm\Delta p_{i}}^2:  \\
& \qquad  p_{k,i+1} = p_{k,i}, \  \Delta p_{i+1} = \Delta p_{i}/2,
\end{split}\\
\begin{split}
& \mathrm{If} \: \chi_{p_{k,i}\pm \Delta p_{i}}^2 < \chi_{p_{k,i}}^2: \\
& \qquad p_{k,i+1} = p_{k,i} \pm \Delta p_{i}, \ \Delta p_{i+1} = \Delta p_{i},   
\end{split}
\end{eqnarray}
For the $\chi_{p_{i} - \Delta p_{i}}^2 < \chi_{p_{i}}^2$ case, if $p_{i} - 2\Delta p_{i} < 0.1$, then we use $\Delta p_{i} = \Delta p_{i}/2$ for the next update. This iteration is performed until the change in $\chi^2_{mod}$ is less than 0.01 or $n_{iter} = 50$ iterations are reached. \textcolor{black}{The distribution of the required number of iterations for systems in the final sample had a median of 15 with few systems above 25. Therefore, the choice to terminate systems on their 50th iteration is more than justified as these systems are unlikely to converge in a timely manner. These systems are also not included in the final catalog (see \S\ref{sec:selGold}).} The final values of $p_{k,n_{iter}}$ are then submitted to $\texttt{MP\_RVFIT}$ one final time, and the results saved for the catalog. The data saved are described in \S \ref{sec:Catalog}. For a few example Keplerian orbit models see Figure \ref{fig:exRVcurves}.

\subsubsection{From Orbital to Physical Parameter Estimates}

Directly from the orbital parameters, we can calculate the projected semi-major axis of the primary star:
\begin{eqnarray}
a_{\star} \sin i = \frac{KP}{2\pi} \sqrt{1-e^2}.
\end{eqnarray}
From this measurement we can define the mass function of the system:
\begin{eqnarray}
f(m,M_{\star}) = 4\pi^2 \frac{(a_{\star} \sin i)^3}{GP^2} = \frac{(m \sin i)^3}{(M_{\star}+ m)^2}.
\end{eqnarray}
This quantity is saved in the catalog, but we also attempt to estimate the secondary mass directly:
\begin{eqnarray}
m \sin i = \left[f(m,M_{\star}) M_{\star}^2 (1+(m/M_{\star}))^2\right]^{1/3}
\end{eqnarray}
The general case of this equation cannot be solved analytically, but often when dealing with planetary companions, we can make the assumption that $m \ll M_{\star}$, and thus can make the approximation $m \sin i \approx (f(m,M_{\star}) M_{\star}^2)^{1/3}$. For companions with $m \sin i < 0.1 M_{\star}$, this approximation is accurate to within 10\%, but this sample contains higher-mass companions for which we want reasonable mass estimates. In these cases we solve the above equation iteratively, initially assuming $m = 0$, returning the above estimate, and iterating until $m\sin i$ changes by $< 10^{-4} M_{\odot}$. Since we are interested in estimating the minimum mass of the companion, we solve for the $\sin i = 1$ case, and thus use $m \approx m\sin i$ after the first iteration. This iterative method for determining $m$ was tested for a variety of mass ratios and a variety of starting points for $m$ (not just $m=0$). From these tests, we have found this method to be quite robust.

Finally, from the estimate of $m\sin i$, we provide an estimate of the semimajor axis, $a$, of the secondary:
\begin{eqnarray}
a = a_{\star} \sin i \frac{M_{\star}}{m \sin i}.
\end{eqnarray}

\subsubsection{Quality Control and Selection of Best Fits} \label{sec:fitCrit}

Finally we compile the three best models from the run with no global linear fit and the three best models from the linear fit run, and compare them to select the best overall fit. Ideally the phase and velocity coverage of the model are uniformly sampled by the data, and we aimed to preferably select models that are as close to this ideal as possible.  A useful way to quantify the phase coverage of the data is the uniformity index \citep{Madore2005}:
\begin{equation}
U_{N} = \frac{N}{N-1} \left[ 1 - \sum_{i=1}^{N} \left( \phi_{i+1} - \phi_{i} \right)^2, \right]
\end{equation}
where the values $\phi_i$ are the sorted phases associated with the corresponding Modified Julian Date (MJD) of the measurement $i$, and $\phi_{N+1} = \phi_1 + 1$. This statistic is normalized such that $0 \le U_N \le 1$, where $U_N = 1$ would indicate a curve evenly sampled in phase space.  Using a similar derivation, we also define an analogous ``velocity" uniformity index with the same properties as $U_N$:
\begin{equation}
V_{N} = \frac{N}{N-1} \left[ 1 - \sum_{i=1}^{N} \left( \nu_{i+1} - \nu_{i} \right)^2 \right].
\end{equation}
We define a ``velocity phase,'' $\nu_i = (v_i - v_{min})/(v_{max}-v_{min})$, 
to have the same properties as $\phi_{i}$ above, where the values $\nu = 0$ and $\nu = 1$ indicate the minimum and maximum velocities of the \textit{model}, $v_{min}$ and $v_{max}$. The values of $v_{i}$ are the radial velocity measurements, sorted by their value, with the adopted global velocity trend subtracted. For models that do not apply a global linear trend, the trend subtracted is the average of the raw velocities: $v_{i} = v_{raw,i} - \bar{v}_{raw}$. Measured velocities below the minimum or above the maximum are assigned $\nu = 0$ and $\nu = 1$, respectively. The purpose of this metric is to prevent the pipeline from selecting an extremely eccentric orbit when the data do not support such a model. Values of $U_N$ and $V_N$ are given in the example RV curves of Figure \ref{fig:exRVcurves}.

Combining the above statistic with the traditional reduced $\chi^2$ goodness-of-fit statistic ($\chi^2_{red}$), we define the modified $\chi^2$ statistic, 
\begin{eqnarray}
\chi_{mod}^2 = \frac{\chi_{red}^2}{\sqrt{U_N V_N}}, 
\end{eqnarray}
by which the models are ranked. In the case that $U_N=0$ or $V_N=0$, $\chi^2_{mod}$ would be recorded as a floating-point infinity and automatically be ranked below all other fits. However, there are some conditions where the fit is unacceptable, but still may be selected as the best fit using the above metric. Therefore, we defined criteria that split the fits into ``good'' and ``marginal'' fits. Any of the following criteria would warrant a ``marginal'' classification:
\begin{itemize}
\item Periods within 5\% of 3, 2, 1, 1/2, or 1/3 day,
\item Periods, $P$, longer than twice the baseline, $2\Delta T$,
\item Extremely eccentric solutions ($e > 0.934$)\footnote{This is the eccentricity of HD 80606 b, the largest eccentricity in the \url{exoplanets.org} database},
\item Orbital solutions that send the companion into the host star: $a(1-e) < R_{\star}$,
\item Poor phase and velocity coverage ($U_N V_N < 0.5 $).
\end{itemize}
The good and marginal fits are ranked by $\chi_{mod}^2$ separately, and the best fit is the good fit with the lowest $\chi^2_{mod}$. If all of the fits were deemed marginal, then the best fit is the marginal fit with the lowest $\chi^2_{mod}.$ For more details on the verification and performance of the $\texttt{apOrbit}$ pipeline, see Appendix \ref{sec:Verification}.

\section{Building the APOGEE Candidate Companion Catalog} \label{sec:Catalog}
A total of 907 stars were successfully run through the \texttt{apOrbit} pipeline. Of these, the F$\chi^2$ algorithm found significant periodic signals for 749, which were submitted for full Keplerian orbit fitting. In this section, we describe the data available for these stars, and the selection of companion candidates from the best Keplerian orbit fit to these stars. Information on catalog content and access can be found in Appendix \ref{sec:data}.

\subsection{Selecting Statistically Significant Astrophysical RV Variations} \label{sec:selRvvar}
In many cases, the RV variations are within the measurement errors, so the derived semi-amplitude for the orbit may be masked by measurement error. In these cases, we cannot reliably state that the RV variations are astrophysical in nature. However, even astrophysical RV variations may not be due to the presence of a companion. Many stars, especially giant stars, which compose a large part of this sample, can have high levels of intrinsic RV variability. To estimate this stellar RV jitter, we adopted the relation found by \cite{Hekker2008}: 
\begin{eqnarray}
v_{jitter} = 2(0.015)^{\frac{1}{3} \log g} \ \mathrm{km} \, \mathrm{s}^{-1},
\end{eqnarray}
where, again, $\log g$ is the logarithm of the surface gravity in cgs units. We define a total RV uncertainty for each point in the model fit by combining this quantity with the RV measurement uncertainties, $\sigma_v$:
\begin{eqnarray}
v_{unc} = \sqrt{\sigma_v^2 + v_{jitter}^2}.
\end{eqnarray}

We use the following criteria to select statistically significant companion candidates:
\begin{eqnarray}
\frac{K}{\tilde{v}_{unc}} \ge 3+3(1-V_N)e, 
\end{eqnarray}
where $\tilde{v}_{unc}$ is the median RV uncertainty of the model fit, $K$ is the RV semi-amplitude of the best-fit model for the star, and $V_N$ is the velocity uniformity index described in \S\ref{sec:fitCrit}. We include the $(1-V_N)e$ term to increase the significance criteria for eccentric systems, particularly those that have poor velocity coverage. Thus, a perfectly covered eccentric orbit ($V_N$ = 1) would be treated the same as a circular orbit ($e=0$).  Using these criteria, 698 stars are selected as statistically significant companion candidates.

\subsection{Refining the Catalog: Defining The Gold Sample} \label{sec:selGold}
In an effort to minimize the number of false positives in this sample and reduce the number of systems with incorrectly-derived orbital parameters (see Appendix \ref{sec:Verification}), we eliminate candidates that do not satisfy the following criteria:

\begin{itemize}
\item None of ``marginal fit'' criteria described in \S \ref{sec:fitCrit} are met.
\item The Keplerian fits must be reasonably good, which we quantify as the criteria:
\begin{eqnarray}
\frac{K}{|\Delta \widetilde{v}|}  &\ge& 3+3(1-V_N)e, \\
\chi^2_{mod} &\le &  \frac{K/|\Delta \widetilde{v}|}{3+3(1-V_N)e},\\
\chi^2_{mod} &\le &  \frac{K/\widetilde{v}_{unc}}{3+3(1-V_N)e},
\end{eqnarray}
where $|\Delta \widetilde{v}|$ is the median absolute residuals of model fit.
From simulations and visual inspection of orbits, orbits with large median $K/v_{unc}$ or $K/\Delta v$ reproduced the correct parameters and had reasonable fits at much larger values of $\chi^2_{mod}$ than orbits with lower values. A major exception to this trend were large $K/\widetilde{v}_{unc}$ orbits due to high $e$ or orbits with poor velocity sampling (low $V_N$), so the metric above includes terms to penalize fits with high eccentricity ($1-e$ term) or low $V_N$ (which inflates $\chi_{mod}^2$) Therefore this ``good fit" limit is stricter for such systems by employing the $\chi^2_{mod}$ metric discussed above. Previous cuts also guaranteed that no systems with $\chi^2_{mod} \le 1$ are excluded because of this metric.
\item The best fit must not require the maximum number of period iterations to converge; as described in \S \ref{sec:KepOrbits}. Systems that reach that maximum limit of iterations in the fitter did not converge on a solution, and the orbital parameters output are likely to be unreliable.
\end{itemize}
As mentioned above, many of these criteria were inspired by the testing of simulated systems with known orbital parameters described in Appendix \ref{sec:Verification}. Using these refined criteria, 382 stars (55\% of the statistically significant RV variable sample) were selected to be a part of the ``gold sample,'' which represent the best-quality companion candidates detected by APOGEE. This is not to say that the other $45\%$ of the statistically significant RV variable sample do not have companions, and there very well may be accurately reproduced companions from the non-gold sample. However, the likelihood of either false positives or poorly-characterized systems is much higher for the non-gold sample than for the gold sample, hence we only present the 382 stars in the gold sample here.

\section{Census of Gold Sample Companion Candidates and Discussion of Initial Results} \label{sec:Results}
In this section, we present a census of the 382 companion candidates in the catalog. \textit{Of these, 376 are newly discovered small separation companion candidates.} Table \ref{tab:Census} provides a broad overview of the distributions of the companion candidates in terms of companion type (planet, BD or binary), host star type (e.g., giant vs. dwarf), and approximate Galactic environment (disk versus halo). We discuss each of these distributions and their implications in more detail in the subsections below. From this point on, we use $\langle m \rangle$ to indicate the maximum-likelihood value of the companion mass, $m$, based on the expectation value of $i$, defined as $\langle\sin i\rangle = \int_0^{\pi/2} P(i)\sin i \, di = \int_0^{\pi/2} \sin^2 i \, di = \pi/4$. Therefore, $\langle m\rangle = (4/\pi) m\sin i$, and we use this number to differentiate between companion types to account for inclination effects in a statistical manner.

\begin{deluxetable}{rcccc}
\centering
\tablewidth{0.98\columnwidth}
\tabletypesize{\scriptsize}
\tablecaption{A Census of APOGEE Gold Sample Companion Candidates}
\tablecolumns{5} 
\tablehead{\colhead{Population} & \colhead{Binaries\tablenotemark{a}} & \colhead{BDs\tablenotemark{b}}&  \colhead{Planets\tablenotemark{c}} & \colhead{Total}}
\startdata
\sidehead{\textbf{Host Star Classification}\tablenotemark{d}}
Red Clump (RC)  & 18 & 5 & 0& 23\\
Red Giant (RG) & 115  & 56 & 9 & 180  \\
Subgiant (SG) &  9  & 10  &  3  & 22  \\
Dwarf (MS) & 71 & 41 & 45 & 157 \\
PMS & 1 & 1 & 0 & 2 \\
\sidehead{\textbf{Host Star Metallicity}}
[Fe/H] $\ge 0$  & 70 & 36  & 13 & 119 \\
$-0.5 \le$ [Fe/H] $< 0$ & 118 & 62 & 42 & 222  \\
$\textrm{[Fe/H]} < -0.5$ & 25 & 14  & 2 & 41  \\
\sidehead{\textbf{Galactic Environment}\tablenotemark{f}}
Thin Disk  & 180 & 91 & 56 &  327\\
Thick Disk  & 31 & 18 & 1 &  50\\
Halo  & 2 & 3 & 0 &  5\\
\hline\\
\textbf{Catalog Totals} & 213  &112  & 57 & 382
\enddata
\tablenotetext{a}{We define likely stellar-mass binaries as having a companion with $ \langle m \rangle > 0.08 M_{\odot}$}
\tablenotetext{b}{Brown dwarf companions: $0.013 M_{\odot} < \langle m \rangle \le 0.08 M_{\odot}$}
\tablenotetext{c}{Planetary-mass companions: $\langle m \rangle \le 0.013 M_{\odot}$}
\tablenotetext{d}{Host star classification and abbreviations discussed in \S \ref{sec:hostStarSort}}
\tablenotetext{f}{To truly distinguish between Thin and Thick Disk populations, a full analysis of the chemistry and kinematics of the stars would be needed. Here we simply present a census of companion as a function of height above the midplane, and use these criteria: Thin Disk = $|Z| < $ 1 kpc, Thick Disk = $1\, \mathrm{kpc} \le |Z| <  5\, \mathrm{kpc}$, Halo = $|Z| \ge $ 5 kpc.}
\label{tab:Census}
\end{deluxetable}

\subsection{Orbital Distribution of Companion Candidates}
\begin{figure*}
\centering
\includegraphics[width=\textwidth]{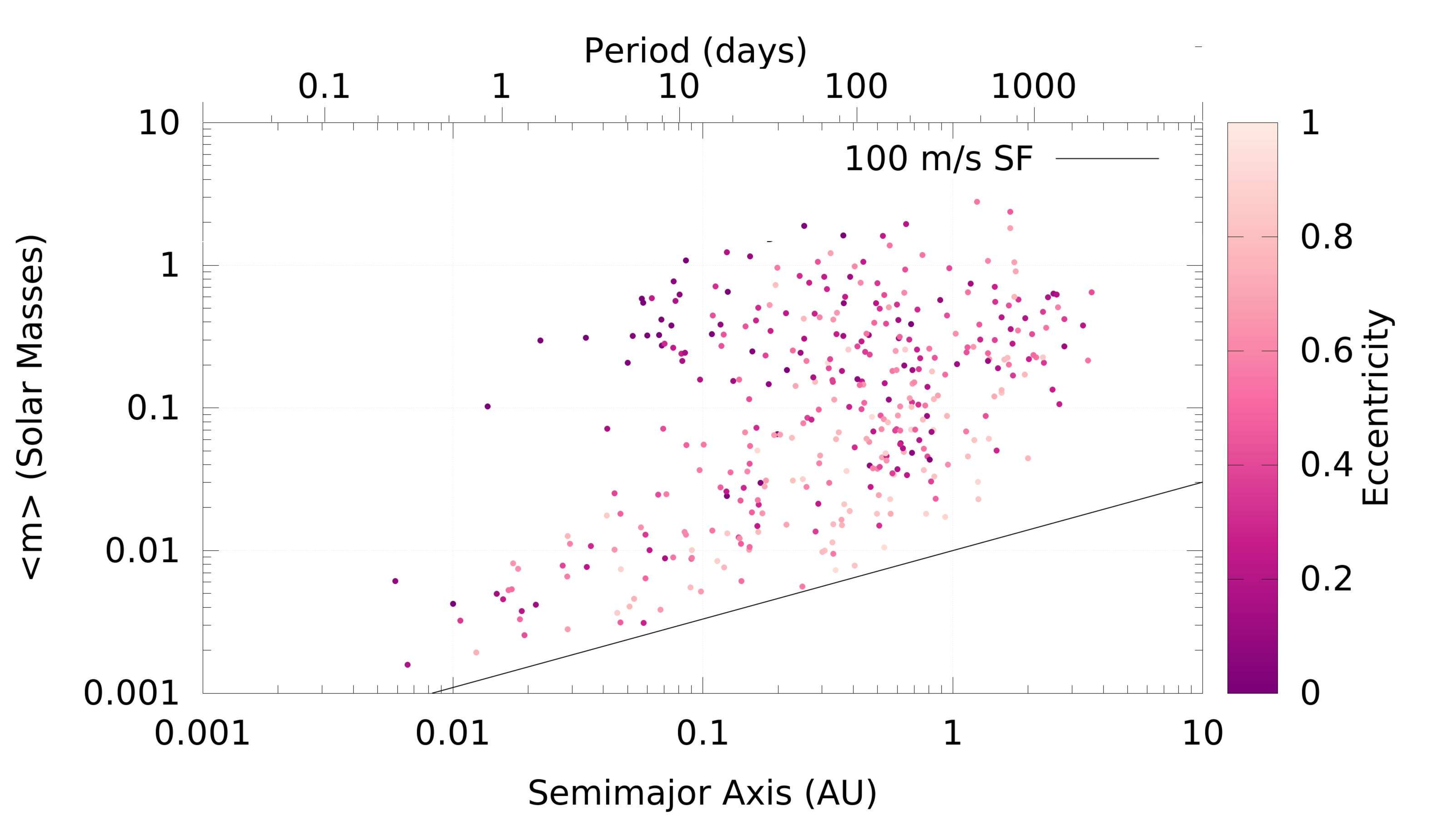}
\includegraphics[width=\textwidth]{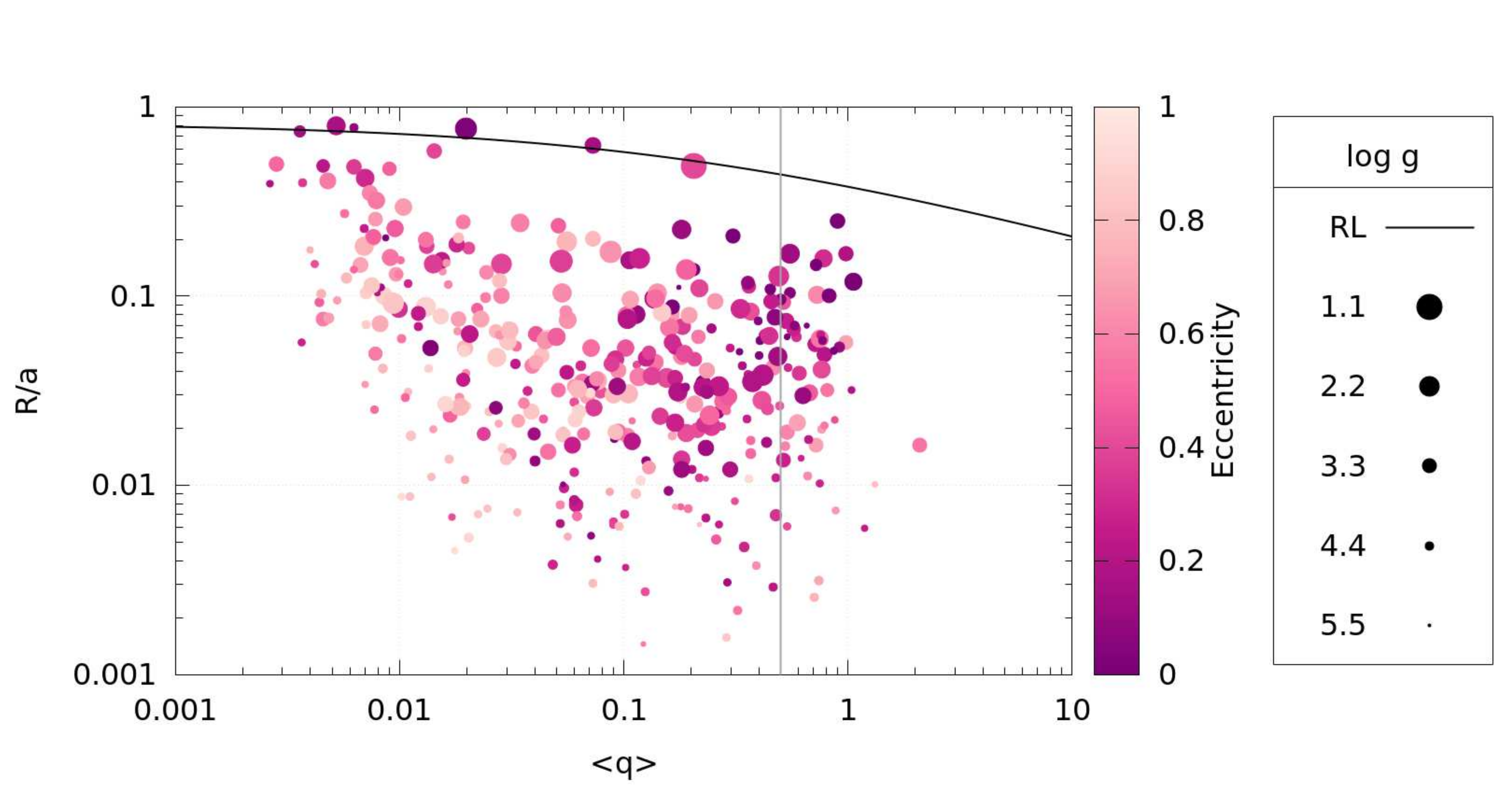}
\caption{\textit{Top Panel:} Orbital distribution of companion candidates in the 382-star gold sample with minimum orbital semi-major axis ($a$) in AU on the abscissa and maximum-likelihood companion mass ($\langle m \rangle = (4/\pi)m\sin i$) in $M_{\odot}$ on the ordinate. The top horizontal axis gives the approximate period for the companion in days as well. Color represents the orbital eccentricity of the companion, with dark magenta representing circular orbits. The black line is the sensitivity function (SF; Equation \ref{eq:sensLimit}) for 100 m s$^{-1}$ RV precision. Systems below this line would generally be undetectable by APOGEE. \textit{Bottom Panel:} Orbital distribution of companion candidates with $\langle q \rangle = \langle m \rangle/M_{\star}$ on the abscissa, and $R_{\star}/a$ on the ordinate. Color again represents eccentricity, and point size indicates the surface gravity ($\log g$) of the host. The grey vertical line marks systems with $\langle q \rangle > 0.5$, and the black line across the top of the panel indicates the Roche limit (RL; Equation \ref{eq:RL}) of the host star.}
\label{fig:orbDist}
\end{figure*}

The top panel of Figure \ref{fig:orbDist} presents the overall distribution of $\langle m \rangle$ and orbital semi-major axis of the candidate companions in the gold sample. In this figure, there appears to be two distinct companion mass regimes in which the candidates lie, and thus suggests different companion formation channels.  The upper regime is the binary star track, where the companion likely formed with (or shortly after) the primary from fragmentation of the cloud or disk from which the primary formed. The lower regime is the ``planet'' track, where the companion likely formed after the primary either through core accretion or gravitation instability in the disk surrounding the protostar. The trend of the lower planetary boundary mimics the sensitivity of the APOGEE survey (see Equation \ref{eq:sensLimit} with $\tilde{\sigma}_v = 100$ m s$^{-1}$). However, the trend of the planet track's upper boundary cannot be explained by a selection or sensitivity effect. One interpretation of the gap between the two regimes is a manifestation of the BD desert in the data, but the two tracks appear to merge at larger semimajor axes ($a > 0.1-0.2$ AU). The implications of this are discussed below.

\subsubsection{Combing the Brown Dwarf Desert}
\begin{figure*}
\centering
\includegraphics[width=\textwidth]{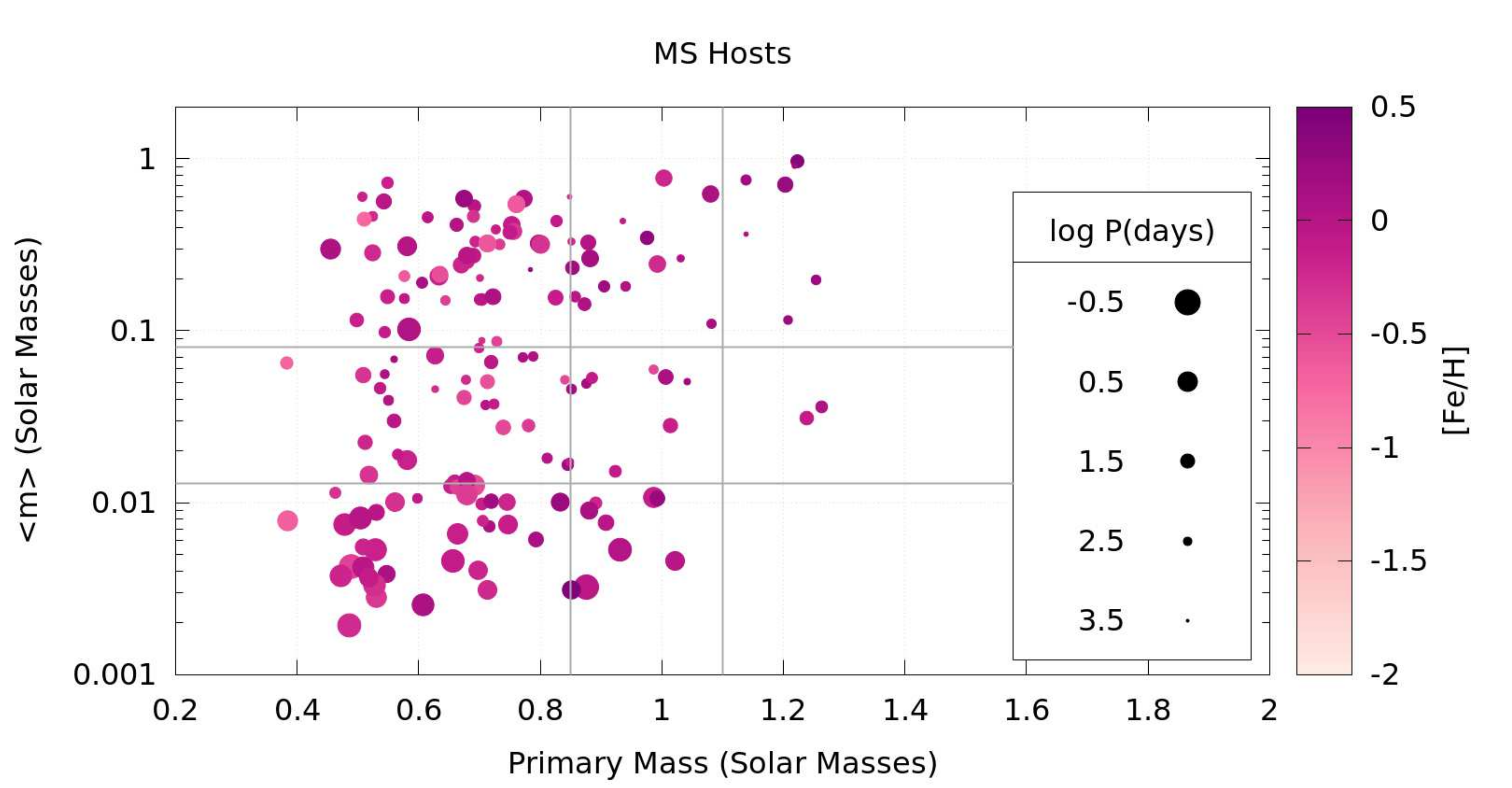}
\includegraphics[width=\textwidth]{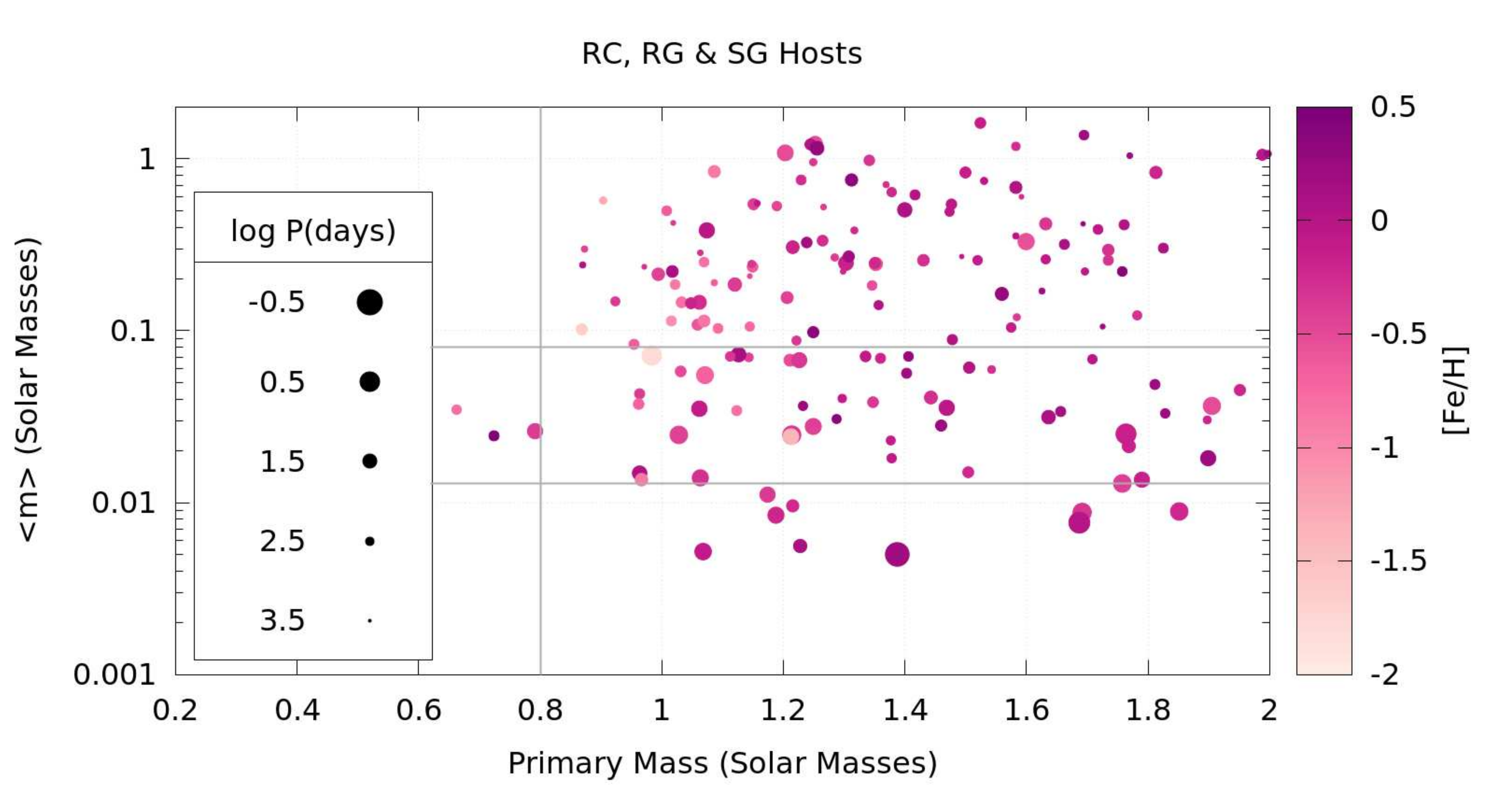}
\caption{Modification of Figure 3 from \cite{Guillot2014}, with $M_{\star}$ on the abscissa and maximum-likelihood companion-mass (\textcolor{black}{$\langle m \rangle$}) on the ordinate. Color represents the host stars' metallicity, and point size represents the period of the companion in log days. Larger points here indicate companions that are more likely to be undergoing tidal interaction with their host star. \textit{Top Panel:} Stars in the gold sample selected as MS stars. The vertical lines mark nominal G dwarfs ($0.85 < M_{\star}/M_{\odot} < 1.1$), and the horizontal lines mark the BD mass regime ($0.013 < \langle m \rangle /M_{\odot} < 0.08$). \textit{Bottom Panel:} Remaining stars in the gold sample with $M_{\star} < 2 M_{\odot}$. The horizontal lines again mark the BD mass regime, and the vertical line marks $M_{\star} = 0.8 M_{\odot}$. It would be a reasonable expectation that a giant star above this mass evolved from a star earlier than a G dwarf since solar-like stars loose about one third of their mass on the RGB.}
\label{fig:typeDist}
\end{figure*}

The top panel of Figure \ref{fig:orbDist} indicates that this sample reproduces the BD desert, but only for orbits with $a<0.1-0.2 \mathrm{AU}$ ($P < 10-30$ days), which is significantly less than the 3 AU extent of the desert as \textcolor{black}{stated} in \cite{Grether2006}. \textcolor{black}{However, their sample mostly considered solar-like dwarf hosts, while this sample contains stars with a variety of spectral types, as well as many evolved stars. From the top panel of Figure \ref{fig:typeDist}, it appears that the relative number of BD companions \textcolor{black}{decreases as host mass increases} for MS hosts. Likely M dwarfs (MS with $M_{\star} < 0.6 M_{\odot}$) have roughly equal numbers of BD and stellar-mass companions, while K dwarfs (MS with $0.6 < M_{\star}/M_{\odot} < 0.85 $) have roughly half the number of BD candidate companions as stellar-mass candidate companions. The G dwarfs (MS with $0.85 < M_{\star}/M_{\odot} < 1.1 $) show a similar relative number of BD companions compared to stellar-mass companions, but they are less uniformly distributed throughout the BD mass regime compared to the lower mass BD candidate hosts, suggesting a higher probability that many of these BD candidates are scattered into the BD mass regime by inclination effects. These results leads one to believe the interpretation of \cite{Duchene2013} that the BD desert is simply a special case for solar-mass stars of a more general lack of extreme mass ratio ($q \lesssim 0.1$) systems. For example, if, in general, systems with $q < 0.08$ are rare (i.e., a BD companion around a 1 $M_{\odot}$ companion), then a relatively high-mass BD companion ($m > 0.04 M_{\odot}$) orbiting a $0.5 M_{\odot}$ star should be a more common occurrence.}

\textcolor{black}{Out of the 112 BD companion candidates in this sample, 71 orbit evolved stars. All but two of the giant (RC and RG) hosts have masses $>0.8 M_{\odot}$ and only one of the SG hosts has a mass $ <1 M_{\odot}$.  Considering that stars like the Sun lose up to a third of their mass on the RGB, it is a reasonable assumption that a vast majority of the evolved stars in this sample descended from main-sequence F (or earlier) dwarfs. As can be seen from the bottom panel of Figure \ref{fig:typeDist}, the evolved stars have roughly half the number of BD candidate companions as stellar-mass candidate companions, and the BD-mass candidates are distributed throughout the BD-mass regime, similar to the K dwarf distribution. If the evolved stars are indeed evolved F dwarfs, and we follow the progression from above, one would expect these stars to have a smaller relative number of BD companions compared to even the G dwarfs.} 
However, it has been previously suggested that the BD desert observed for Solar-like stars may cease to exist for F dwarf stars \citep{Guillot2014}. Their proposed explanation of this effect is that G dwarfs are more efficient at tidal dissipation. In general, compared to Jupiter-mass planets, more massive small separation companions undergo stronger tidal interaction with their host star through angular momentum exchange. Stellar-mass companions, however, have sufficient orbital angular momentum to remain in a stable orbit, which explains the demise of small separation BD-mass but not stellar-mass companions. However, F (and earlier) dwarfs are known to remain rapid rotators ($v_{rot} \sim 20-100$ km s$^{-1}$) throughout their main-sequence lifetimes due to their smaller outer convective zones leading to weaker magnetic breaking. This means F dwarfs are also less efficient at extracting angular momentum from an orbiting companion. Therefore, rapid rotators such as F dwarfs inhibit tidal dissipation, which explains this ``F dwarf oasis'' for BD companions. The dynamical model presented in Figure 4 of \cite{Guillot2014} shows that a companion in the BD-mass regime on an initial 3-day orbit around a $1 M_{\odot}$ star will survive for $<40\%$ of the star's main sequence lifetime ($\lesssim 4$ Gyr), while the same companion around a host star with $M_{\star} > 1.2 M_{\odot}$ will survive for at least the entirety of the host star's main sequence lifetime ($\sim 6.5$ Gyr for a $1.2 M_{\odot}$ star). The presence of a large number BD companions orbiting the evolved stars in this sample strongly supports this ``F dwarf oasis'' hypothesis. 

\textcolor{black}{However, the tidal effects explanation would only strongly affect the closest-in companions. Since the rotation period of a G dwarf is $P_{\star} = 30$ days (compared to a few days for an F dwarf), tidal dissipation could only explain BD companions with orbital periods less than 30 days, and the majority of the BD candidate companions in this sample have periods significantly greater than that. Therefore, tidal dissipation can only explain the BDs (or lack thereof) with orbits within 0.2 AU. Curiously, this sample reproduces the BD desert out to approximately 0.2 AU, suggesting this mechanism may indeed play a role in shaping the BD desert. Another possible explanation for the presence of BD candidate companions is Roche lobe overflow of the star as it evolves off the main sequence onto an orbiting planetary-mass candidate, allowing it to grow to BD mass as the star evolves up the RGB.
\cite{Eggleton1983} gives the following approximation for the Roche lobe of a primary donor star with mass $M_1$ orbited by a companion with $M_2$:
\begin{equation}
\frac{r_1}{a} = \frac{0.49q^{-2/3}}{0.6q^{-2/3}+\ln(1+q^{-1/3})},
\label{eq:RL}
\end{equation}
where, $q = M_2/M_1$, $a$ is the separation of the two bodies, and $r_1$ is the Roche lobe radius of the potential donor. In the bottom panel of Figure \ref{fig:orbDist}, we mark the Roche lobe as a function of $\langle q \rangle$. As an interesting note, it appears there are seven stars in this sample that are currently at or near Roche lobe overflow, three of which are currently of planetary mass, and three of which are BD mass. These systems will all be the subject of further scrutiny. In general, for a 1-10 $M_{Jup}$ planet to cause a $\sim1 M_{\odot}$ primary to overflow its Roche lobe, the radius of the primary would have to exceed $\sim 70-80 \%$ of the separation between the two bodies. This would not be an unreasonable expectation for a companion orbiting within 1 AU, as solar-mass stars can achieve radii approaching 1 AU at the tip of the RGB. Therefore, this mechanism may be a way to explain the relatively large number of BD companion candidates orbiting the evolved stars in this sample.}
Overall, this catalog's large number of systems with short-period BD companion candidates challenges the notion of the BD desert as we know it, and certainly warrants further investigation.

\subsubsection{Eccentricity Distribution} \label{sec:ecc}
In Figure \ref{fig:orbDist}, we also see the distribution of orbital eccentricities. As expected, the smallest-separation ($a<0.1$ AU) stellar-mass companions all have circular orbits. The circularization cutoff period increases with the age of the system with 5-10 Gyr systems having cutoff periods of 12-20 days \citep{Mathieu2004}. All of the binary companions in this catalog with $a < 0.1$ AU have $P < 20$ days. Therefore, the distribution of eccentricities for the binary systems in this sample, with circular orbits at small separation, and eccentric orbits at large separations is not unexpected. The closest ($a < 0.01$ AU) planetary-mass companions appear to have also circularized, as expected, but a surprising result is the relatively large fraction of eccentric orbits for relatively close-in planetary-mass candidate companions. For the RG and RC hosts, one interpretation of these eccentricities is ongoing tidally-induced migration (see \S \ref{sec:evolHosts} for further discussion of this). However, the majority of the small-separation planetary and BD candidate companions orbit dwarf and SG stars. For these systems, their higher eccentricities may be further evidence for the mechanism suggested by \cite{Tsang2014} whereby stellar illumination heating a gap cleared by a forming planet may excite the eccentricity of the planet in the gap. 

\subsubsection{High Mass Ratio Systems} 
\textcolor{black}{Of this catalog's candidate companion systems, there are 50 systems with a mass ratio $\langle q \rangle = \langle m \rangle/M_{\star} \ge 0.5$ (see bottom panel of Figure \ref{fig:orbDist}). One would expect that these systems would manifest themselves as SB2s, but these systems show no strong indication of such behavior in their APOGEE spectra. Of these, 24 are RG stars, which would explain their lack of SB2 behavior, as their companion is likely still on the main sequence, and thus the flux ratio would be too large. However, this still leaves 26 MS and SG hosts, of which one explanation is that they host massive compact objects, such as stellar remnants. These seven companions have $0.3 M_{\odot} <\langle m \rangle < 1.2 M_{\odot}$, which would indicate these systems might host white dwarf companions, eight of which may be low-mass ($\langle m \rangle < 0.45 M_{\odot}$) He-core white dwarfs \citep{Liebert2005}. Furthermore, two of the systems with a RG host have $\langle q \rangle > 1$, indicating the companion has already completed its evolution, and the recovered $\langle m \rangle$ of the companions ($2.8$ and $1.6 M_{\odot}$) indicates they may be neutron stars.} 

\subsection{Host Star Distribution}
\begin{figure*}[t]
\centering
\includegraphics[width=\textwidth]{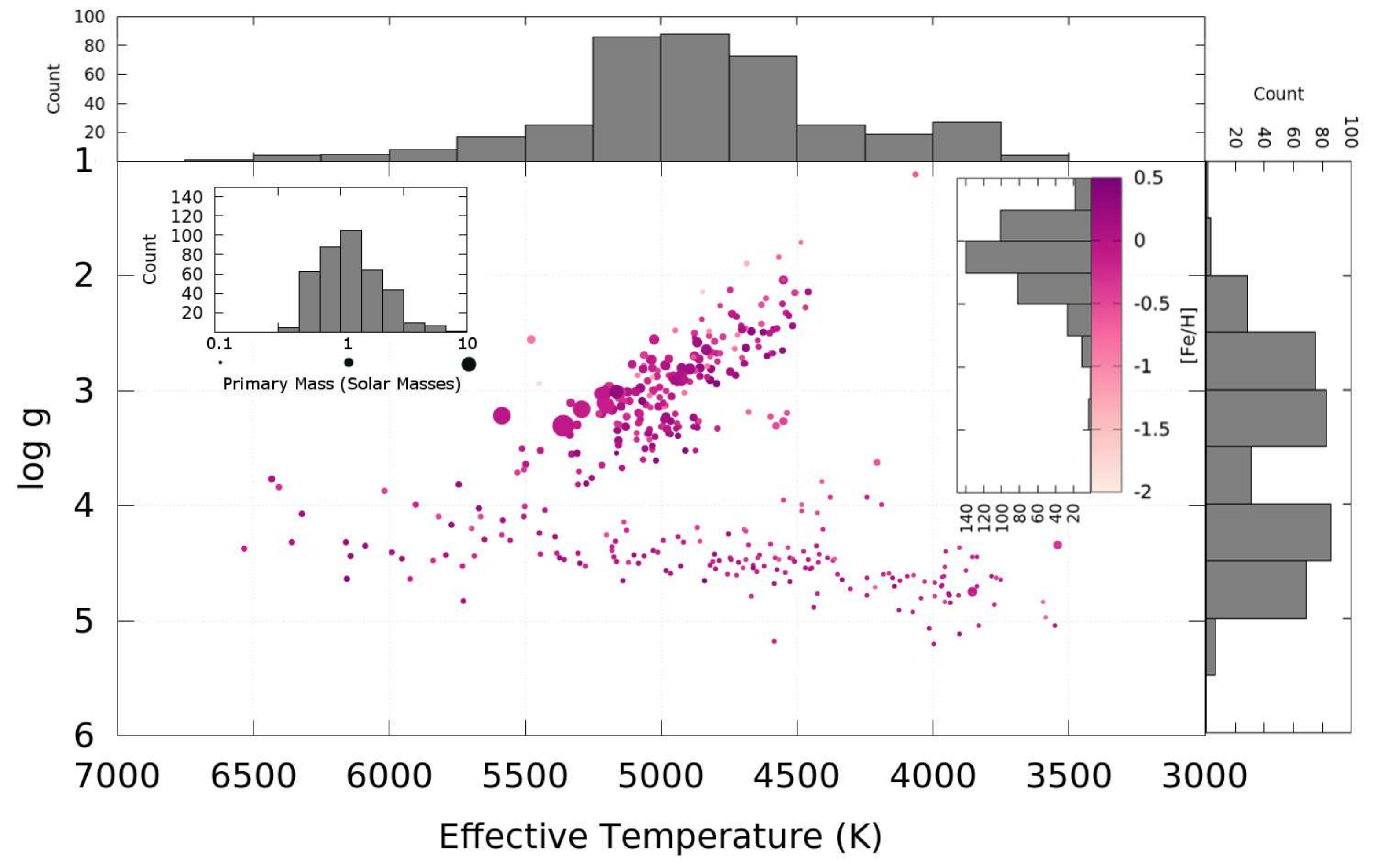}
\caption{A spectroscopic HR diagram of the companion candidate hosting stars, with the host stars' $T_{\rm eff}$ and $\log g$ as the abscissa and ordinate. The points are color-coded by host star metallicity ([Fe/H]), and point size indicates the primary mass in Solar masses. The stars along the bottom of the figure are the dwarf stars, and stars along the line connecting $(T_{\rm eff},\log g$) = (5500 K, 3.5) and (4000 K,1) are the giants. Histograms of the effective temperature ($T_{\rm eff}$, top panel), surface gravity ($\log g$, right panel), metallicity ([Fe/H], inset with color bar), and primary mass (inset with size legend) of the host stars in this gold sample are also shown.}
\label{fig:hostDist}
\end{figure*}

Solar type stars (i.e., G dwarfs) have been the primary focus of exoplanet and stellar multiplicity studies. Out of the 382 stars in this sample, only 36 are solar-type (MS with $5000 \mathrm{K} < T_{\rm eff} < 6000 \mathrm{K}$) stars. 
Figure \ref{fig:hostDist} reveals that, in addition to the solar-type stars, this sample contains cool dwarfs, subgiant and giant stars, which allows us to probe many different stellar types and stages of stellar evolution. Figure \ref{fig:hostDist} also presents distributions of the stellar parameters of the host stars in this sample.

\subsubsection{The Fate of Companions: Exploring Evolved Host Stars} \label{sec:evolHosts}
Tidal dissipation is thought to play an important role in the destruction of planetary systems as a star evolves off the main sequence and expands \citep{Penev2012a}. This sample contains 225 $a < 3$ AU candidate companions to evolved stars, indicating either many initial small separation companions survive engulfment or farther-orbiting planets undergo increasing tidal migration as its host ascends the giant branch, bringing the companion closer to its host star. The nine candidate planetary-mass ($\langle m\rangle < 0.013 M_{\odot}$) companions orbiting giant stars in this sample would be a $ 20\%$ increase in the number of currently known giant stars hosting a planet \citep[$\sim50$ according to the tabulation by][]{Jones2014b}. As \cite{Jones2014b} mentions, there is a small separation cut-off for RG hosts. The current record-holder for smallest separation of an RV-detected planet RG host is HIP 67851b with $a=0.539$ AU \citep{Jones2014}. The shortest period planet orbiting a giant star, Kepler 91b, is on a 6-day orbit \citep{Lillo-Box2014a}. Most of the candidate planets orbiting giants lie between these two systems, with a few candidates closer than Kepler 91b. 

\begin{figure*}[t]
\centering
\includegraphics[width=\textwidth]{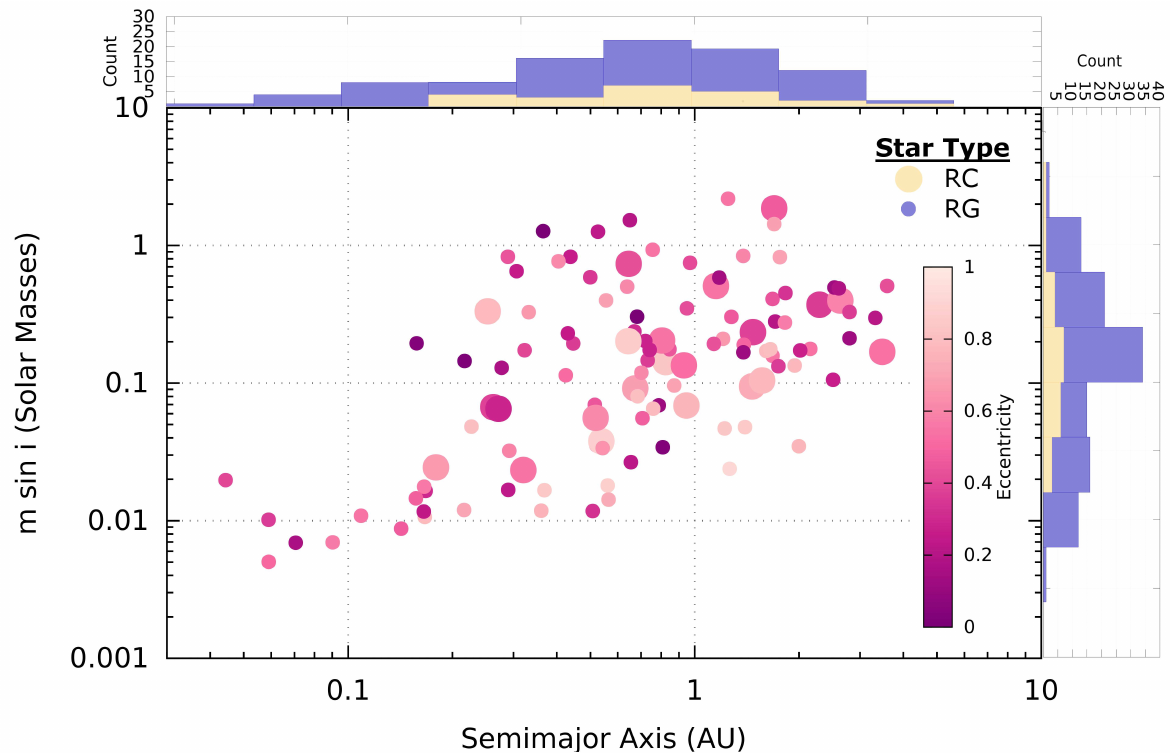}
\caption{Orbital distribution of companion candidates to RC stars (large points) and RG stars with similar stellar parameters as this sample's RC stars (small points). Minimum orbital semi-major axis in AU is on the abscissa and minimum companion mass in $M_{\odot}$ is on the ordinate. Color again represents the orbital eccentricity of the companion. The panels above and to the right of main plot show the msini and semi-major axis distribution for the RG comparison sample (purple histogram) and RC (gold histogram) hosts.}
\label{fig:orbDistGiants}
\end{figure*}

Of the evolved stars in this sample, 23 are verified Red Clump (RC) stars \citep{Bovy2014}. RC stars are metal-rich stars which have passed through the tip of the red giant branch (RGB) and have contracted due to the ignition of core helium burning. It is expected that stars like the Sun may reach radii up to 1 AU when they reach the tip of the RGB. Therefore, the presence of companion candidates orbiting RC stars at $a<1 \, \mathrm{AU}$ in this catalog (see Figure \ref{fig:orbDistGiants}) is a surprising discovery. To investigate this further we compared the RC stars to RGs in this sample, but we consider only RGs with [Fe/H] $> -0.42$ ([Fe/H] of the most metal-poor RC in this sample) and $2.4 \le \log g \le 3.3$ (the $\log g$ range of the RC stars) to eliminate possible effects from RV sensitivity issues. This also allows us to compare stars approximately half way up the giant branch to stars that have already passed through the tip of the RGB, and have achieved their largest extent. These 92 RG stars have 62 stellar-mass, 25 BD-mass, and 5 planet-mass companion candidates compared to 18, 5 and 0 for the 23 RC giants. 

A cursory look at these numbers (and Figure \ref{fig:orbDistGiants}) shows a lack of smaller companions for Red clump stars, as well as a companion candidates found at smaller separations for the 92 RG stars when compared to RC stars (0.07 AU vs. 0.2 AU at the low-mass end). It is also interesting to note that no companion candidates have circular orbits among RC hosts (smallest $e = 0.284$). This all points to the role of the tidal migration and destruction of companions, particularly planetary-mass companions. However, any tidally induced migration of companions will be much weaker than when the star was in the RGB phase. Therefore any companions with $a < 1$ AU around an RC star likely would have to have survived inside the star's envelope during its RGB phase. Most of the RC hosts with $a < 1$ AU candidate companions are likely post-common envelope systems, and thus may have experienced drag-induced migration to bring them to their current orbit. These systems certainly warrant deeper investigation.

\subsubsection{Metal-Poor Companion Hosts} \label{sec:mpHosts}
According to the compilation of exoplanets.org \citep{Han2014}, of the confirmed planet-hosting stars with metallicity measurements, only 15 have [Fe/H] $< -0.5$. This sample has 41 stars with [Fe/H]$<-0.5$, and of these, two host candidate planetary-mass companions and 14 host candidate BD companions. The most metal-poor stars in this sample approach [Fe/H] $= -2$. While there are no candidate planets among the most metal-poor ([Fe/H] $< -1$) hosts (5 stars), there are two companions in the BD mass regime. The smaller fraction of the lowest-mass companions detected among the most metal-poor stars in this sample is not surprising as the RV uncertainties are higher for metal-poor stars, as described in equation \ref{eqn:RVprec}. Also, it is not too surprising to find metal-poor stars hosting binary companions, given the \cite{Carney2003} result. However, finding a population of metal-poor stars potentially hosting BD companions is surprising in the context of the core accretion model of companion formation, and may suggest an alternate formation mechanism for these companions. 

\subsection{Galactic Distribution of Candidate Hosts}
Most surveys for stellar and substellar companions have focused on stars in the solar neighborhood, especially with the recent interest in M dwarf planet hosts. In contrast, only three of the sources in this catalog are within $100$ pc of the Sun, where the vast majority of known planets with distance measurements have been found. In a Galactic context, this sample is truly complementary to previous studies. The current most-distant known planet host is the microlensing source OGLE-2005-BLG-390L at 6.59 kpc \citep{Beaulieu2006}. The most-distant planetary-mass ($\langle m \rangle = 7.26 M_{Jup}$) candidate companion in this catalog orbits the slightly metal-poor ([Fe/H]$ = -0.34$), RG ($\log g = 2.5$) star 2M05445028+2847562, which lies at a comparable distance of 6.13 kpc. 

\begin{figure*}
\centering
\includegraphics[width=\textwidth]{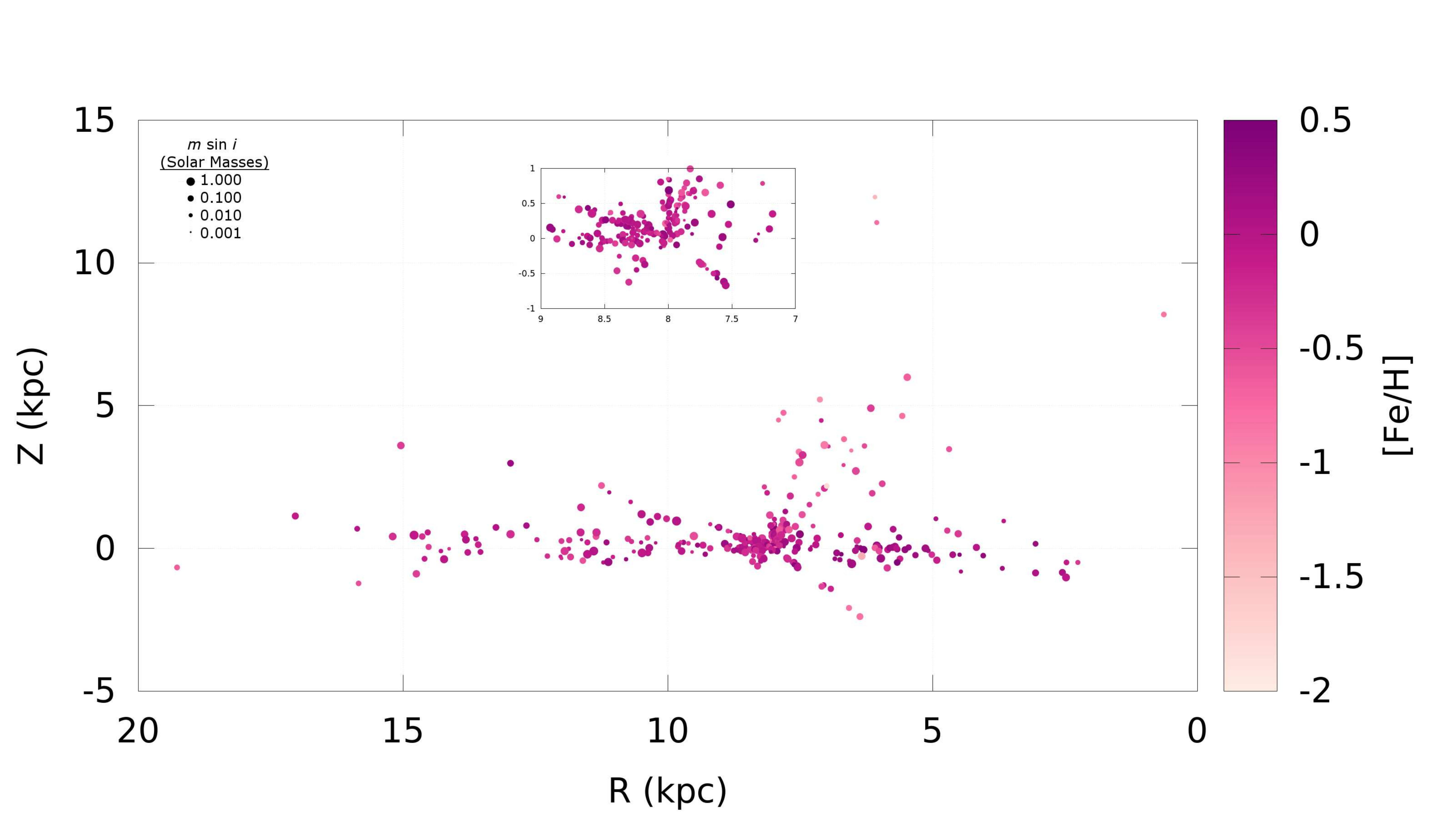}
\includegraphics[width=\textwidth]{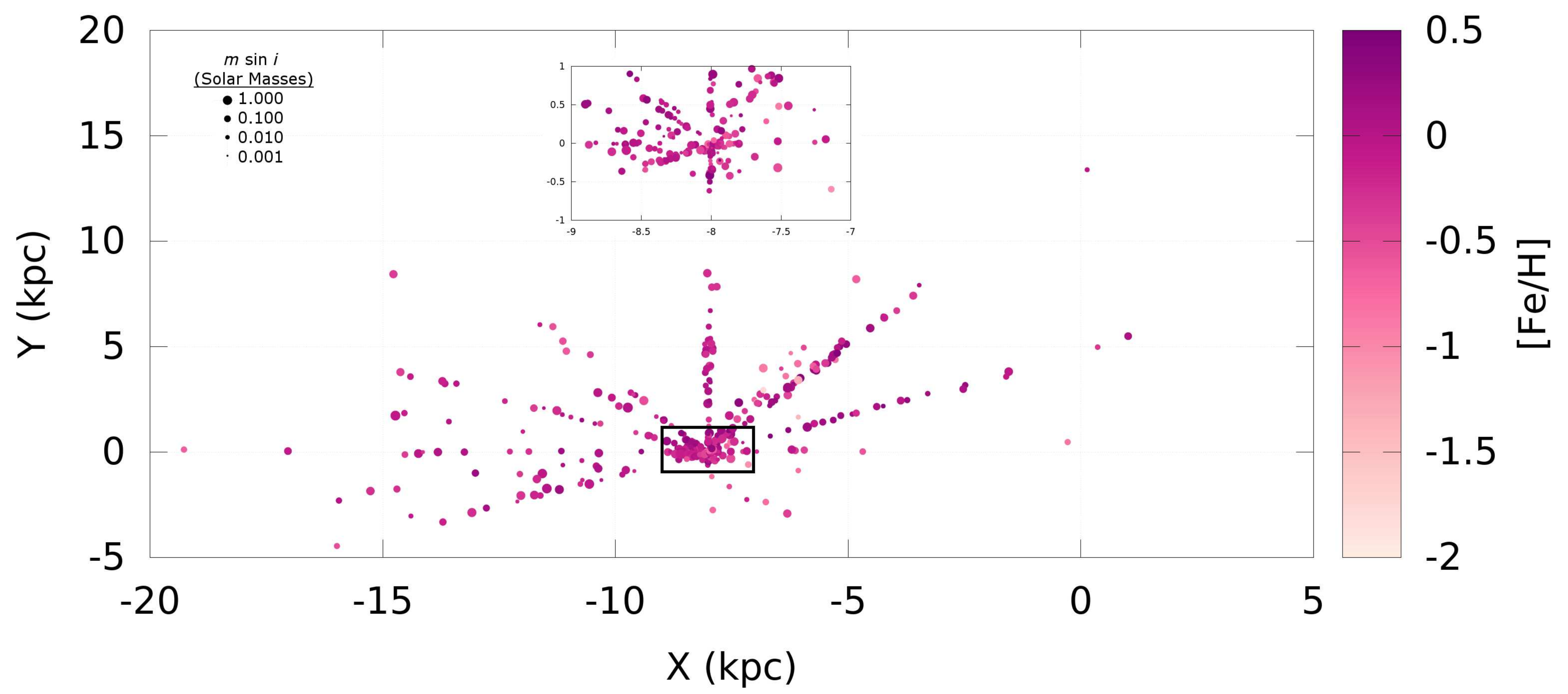}
\caption{The Galactic distribution of companion candidate hosts in this catalog.
\textit{Top Panel:} Distribution in Galactocentric $R$ and $Z$, where $R$ is the radial distance from the Galactic Center, and $Z$ is the height above the Galactic midplane. The color of the points indicates the metallicity ([Fe/H]) of host star, and the point size indicates $m \sin i$ of the companion candidate orbiting the star. The inset panel shows a detailed view of the solar neighborhood, which is indicated by the black box in the main plot ($7 \, \textrm{kpc} < R < 9 \, \textrm{kpc}$, $|Z| < 1 \, \textrm{kpc}$).
\textit{Bottom Panel:} Distribution in Galactocentric $X$ and $Y$ rectilinear coordinates, where $(X,Y) = (0,0)$ and $(-8,0)$ kpc are the locations of the Galactic Center and the Sun respectively, and $Y>0$ is in the direction of the Sun's orbit. The color and size of the points indicate the same data as they do in the top panel. Again, the inset panel shows a detailed view of the solar neighborhood, which is indicated by the black box in the main plot ($7 \, \textrm{kpc} < X < 9 \, \textrm{kpc}$, $|Y| < 1 \, \textrm{kpc}$).}
\label{fig:galDist}
\end{figure*}

Furthmore, this sample has 36 companion candidates farther than this distance. Of these, 12 are BD-mass companions around stars reaching to a distance of 15.7 kpc. Figure \ref{fig:galDist} demonstrates the Galactic reach of this catalog's companion candidate hosts. A large majority of this sample (327 stars) resides in the Galactic Thin Disk ($|Z| < 1$ kpc, but see note f in Table \ref{tab:Census}), but these disk stars reach from inner disk ($R \sim 2$ kpc)  to the outer disk ($R \sim 15$ kpc). From this preliminary analysis, it is safe to say that companions of all types are ubiquitous across the thin disk. As we move from the thin disk to the halo, the proportion of higher-mass companions increases. This trend is likely due to the combination of the sensitivity bias that low-mass companions are less likely to be detected around more metal-poor stars (see Equation \ref{eqn:RVprec}), and the Planet-Metallicity correlation.

\section{Future Survey Directions} \label{sec:Future}

The APOGEE-2 survey is a six-year extension of the APOGEE survey as a part of SDSS-IV.  \text{APOGEE-2} continues the survey of the Northern Hemisphere at APO, and implements a new component of the survey at the DuPont Telescope at Las Campanas Observatory (LCO) to cover the Southern Hemisphere. A dedicated search for substellar companions was approved as a goal science program in APOGEE-2. By the end of APOGEE-2, in 2020, we will have acquired $\ge$ 24 epochs of RV measurements of 1074 red giant stars across 5 fields, including fields containing the star cluster NGC 188 as well as a COROT \citep{Borde2003} field. These fields were selected to search for companions because of previous observations from APOGEE-1. Many of these targets will accrue up to a 9-year temporal baseline of observations.

In addition to the planned dedicated fields, we expect many additional APOGEE-2 fields will have candidates discovered serendipitously, as they were with this work. In \text{APOGEE-1},  $\sim 10\%$ of the targeted stars had $\ge 8$ visits, and of those, $\sim 2.6\%$, or a cumulative $\sim 0.26 \%$ of all survey stars, were selected as having companion candidates. \text{APOGEE-2} will bring the cumulative total number of stars observed by the APOGEE instrument to $\sim$500,000 stars. Therefore, assuming a similar detection rate and visit distribution, as well as the 'gold sample" selection criterion used here, we expect to detect a cumulative total of at least $\sim1300$ companion candidates by the end of APOGEE-2.

Several technical improvements to the APOGEE pipelines planned for SDSS-IV will improve RV precision, and as a result, improve our ability to measure the orbital and physical parameters of systems observed in \text{APOGEE-2} as well as the current sample of candidates. One upgrade to ASPCAP of particular importance to our efforts is the implementation of stellar rotational velocity determination to ensure more reliable parameters for dwarf stars and rapidly-rotating giants. In addition, acquiring rotational velocities will allow estimates of the ages of the dwarf star hosts in future catalogs through the age-rotation correlation. We will also reap the rewards of a fully-vetted and improved distance catalog.

We have a significant ongoing observational program to individually investigate the best planetary mass and BD systems in this catalog that includes high-resolution spectroscopy, diffraction-limited imaging, and photometric variability monitoring.  The results of these efforts will be included in future versions of this catalog.

\section{Conclusions} \label{sec:Conclusion}
Through analysis of multiple epochs of APOGEE spectroscopic data, we have identified 382 stars that have strong candidates for stellar and substellar companions, of which 376 had no previous reports of small separation companions. From an initial analysis of this sample we have found:
\begin{enumerate}
\item Two distinct regimes of companions in $m \sin i$ - $a$ space exist that are likely the result of distinct formation paths for stellar-mass and planetary-mass companions, with the gap between the two regimes being a manifestation of the BD desert. However, we find a smaller and ``wetter'' BD desert with the BD desert only manifesting itself for orbital separations of $a < 0.1-0.2$ AU in this sample of candidate companions, much smaller than the \textcolor{black}{3} AU proposed in previous studies. \textcolor{black}{We proposed a few potential explanations of this result: (a) Lower mass MS candidate hosts host a higher relative number of BD-mass candidates than their higher-mass MS counterparts,  lending evidence to the \cite{Duchene2013} interpretation that the BD desert may be a special case of a more general dearth of extreme mass ratio binary systems. (b)} A majority of the candidate BD companions in this catalog orbit evolved F dwarfs, supplying further evidence to the ``F dwarf oasis'' hypothesis proposed for small separation BD companions by \cite{Guillot2014}. \textcolor{black}{(c) The possibility of planetary-mass candidates orbiting within $\sim$ 1 AU initiating Roche lobe overflow of their hosts as it ascends the giant branch, allowing planetary-mass companions to grow to BD mass.}
\item A significant number of small-separation eccentric systems which may be evidence for ongoing tidal migration among the giant hosts and the eccentricity-pumping mechanism proposed by \cite{Tsang2014} for the dwarf hosts.
\item \textcolor{black}{A set high mass ratio candidate systems ($\langle q \rangle > 0.5$), of which 28 show indications of containing a stellar remnant, including two neutron stars, and eight potential He-core white dwarfs.}
\item 225 candidate companions orbiting evolved (RC, RG, and SG) stars. This includes nine new planetary-mass candidate companions around giant stars, which, if confirmed, would be a $> 20\%$ increase from the previously known number given by \cite{Jones2014b}, as well as 3 planetary-mass candidates orbiting subgiant stars. Among the RC stars, 15 host companion candidates orbiting within 1 AU, the maximum expected extent of a RGB star evolved from a Sun-like star, indicating these systems are likely post-common envelope systems.
\item A population of 41 metal-poor ([Fe/H] $< -0.5$) candidate companion hosting stars, of which 2 host planetary-mass candidates, and 14 host BD candidates. These systems challenge the planet-metallicity correlation, and thus the core accretion paradigm of companion formation. It is possible the formation pathway for these companions may closer mimic that of binary systems or a gravitation instability scenario.
\item To first order, companions of all kinds are prevalent throughout the disk ($2 \, \textrm{kpc} < R < 15 \, \textrm{kpc}$, $-2 \, \textrm{kpc} < Z < 2 \, \textrm{kpc}$), with planetary-mass companions found out to distances of $\sim$ 6 kpc, and BD-mass companions to distances of $\sim 16 $ kpc.
\end{enumerate}

A campaign is underway to confirm and further characterize the nature of the candidate companion systems reported here. This effort will be augmented with SDSS-IV APOGEE-2 observations. Between APOGEE-1 targets obtaining additional visits and new APOGEE-2 targets obtaining a large number of visits, we expect APOGEE's sample of candidate companions to at least triple by the end of SDSS-IV.

\acknowledgments

The research described in this paper makes use of Filtergraph, an online data visualization tool developed at Vanderbilt University through the Vanderbilt Initiative in Data-intensive Astrophysics (VIDA). Particular thanks to Dan Burger who quickly answered questions and solved problems as they occurred.

This research has made use of the NASA Exoplanet Archive, which is operated by the California Institute of Technology, under contract with the National Aeronautics and Space Administration under the Exoplanet Exploration Program.

This research has made use of the Exoplanet Orbit Database and the Exoplanet Data Explorer at \url{exoplanets.org}.

Funding for SDSS-III has been provided by the Alfred P. Sloan Foundation, the Participating Institutions, the National Science Foundation, and the U.S. Department of Energy Office of Science. The SDSS-III web site is \url{http://www.sdss3.org/}.

SDSS-III is managed by the Astrophysical Research Consortium for the Participating Institutions of the SDSS-III Collaboration including the University of Arizona, the Brazilian Participation Group, Brookhaven National Laboratory, Carnegie Mellon University, University of Florida, the French Participation Group, the German Participation Group, Harvard University, the Instituto de Astrofisica de Canarias, the Michigan State/Notre Dame/JINA Participation Group, Johns Hopkins University, Lawrence Berkeley National Laboratory, Max Planck Institute for Astrophysics, Max Planck Institute for Extraterrestrial Physics, New Mexico State University, New York University, Ohio State University, Pennsylvania State University, University of Portsmouth, Princeton University, the Spanish Participation Group, University of Tokyo, University of Utah, Vanderbilt University, University of Virginia, University of Washington, and Yale University.

D.L.N. was supported by a McLaughlin Fellowship at the University of Michigan. 
J. K. C. was supported by an appointment to the NASA Postdoctoral Program at the Goddard Space Flight Center, administered by Universities Space Research Association through a contract with NASA.
Szabolcs Meszaros has been supported by the J{\'a}nos Bolyai Research Scholarship of the Hungarian Academy of Sciences. 
C. A. P., D. A. G. H., and O. Z. acknowledge support provided by the Spanish Ministry of Economy and Competitiveness (MINECO) under grants AYA2014-56359-P, RYC-2013-14182, and AYA-2014-58082-P.

We would like to extend our gratitude to Phil Arras (Virginia) and Kaitlin Kratter (Steward Observatory) for their most useful theoretical insights on our results.

{\it Facilities:} \facility{APO}.

\appendix

\section{Verification and Performance} \label{sec:Verification}
As with all survey reduction pipelines, the goal is to balance speed and accuracy. Our code is reasonably fast, with a typical star taking 30-60 seconds for a complete fit to be performed, as described above. Below, we describe the efforts to verify the accuracy of the \texttt{apOrbit} pipeline.

\subsection{Simulated Systems}
RV curves were generated for a suite of simulated systems to verify the output of the \texttt{apOrbit} pipeline. The simulations mimic the observations of candidate planet hosting stars by the APOGEE survey \citep[see \S 2.8 of][]{Majewski2015}, and can be used to investigate the types of systems that can be identified and characterized in the APOGEE-1 survey. 

\subsubsection{Generation of Simulated Systems}
We simulated 9000 planetary systems with random characteristics. The masses of the primary stars were drawn from the distribution of estimated masses for the actual candidate substellar hosts in the APOGEE data. The companion masses and periods were drawn from the distributions specified by \cite{Tabachnik2002}, using a mass range of 1-100 $M_{Jup}$ and periods from 0.1 to 2000 days. Eccentricities were drawn from a uniform distribution with a maximum of $e=0.934$, which corresponds to the eccentricity of HD 80606 b, the largest eccentricity in the \url{exoplanets.org} database. Companions with $P<5$ days were assumed to have circular orbits.  The radii of the APOGEE candidate host stars were also estimated, and planets with orbital separations less than 5 $R_\star$ were considered unphysical because the tidal decay of planetary orbits becomes relevant at such small separations. The longitude of periastron and the orbital phase of periastron passage relative to a reference date were drawn from uniform distributions.

With the orbital characteristics of the simulated companions defined, we simply used the $\texttt{helio\_rv}$ code in the IDL astronomy user’s library\footnote{\url{http://idlastro.gsfc.nasa.gov}} to calculate the measured heliocentric RV for each system on a set of observation dates.  The observation dates for each system were designed to mimic the way the survey proceeded.  The observations for each star were spread randomly over a 3.2 year time period assuming the telescope was on-sky for 15 days followed by 14 days off sky since APOGEE observed primarily during bright time. The simulated ‘measured’ RVs consisted of the actual motion of the star at the time of observation plus two sources of noise, drawn from Gaussian distributions. The first is simply measurement noise, which nominally has $\sigma_v = 100$ m s$^{-1}$ but is increased to $\sigma_v= 130$ m s$^{-1}$ for $20\%$ of the visits to simulate poor observing conditions. The second noise source is intrinsic stellar atmospheric RV jitter, with an amplitude drawn from the distribution in \cite{Frink2001}.

A second data set of 9000 simulated system were generated with much of the same parameters as the first, except that it had mass ratios approaching one, all orbital parameters were drawn from a uniform distribution, and it was much sparser in the lower-mass companion regime. We combined these two data sets to obtain complete coverage of parameter space. From the combined data set, we generated RV curves with 9, 12, 16, and 24 visits selected from the 24-visit parent sample with the 100 m s$^{-1}$ RV uncertainty level, as well as a set where the base uncertainty level is inflated to 1 km s$^{-1}$ to emulate RV measurements from the metal-poor host stars in this sample.

\subsubsection{Determination of Quality Criteria and False Positive Analysis} \label{sec:simRes}
The full test suite of simulated systems was run through the $\texttt{apOrbit}$ pipeline each time an update to the fitting algorithms was implemented. Many of these updates were inspired by the simulated systems for which the pipeline failed to reproduce the correct orbit in the previous run. In addition, many of the criteria used to select the RV variable and gold candidate sample, described in \S \ref{sec:Catalog}, were inspired by these results. Notable failures in previous runs that led to new selection criteria for candidate companions included:

\begin{itemize}
\item \textbf{Long-Period Systems:} The longest-period simulated systems demonstrated the  largest scatter in their results. This inspired the use of the phase uniformity index (see \S \ref{sec:fitCrit}), as well as a procedure to reject any solutions for which the period was longer than twice the baseline (\S \ref{sec:selGold}).
\item \textbf{Highly Eccentric Systems:} The code  had the most difficulty reproducing the orbital parameters of systems with high eccentricity ($e > 0.9$). However, these systems are extremely rare, so this is not a major issue. Nevertheless, this result still led to the decision to reject all orbital solutions with $e > 0.934$ (\S \ref{sec:fitCrit}), which is the planetary system with the largest known eccentricity anyway. Even with this cut, however, the more eccentric the system, the more trouble the code had in recovering the correct orbital parameters. In particular, systems with low numbers of visits had the most issues. This result inspired the use of the velocity uniformity index in the fitting code (\S \ref{sec:fitCrit}), and it led to the decision to implement more stringent significance cuts for eccentric systems (\S \ref{sec:selRvvar}).
\item \textbf{One-day Aliased Systems:} Early tests of the code on simulated systems revealed a tendency for solutions to cluster around integer fractions of one day, despite the initial period selection avoidance of such periods. This inspired the decision to reject any periods within $5\%$ of 1/3, 1/2,1, 2, or 3 days (\S \ref{sec:fitCrit}).
\end{itemize}

\begin{figure*}
\includegraphics[height=\textwidth,angle=-90]{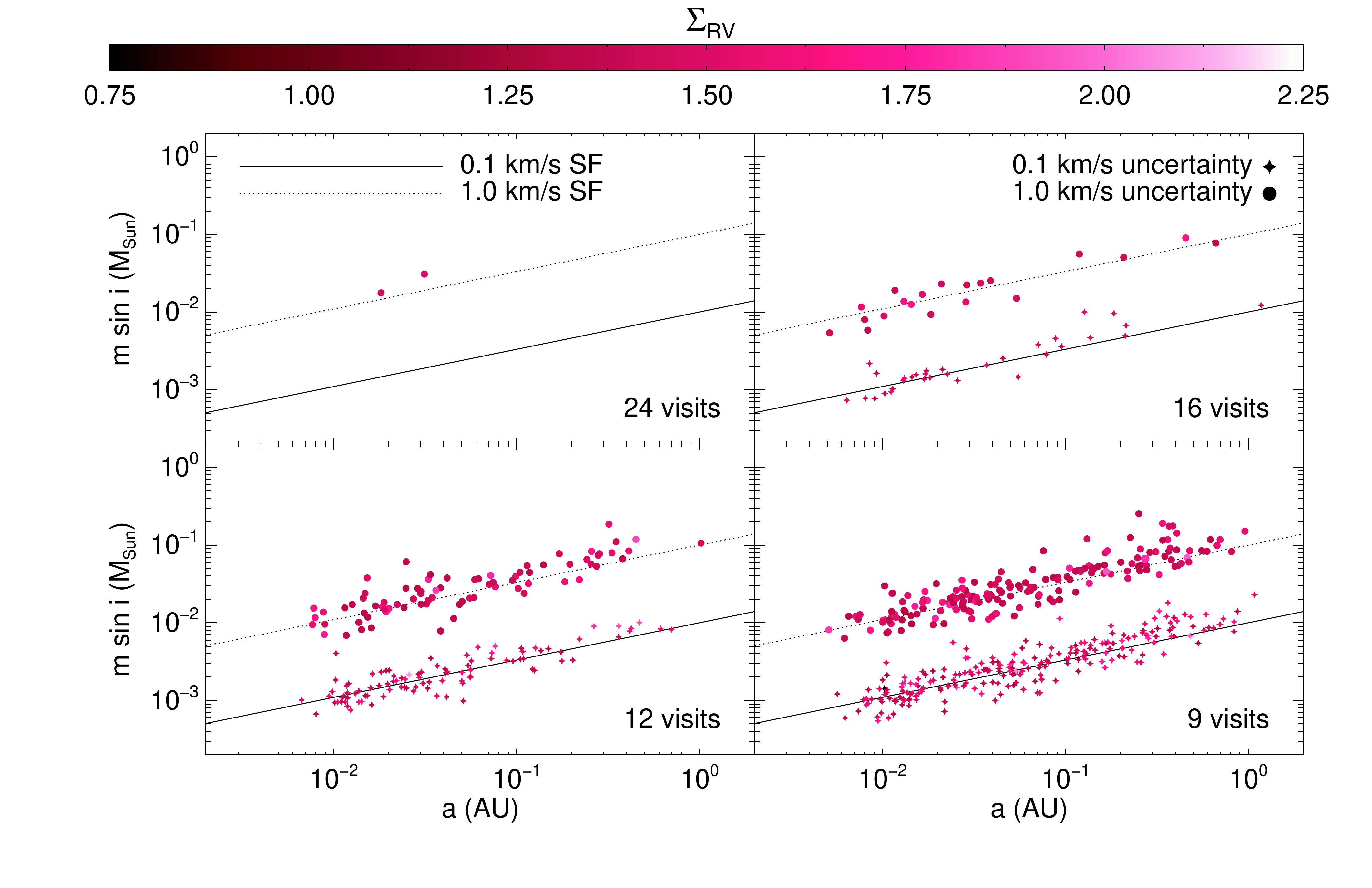}
\caption{Distribution of false positive companions in recovered $m\sin i - a$ space, with color representing the $\Sigma_{RV}$ statistic (see equation \ref{eq:rvvar}). Each set of points (unique color and shape) is drawn from a sample of 18,000 simulated systems with 9 (bottom right panel), 12 (bottom left panel), 16 (top right panel), and 24 (top left panel) simulated RV measurements based on stellar jitter and random measurement error. Symbol shape indicates the uncertainty level used for the simulation, with circles indicating 1 km s$^{-1}$ and four-point stars indicating 0.1 km s$^{-1}$ uncertainties. The solid and dotted lines show the approximate sensitivity function (SF; see Equation \ref{eq:sensLimit}) for 0.1 and 1 km s$^{-1}$ RV uncertainties. False positive signals such as these are removed from the sample via the velocity cut described in \S \ref{sec:selRVs}.}
\label{fig:falsePos}
\end{figure*}

These simulations were also used to understand how RV noise from the star or measurement error can potentially lead to a false positive candidate companion. To accomplish this we ran the simulated systems through the \texttt{apOrbit} pipeline following the procedures laid out in \S \ref{sec:orbFit} using just the RV signals from the star's atmospheric jitter and random measurement errors. We ran the simulations using four different numbers of visits (9, 12, 16, and 24) and two uncertainty levels (0.1 and 1 km s$^{-1}$), and selected candidates using the criteria described in \S \ref{sec:Catalog}. The results of the false positive tests are summarized in Figure \ref{fig:falsePos}. For systems with 24 visits, out of 18,000 simulated systems only two systems at the 1 km s$^{-1}$ uncertainty level registered as false positives. The raw number of false positives increased dramatically from 24 to 16 visits, and the higher uncertainties lead to a higher rate of false positives. Most false positives clustered around the sensitivity limit corresponding to the RV measurements from which they were derived (see Equation \ref{eq:sensLimit} below), and are generally assigned eccentric orbits ($e > 0.5$). This information, along with visual inspection of these fits, led to a pre-cut based on the velocity variations of the star which was based on the value of the statistic described in \S \ref{sec:selRVs} for these systems.

\subsubsection{Sensitivity Limit, Parameter Accuracy, and Recovery Rate} \label{sec:simAnal}
\begin{figure*}
\center
\includegraphics[width=0.75\textwidth]{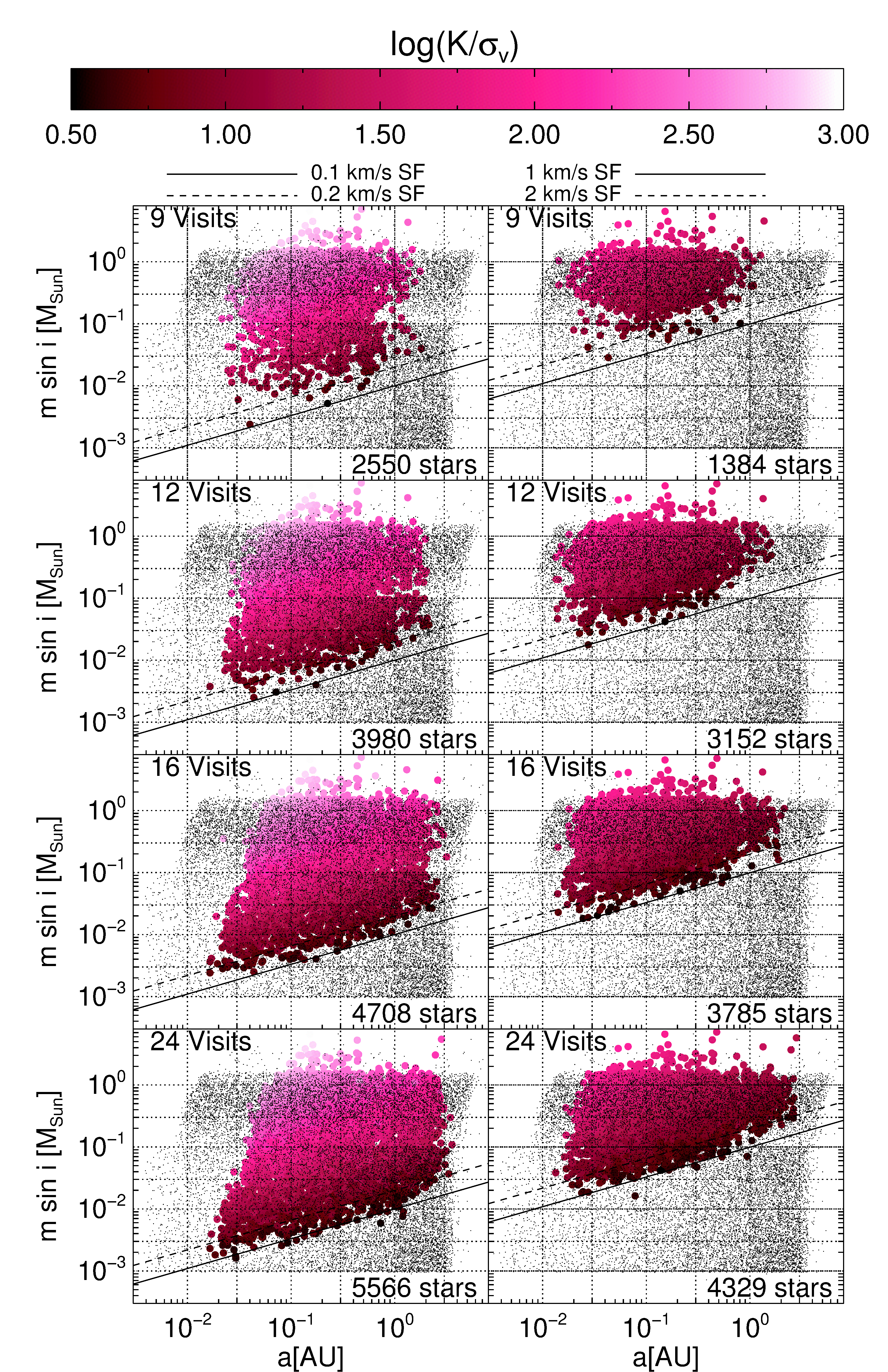}
\caption{The grey dots are the locations of the 18,000 simulated systems in actual $m\sin i$ - a space. The colored circles indicate the correctly recovered systems that were selected as candidates using the same metric as the ``gold sample'', color-coded by recovered $\log(K/\sigma)$, with the number of systems correctly recovered indicated in the bottom right corner of each panel. The black solid and dashed lines mark the sensitivity function (SF; see Equation \ref{eq:sensLimit}) for the baseline and twice the baseline RV uncertainties of 0.1 km s$^{-1}$ (left column) and 1.0 km s$^{-1}$ (right column) used by the simulations. The simulations presented here emulate stars with, from top row to bottom row, 9, 12, 16, and 24 visits.}
\label{fig:compSim}
\end{figure*}

In the results presented here, we ran the 18,000 simulated systems through the $\texttt{apOrbit}$ pipeline as described in \S \ref{sec:KepOrbits} for the four visit levels ($n_{RV} = 9, 12, 16, 24$) and two uncertainty levels ($\sigma_v = 0.1,1$ km s$^{-1}$), and selected companion candidates in the same manner as the gold sample, as described in \S \ref{sec:selRVs}, \ref{sec:fitCrit}, and \ref{sec:Catalog}. The systems correctly recovered ($P$ and $K$ recovered within 10$\%$, and $e$ recovered within 0.1) are shown in Figure \ref{fig:compSim}. Systems with large $K$ but small errors can lead to larger values of $\chi^2$ for a fit that still produces the correct orbital parameters. Unfortunately, removing the $\chi_{mod}^2$ constraint allows many systems with incorrect solutions to pass, so we err on the side of caution and keep it in place. Due to limits of the period search and the other constraints on the period, $a$, and $\chi_{mod}^2$ of the fit described in \S \ref{sec:selGold}, we expect to be able to recover orbits for companions having $0.01 \mathrm{AU} \lesssim a \lesssim 3 \mathrm{AU}$, depending on the baseline, number of visits, and the RV uncertainty level (see Figure \ref{fig:compSim}). By fitting a trendline to the simulated systems with $2.8 \le K/\tilde{\sigma}_v \le 3$ for various values of $\sigma_v$, we also find that the lower limit on detectability, which we will refer to as the sensitivity function(SF), of $m \sin i$ can be written as:
\begin{eqnarray}
\log (m \sin i) = 0.48\log(a) - C, \label{eq:sensLimit}
\end{eqnarray}
where the constant offset, $C$, depends on the sensitivity level (which we interpret as the median RV uncertainty, $\tilde{\sigma}_v$):
\begin{eqnarray}
C/\log(M_{\odot}) = 2.0 - 0.3\log_2(\tilde{\sigma}_v/100 \, \mathrm{m\,s^{-1}}).
\end{eqnarray}

\begin{figure*}
\center
\includegraphics[width=0.33\textwidth]{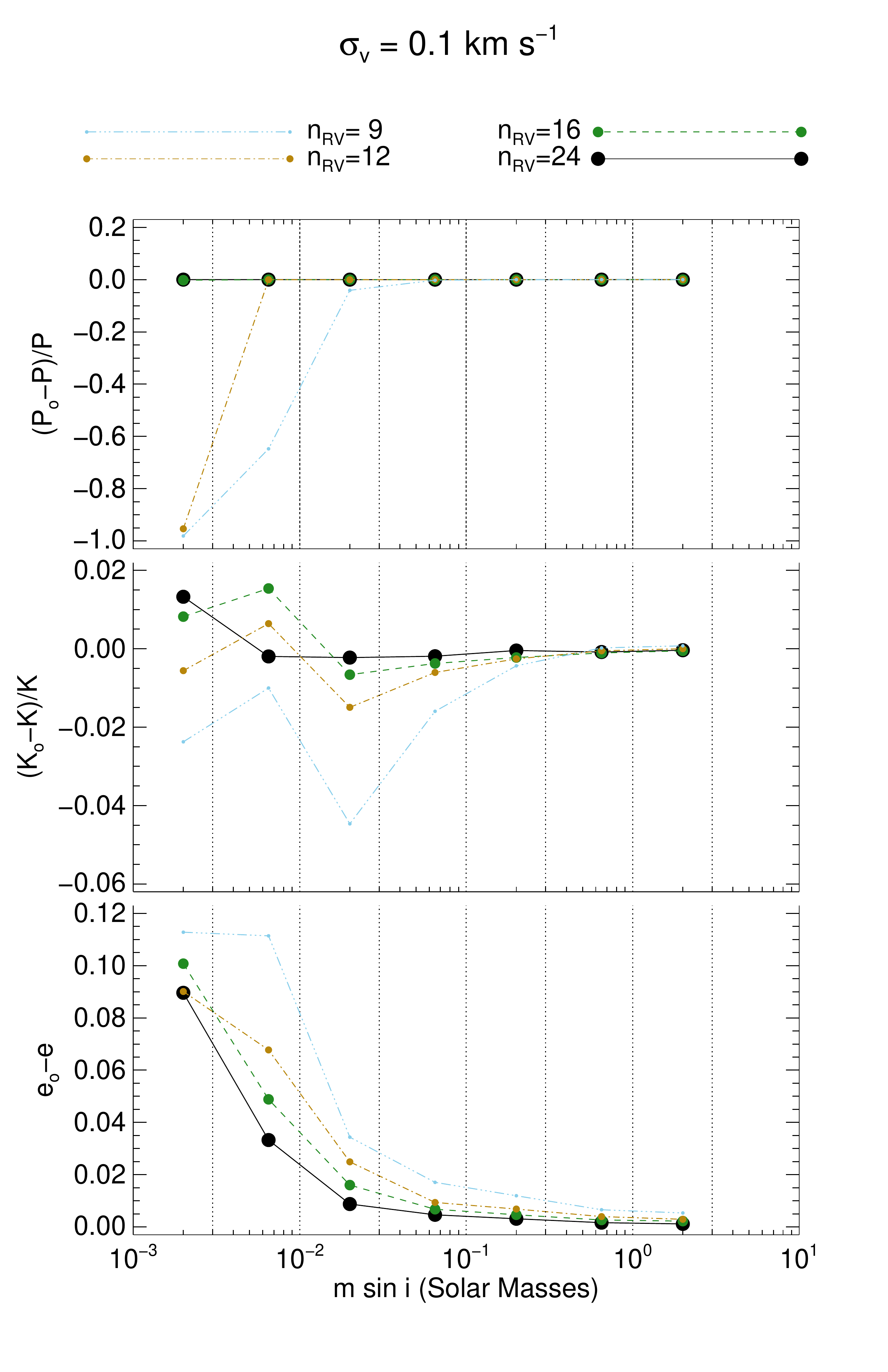} 
\includegraphics[width=0.33\textwidth]{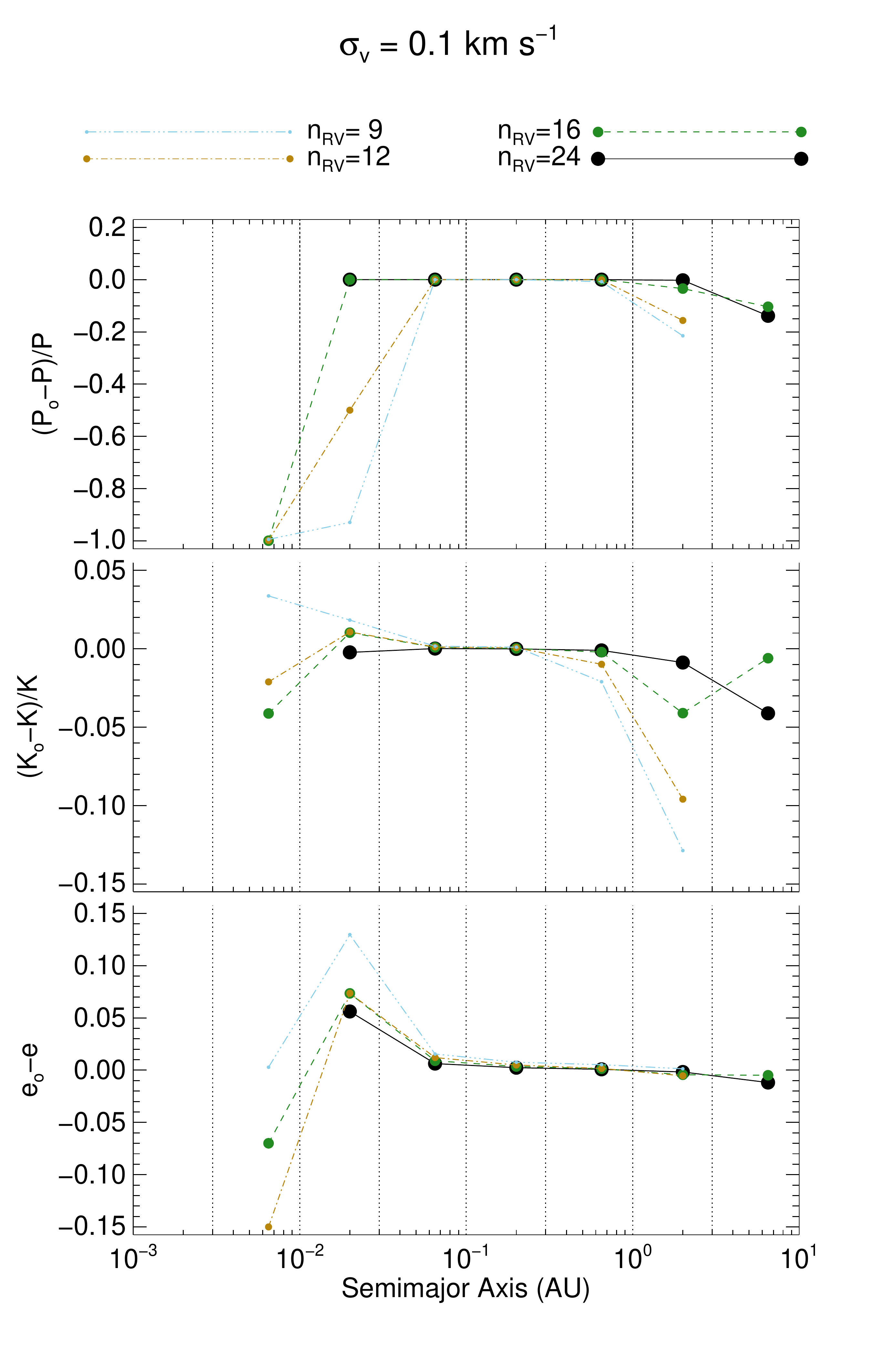}  
\includegraphics[width=0.33\textwidth]{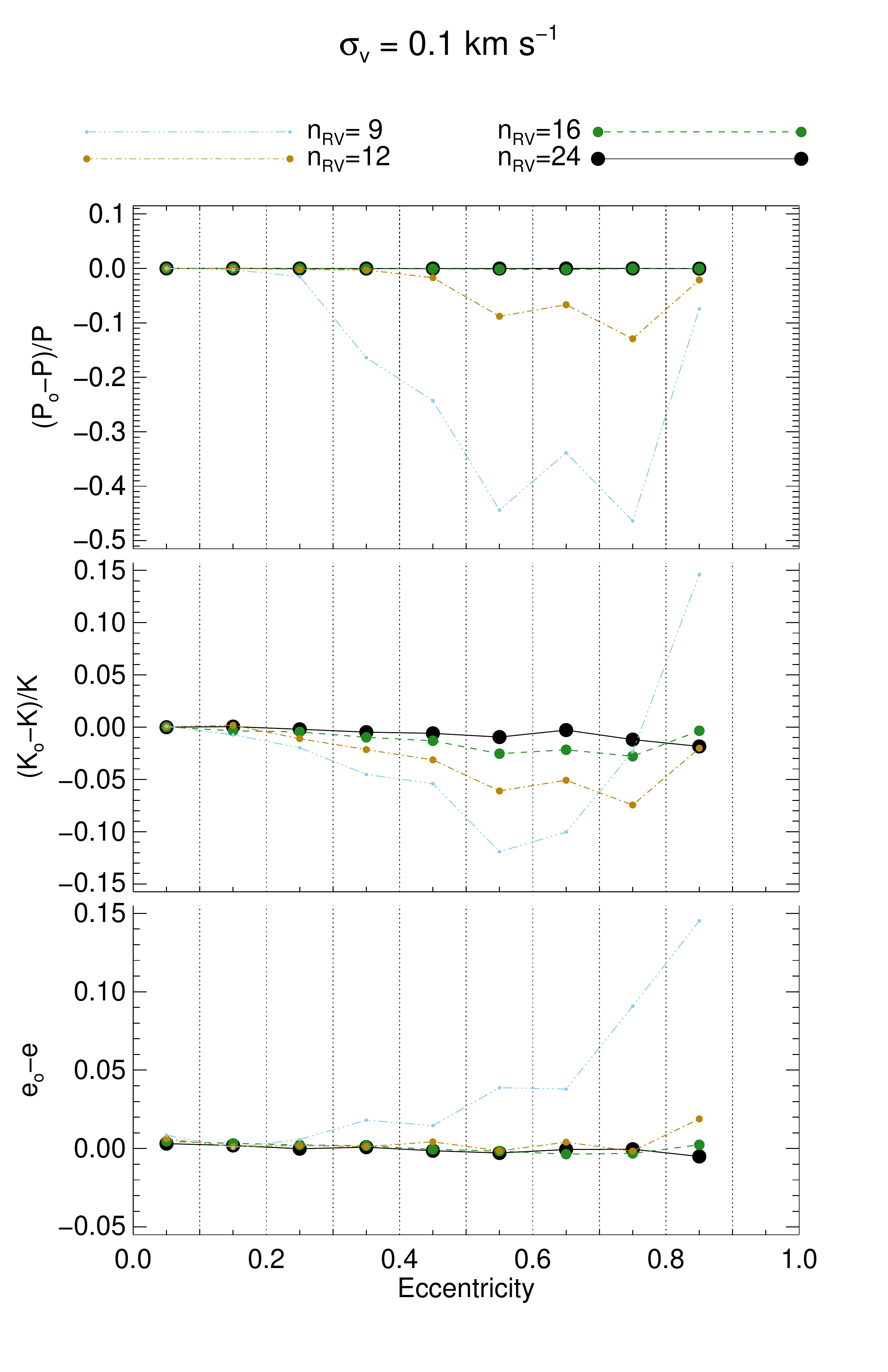} \\
\includegraphics[width=0.33\textwidth]{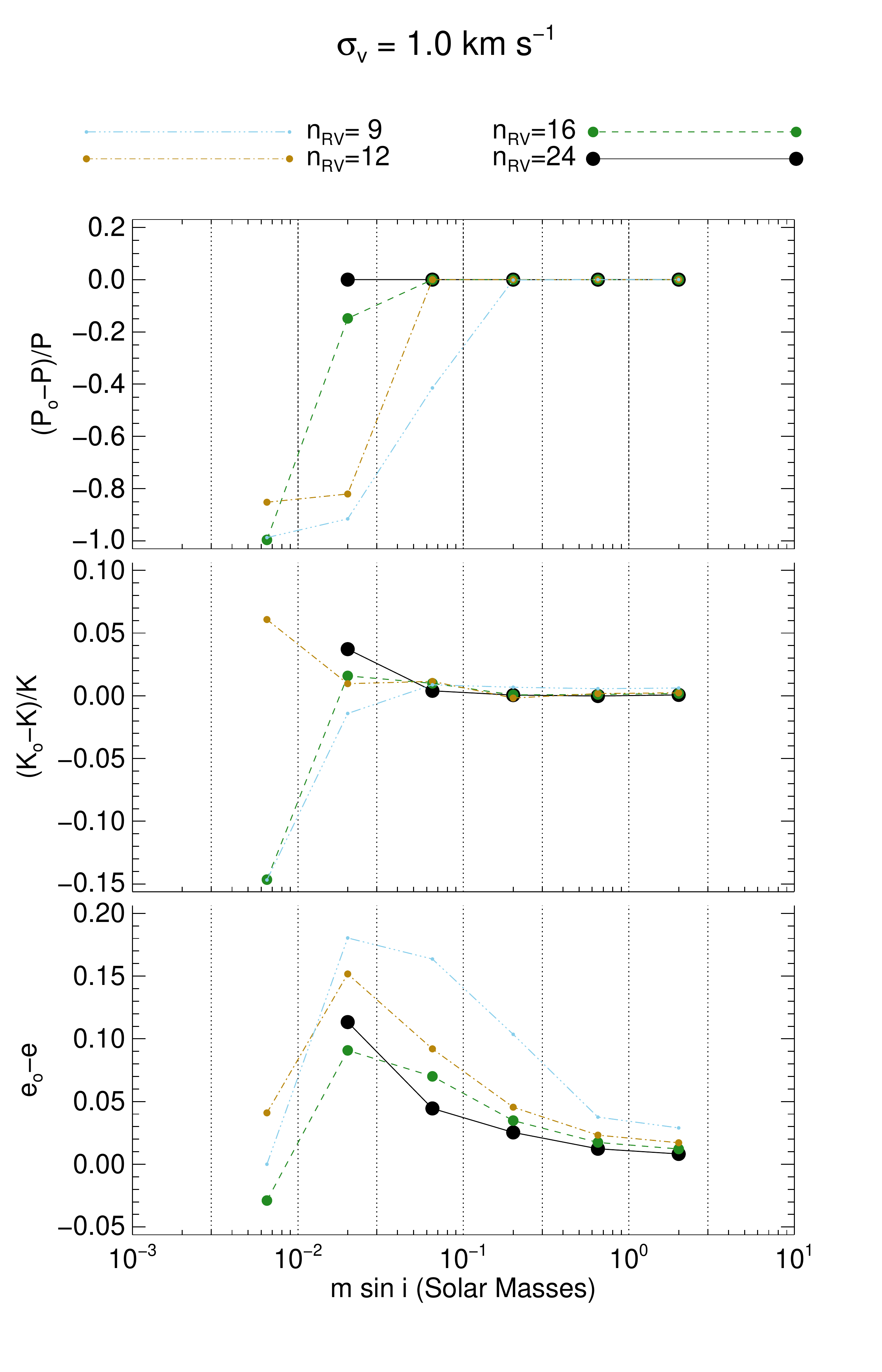}
\includegraphics[width=0.33\textwidth]{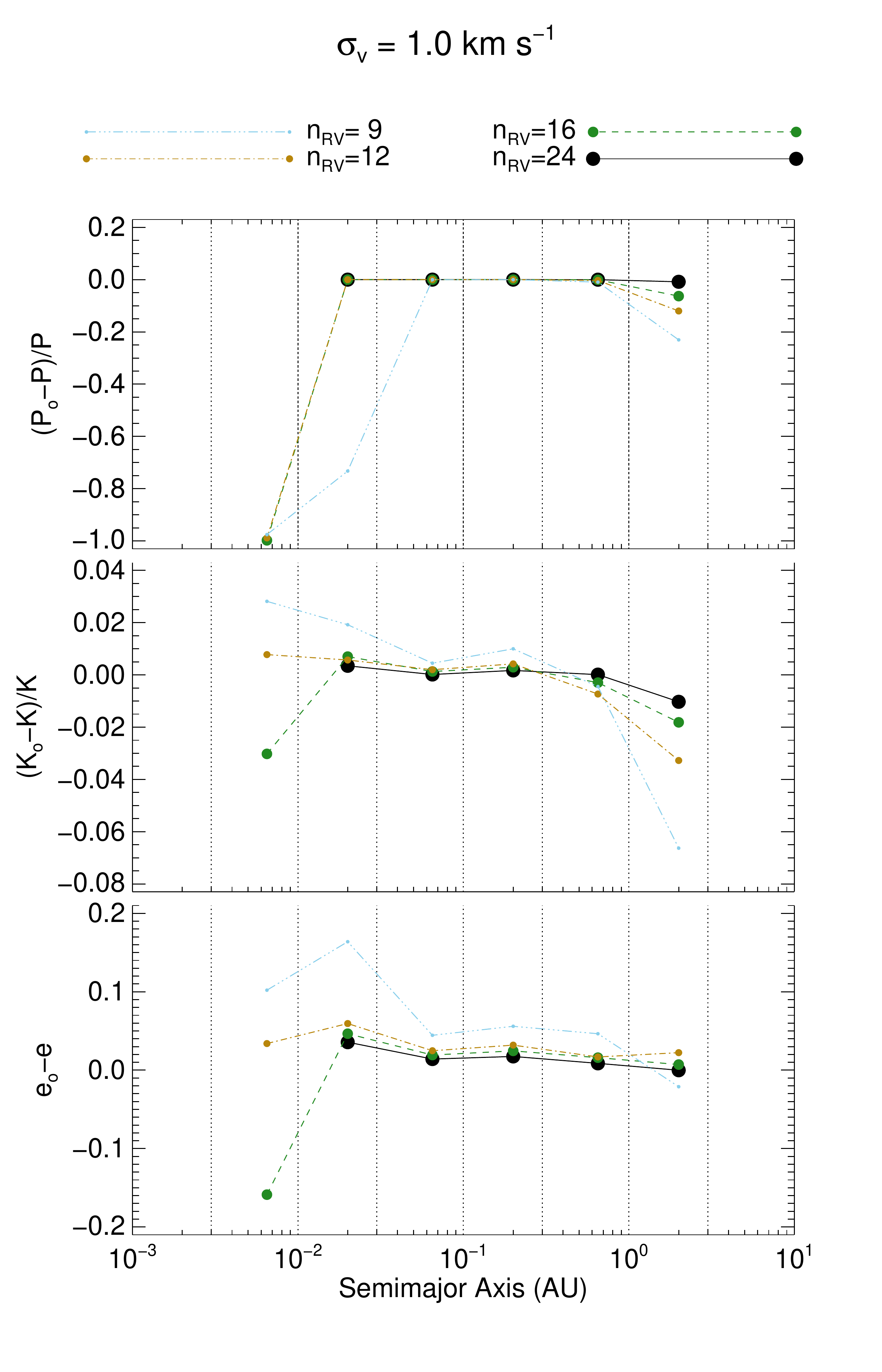} 
\includegraphics[width=0.33\textwidth]{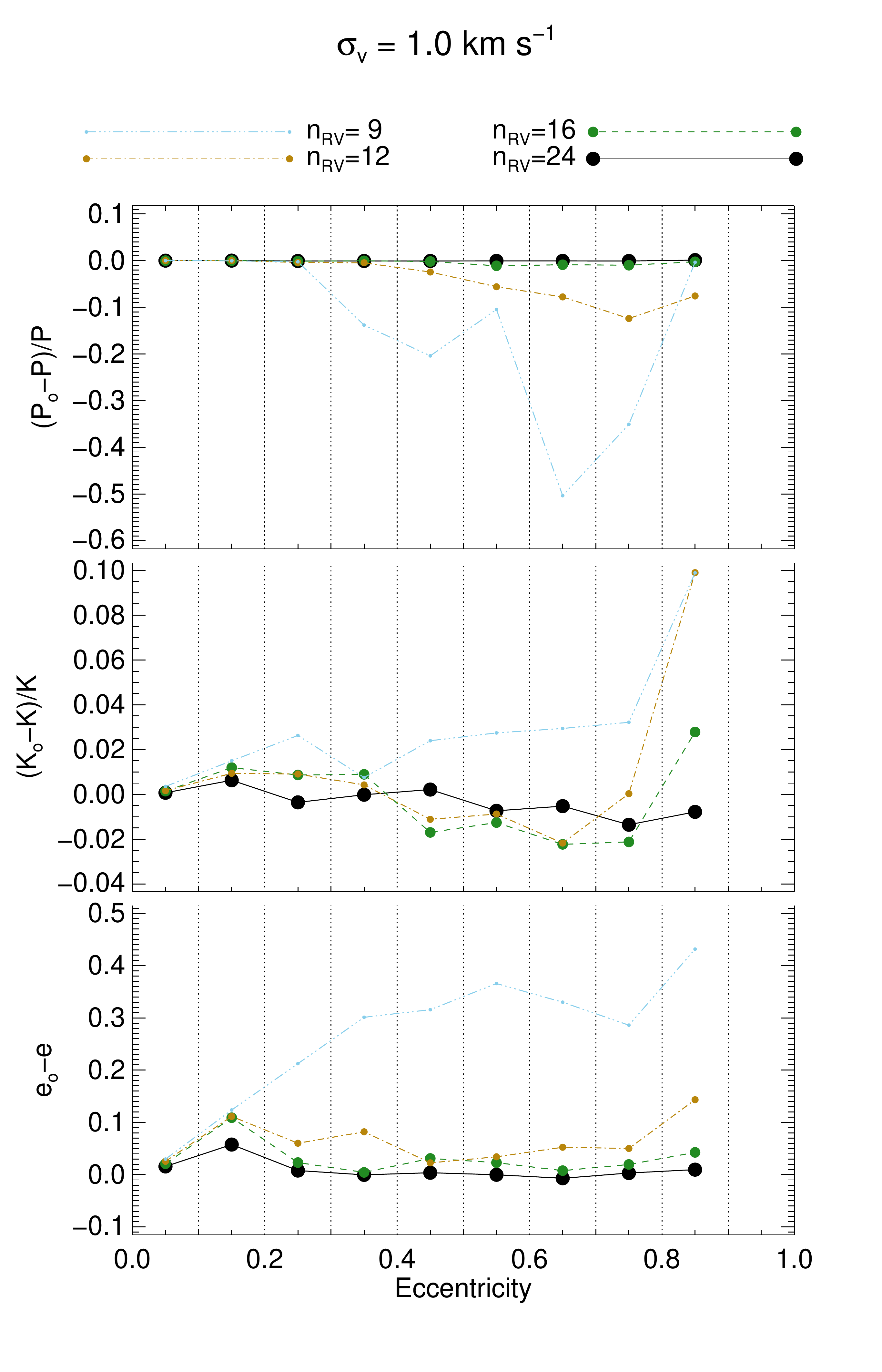} 
\caption{The recovered simulated systems binned by their recovered companion mass ($m \sin i$; left column), semimajor axis ($a$; center column), and eccentricity ($e$; right column). In each plot, the ordinates are the fractional error in period (top panel), fractional error in semiamplitude (middle panel), and error in eccentricity (bottom panel),  where $X_o$ indicates the recovered value of parameter with true value $X$. The top row of plots present the results using a base RV uncertainty of $\sigma_v = 0.1$ km s$^{-1}$, and the bottom row shows $\sigma_v = 1$ km s$^{-1}$.  Cyan, tan, green, and black points (dash triple-dotted, dash dotted, dashed, and solid lines) are from simulations with 9, 12, 16, and 24 visits, respectively. The vertical dotted lines mark the bins used, and for any bin with $<3$ stars, the point is excluded.}
\label{fig:compSim_errAnal}
\end{figure*}

We then compared each observed/recovered orbital parameter, $X_o$, with the actual parameters for the system, $X$, to determine how accurately the parameters are recovered as a function of parameter space. These results are summarized in Figure \ref{fig:compSim_errAnal}. For a large portion of the parameter space, the selected candidates reproduce the correct orbital parameters quite well. The systems that give the most trouble appear to be the low-mass companions, companions at large separations, and companions with highly eccentric orbits. Unsurprisingly, parameter recovery is overall worse for stars with fewer visits and higher RV uncertainties, but the drop in performance was not as dramatic between the 24 visit and 16 visit simulations as it was between the 16 visit to 9 visit simulations. However, from these results we can still conclude that in almost all regimes, the recovered orbital parameters are at least characteristic of the true values for the system.

\begin{figure*}
\center
\includegraphics[width=0.33\textwidth]{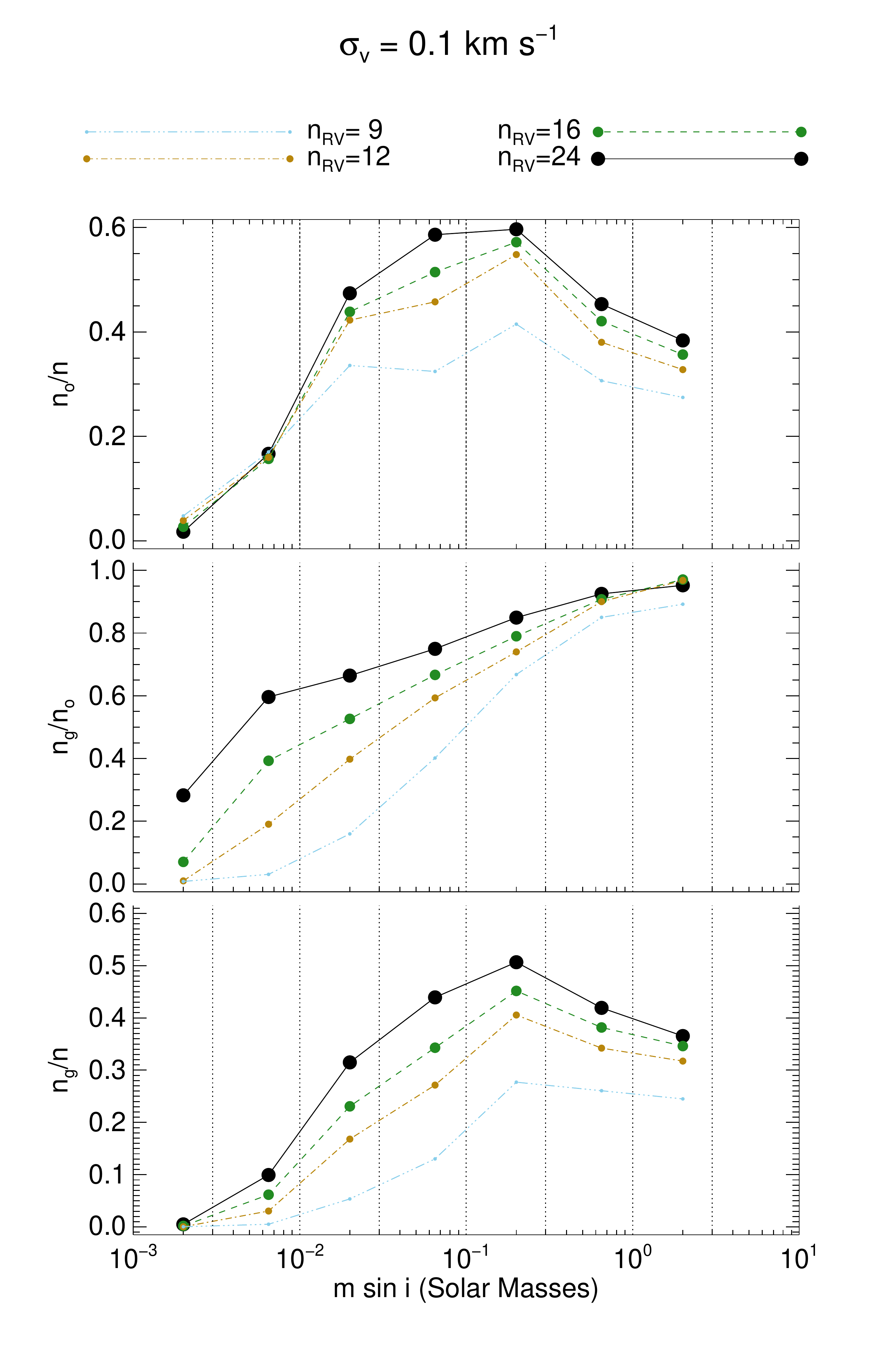} 
\includegraphics[width=0.33\textwidth]{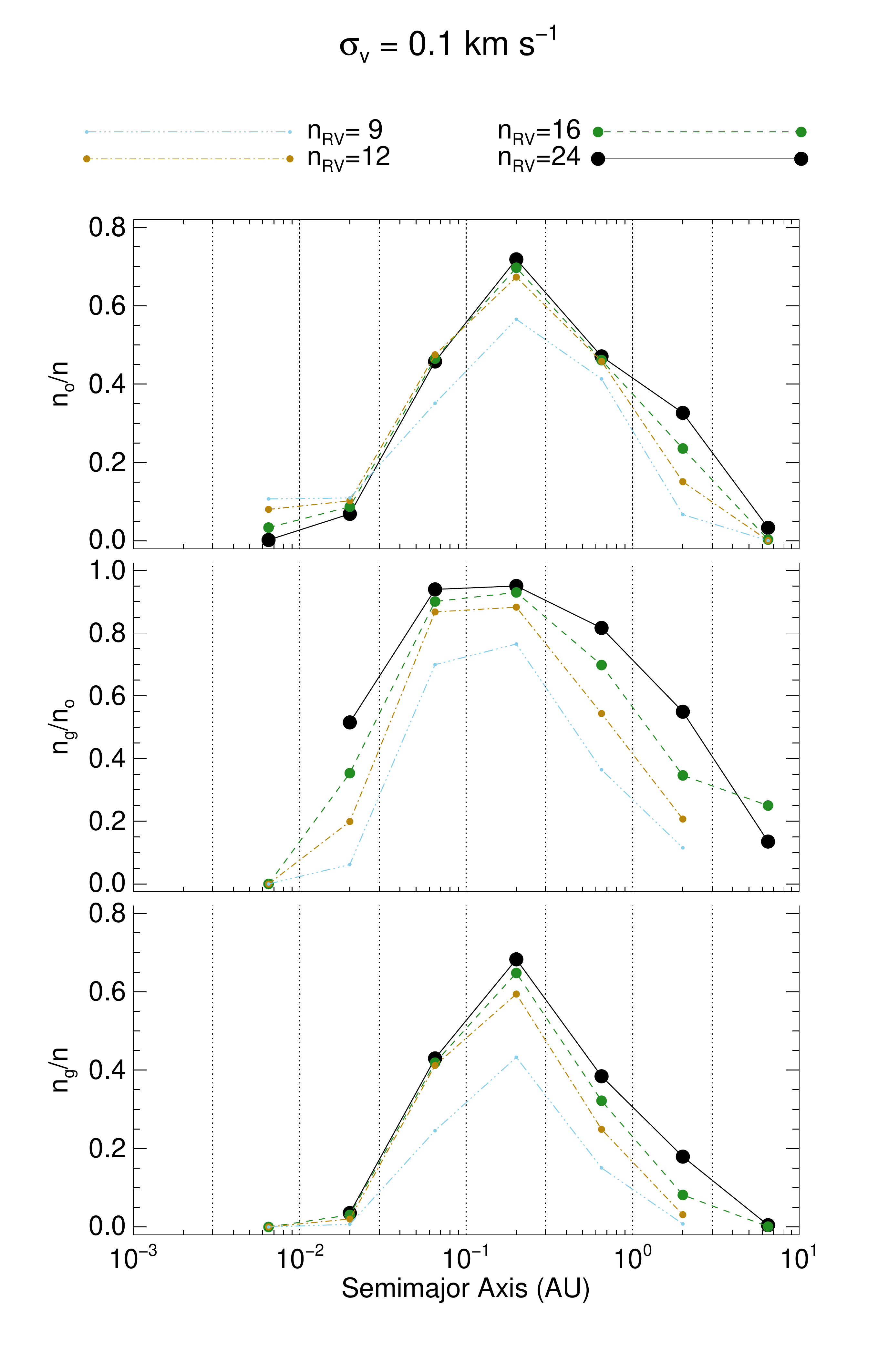}  
\includegraphics[width=0.33\textwidth]{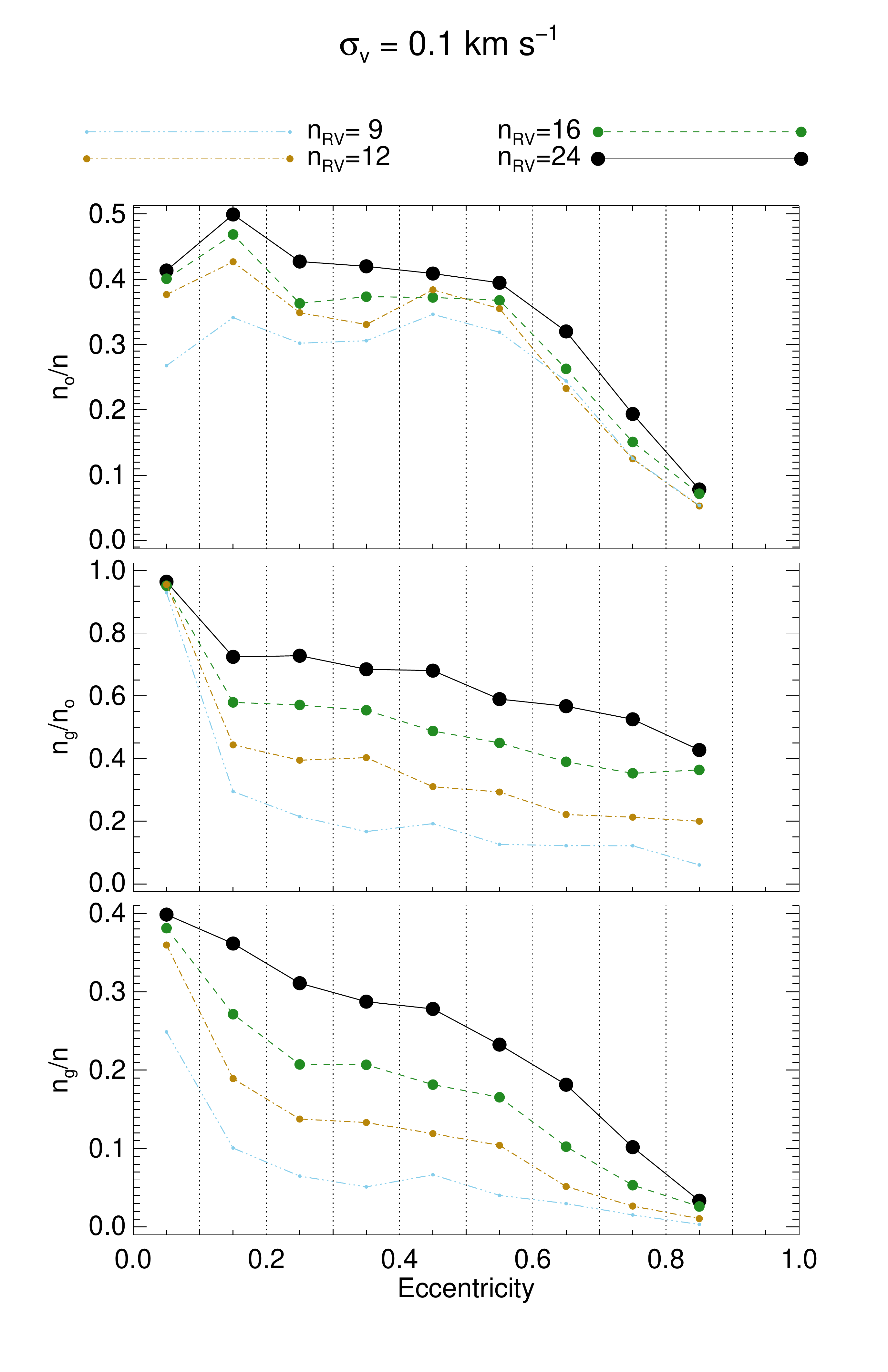} \\
\includegraphics[width=0.33\textwidth]{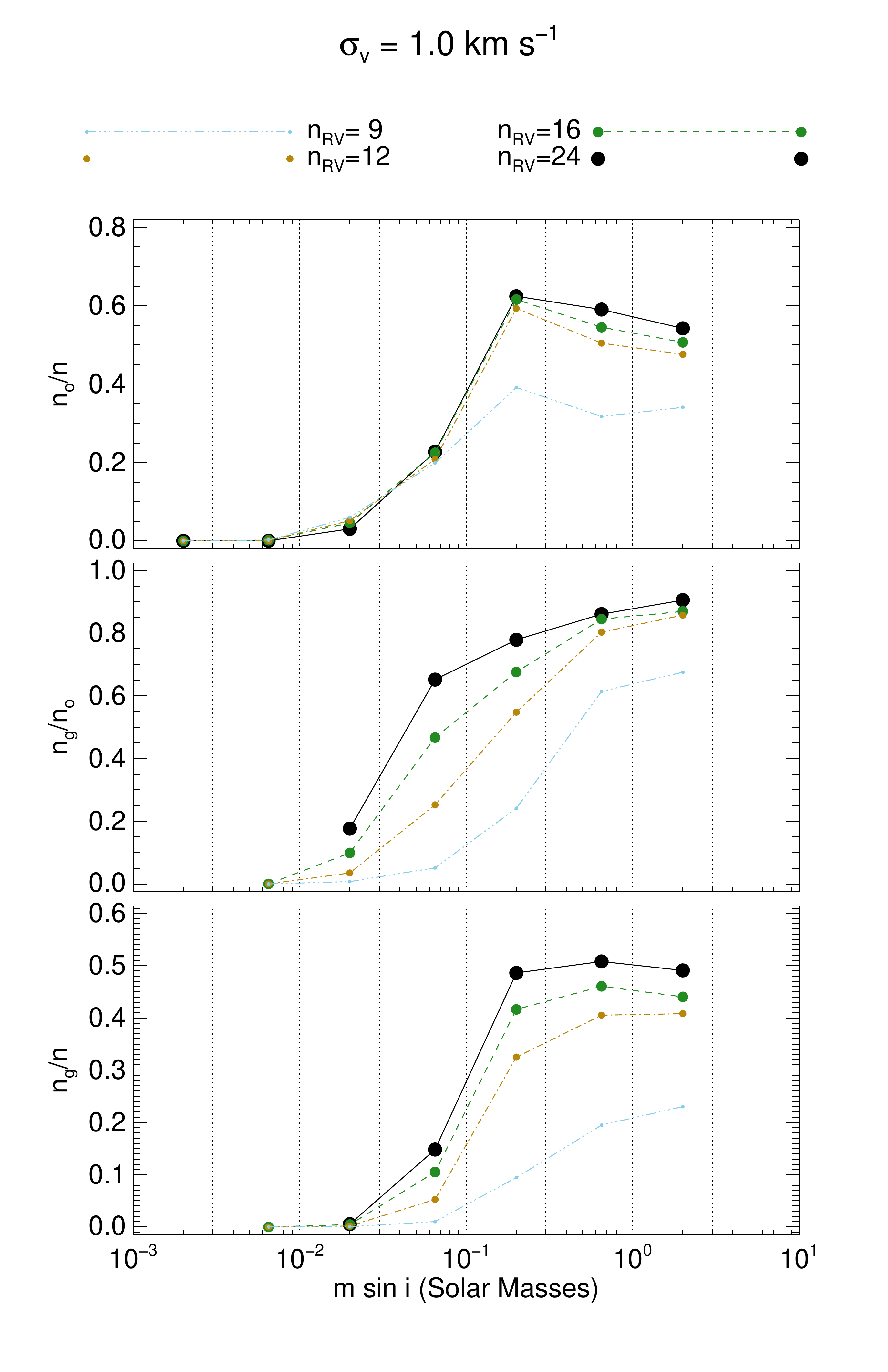}
\includegraphics[width=0.33\textwidth]{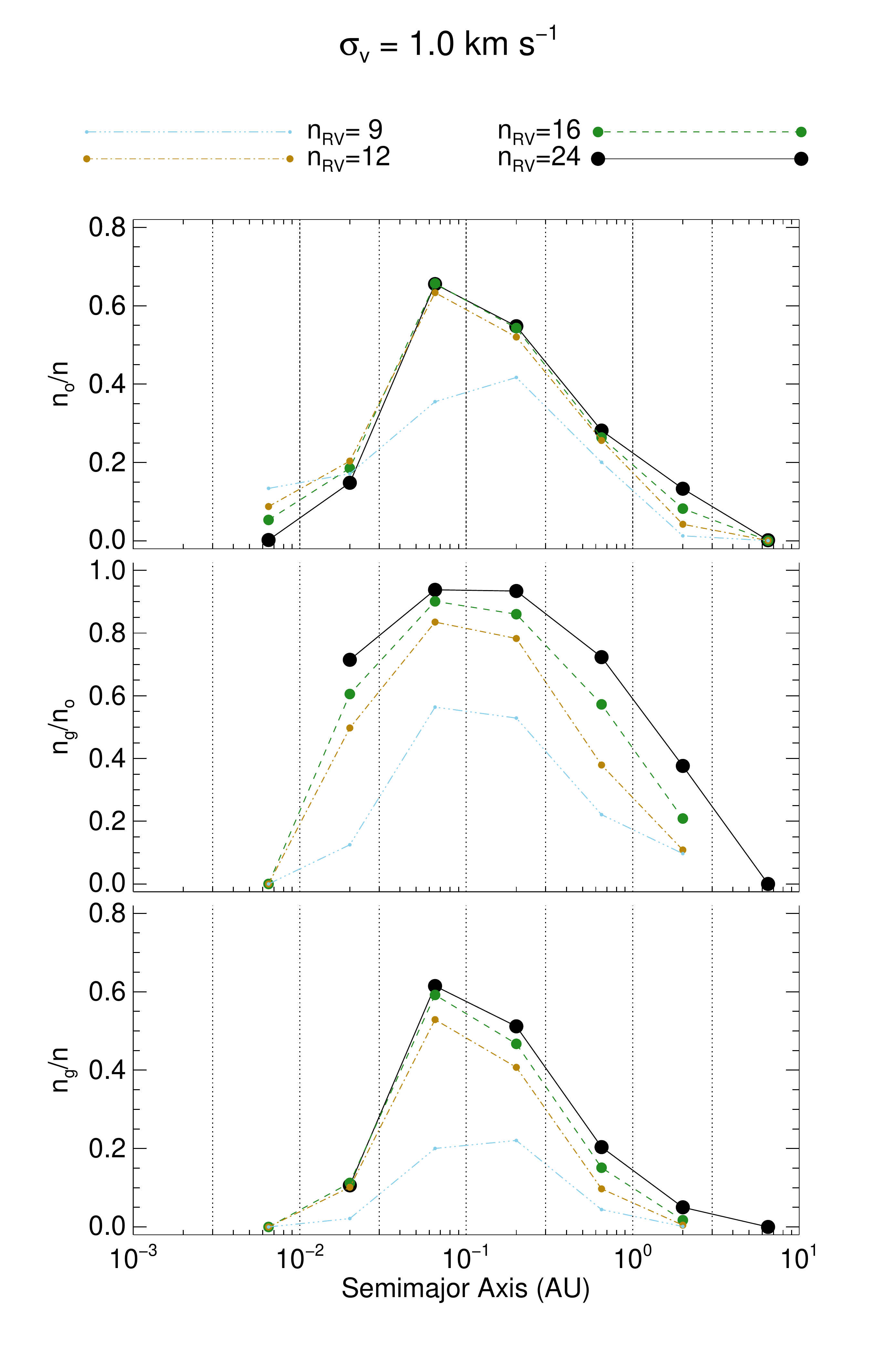} 
\includegraphics[width=0.33\textwidth]{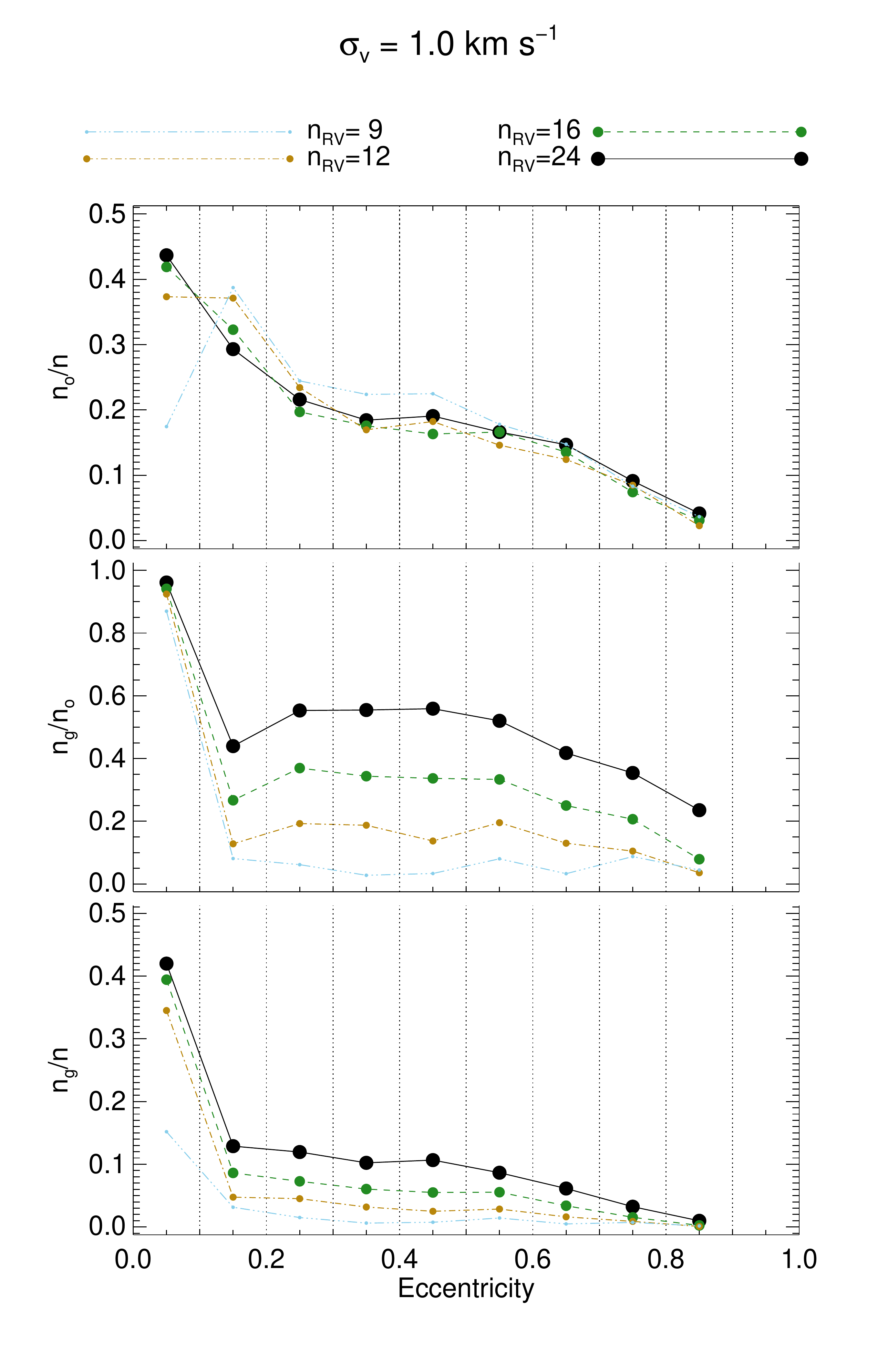}
\caption{The same as Figure \ref{fig:compSim_errAnal}, except on the ordinate in each plot, the top panel shows the fraction of systems recovered in the bin, $n_o$, compared to the total number of simulated systems in the bin, $n$, the middle panel shows the fraction of systems recovered with correct orbital parameters ($P$ and $K$ within 10$\%$ and $e$ within 0.1), $n_g$ compared to $n_o$, and the bottom panel shows $n_g/n$. Here a bin is excluded if the denominator of the ordinate is $<2$.}
\label{fig:recPlots}
\end{figure*}

Finally, we construct the recovery rates across the parameter space covered by this catalog. These are summarized in Figure \ref{fig:recPlots}. Unsurprisingly, recovery rate drops as $n_{RV}$ decreases, and higher RV uncertainties lead to lower recovery rates in general.

\subsection{Comparisons with Systems with Known Companions}

In addition to comparing to the known parameters of simulated RV signals, we also compared our results to the transit periods of 5 Kepler object of interest (KOI) hosts and one non-KOI eclipsing binary (EB) observed by APOGEE \citep{Fleming2015} that also meet the gold sample selection criteria described in \S \ref{sec:Catalog}. This comparison is presented in Table \ref{tab:KOIcomp}. We use the radius of the KOI as determined by \textit{Kepler} transit data to split the sample. We assume a KOI with $R_{KOI} = R_{Jup}$ will have $m\sin i \approx M_{Jup}$, and therefore any KOI with a radius less than this will likely be undetectable by APOGEE. 

For three of the five APOGEE-detectable KOIs and EBs in the gold sample, the transit period is reproduced almost exactly, and for the remaining two, it appears that the $\texttt{apOrbit}$ code simply selected the wrong harmonic for the period. For example, with KOI-1739, if we assume that the 146.0 day period found by APOGEE is the first harmonic of a fundamental period of 73 days, then the first harmonic period would be 219 days, which is much closer to the transit period. Most of the APOGEE-detectable KOIs have been designated as ``False Positives'' by the \textit{Kepler} team, meaning that the companion detected is not a planet, but rather a binary star companion. KOI-1739 is still designated as a candidate, but from APOGEE's RV data, we can conclude that this KOI should have a ``false positive'' disposition, as its companion is almost certainly not of planetary mass according to the analysis presented here. For the APOGEE-undetectable KOI, the APOGEE results from KOI-2598 may be indicative of longer-period companion previously undetected by transit. Further investigation of these this system is certainly warranted.

\begin{deluxetable*}{cccccccc} 
\tabletypesize{\scriptsize}
\tablecaption{KOIs and Kepler EBs Selected as Gold Sample Companion Candidates}
\tablewidth{0.98\textwidth}
\tablecolumns{7} 
\tablehead{
\colhead{APOGEE\_ID} & \colhead{KOI\#} & \colhead{$m \sin i$\tablenotemark{a}}&  \colhead{RV Period\tablenotemark{a}} & \colhead{Transit P\tablenotemark{b}} & \colhead{$R_{KOI}$\tablenotemark{b}} &\colhead{KOI Disposition\tablenotemark{c}} & \colhead{EB?\tablenotemark{d}} \\
\colhead{} & \colhead{(KIC ID)} & \colhead{($M_{Jup}$)}&  \colhead{(days)} & \colhead{(days)} & \colhead{($R_{Jup}$)} & \colhead{} & \colhead{}
}
\startdata

\sidehead{KOI Likely Detectable by APOGEE ($R_{KOI} > R_{Jup}$)}
2M19263602+4242028 & 1739 & 622   & 146.0 & 220.6    & \nodata & Candidate & no \\
2M19335125+4253024 & 3546 & 261   &  4.286 &  4.286  & 2.45 	& False Positive & yes \\ 
2M19290626+4202158 & 6742 & 615   &	63.63  &  63.52 & 2.76	& False Positive & yes\\ 
2M19352118+4207199 & 6760  & 195   &  10.82  & 10.82   & 2.40 & False Positive & yes \\ 
2M19315429+4232516 & (7037405) & 667 & 103.3  &  207.15   & \nodata & \nodata	  & yes \\ 
 \sidehead{KOI Likely Undetectable by APOGEE ($R_{KOI} < R_{Jup}$)}
2M19291780+4302004 & 2598 &  111   &  274.4  & 2.69    & 0.093 & Candidate & no
\enddata
\tablenotetext{a}{As determined by APOGEE RV data (this work).}
\tablenotetext{b}{As determined by \textit{Kepler} transit data \citep{Mullally2015}.}
\tablenotetext{c}{Official \textit{Kepler} KOI disposition, which refers to the companion's status as a planet. If this field is blank, then the object is not a KOI, and the KIC ID is given for the star rather than a KOI number. A ``false positive'' disposition often (and in this case always) indicates a companion that was found not to be of planetary mass. This case is distinct from the definition of false positive we used in the rest of this paper to indicate RV measurements that masquerade as a non-existent companion.}
\tablenotetext{d}{Is the star in the Kepler Eclipsing Binary catalog \citep{Slawson2011,LaCourse2015}?}
\label{tab:KOIcomp}
\end{deluxetable*}

\textcolor{black}{Furthermore, APOGEE recovered the known planet HD 114762b (2M13121982+1731016), with which we compare orbit parameters and host stellar parameters derived and adopted by the \texttt{apOrbit} pipeline to literature values in Table \ref{tab:recPlanet}.  APOGEE's recovered stellar parameters, as well as the recovered period and orbital semi-major axis are in good agreement with the results from \cite{Kane2011}, but APOGEE overestimates the eccentricity of the system, and thus the values of $K$ and $m \sin i$. This is in agreement with our findings in Appendix \ref{sec:caveats}. However, this star was selected for use as a telluric standard, and thus is not included in our gold sample. This result may lead us to reconsider excluding stars selected as telluric standards in future versions of this catalog, especially considering the upgrades described in \S \ref{sec:Future} which will lead to improved stellar parameters and RV determinations for dwarfs.}

\begin{deluxetable}{rcc}
\tabletypesize{\scriptsize}
\tablecaption{Comparison of Recovered Parameters of Known Exoplanet System HD 114762\MakeLowercase{b}}
\tablewidth{\columnwidth}
\tablecolumns{3} 
\tablehead{
\colhead{Parameter} & \colhead{APOGEE Value} & \colhead{Literature\tablenotemark{a} Value}
}
\startdata
\sidehead{\textbf{Host Stellar Parameters}}
$T_{\rm eff}$ & 5466 K& 5673 K\\
$\log g$ & 4.196 & 4.135\\
\lbrack Fe/H\rbrack & -0.832 & -0.774 \\
Distance &  36.5 pc  &  38.7 pc  \\
$M_{\star}$ & 0.86 $M_{\odot}$   & 0.83 $M_{\odot}$\\
$R_{\star}$ & 1.20 $R_{\odot}$  & 1.24 $R_{\odot}$ \\
\sidehead{\textbf{Orbital Parameters}}
$P$ &  85.58 days & 83.92 days \\
$K$ & 925.5 m s$^{-1}$ & 612.5 m s$^{-1}$ \\
$e$ &  0.594  & 0.335 \\ 
$m \sin i$ & 14.72 $M_{Jup}$ & 10.98 $M_{Jup}$ \\
$a$ & 0.361 AU & 0.353 AU 
\enddata
\tablenotetext{a}{\cite{Kane2011}}
\label{tab:recPlanet}
\end{deluxetable}

\section{Catalog Information} \label{sec:data}
For each star in the 382-star gold sample, the following data are available:
\begin{itemize}
\item APOGEE targeting information, 2MASS photometry, proper motions, and reduction flags.
\item Adopted APOGEE stellar parameters ($T_{\rm eff}, \log g,$ [Fe/H]), and estimates of each primary star's mass, radius, and distance, with flags indicating the source/quality of the stellar parameters and mass/radius/distance estimates.
\item Heliocentric RV measurements for each star derived using best-fit ASPCAP synthetic spectra as RV templates.
\item The best-fit orbital and physical parameters of each system's candidate companion.
\end{itemize}
These data are compiled into a $\texttt{FITS}$ table, whose content is described in Table \ref{tab:DataModel}. The catalog is also available as a Filtergraph portal here: \url{https://filtergraph.com/apOrbitPub}. The Filtergraph portal also contains links to webpages containing plots of the RV curves for these systems. Additional data for each star, including spectra and additional photometry, are available publicly via SDSS DR12 \citep{Alam2015}. See \url{http://www.sdss.org/dr12/} for instructions on the access and use of APOGEE DR12 data.

\subsection{Caveats} \label{sec:caveats}
Here we present some caveats regarding the quality of the data in this catalog: 
\begin{itemize}
\item All caveats that apply to all APOGEE data \citep{Holtzman2015} also apply to this catalog. 
\item Many stars with the longest baselines were observed during APOGEE commissioning, during which the instrument did not employ dithering. However, these stars were reobserved at the end of the survey with the standard instrument configuration, and RVs derived from commissioning data have been shown to be of similar quality to main-survey RVs.
\item The stellar parameters derived by ASPCAP for dwarf stars are uncalibrated, but good enough to establish estimates of the star's primary mass, and sufficiently accurate to distinguish between dwarfs and giants.
\item The RV errors output by the APOGEE reduction pipeline may be slightly underestimated. We refer the reader to \S 10.3 of \cite{Nidever2015} where RV uncertainties are discussed more fully.
\item The distances presented here are from a preliminary catalog, and will likely undergo future refinement.
\item The most common source of errors in the orbital parameters is the fitter choosing the wrong harmonic for the period. Therefore, the periods presented here may be an integer number (2 or 3) or an integer fraction (1/2 or 1/3) times the true period for the system. The fitter also had a tendency to inflate the eccentricities of the simulated systems, so the eccentricities, and thus the values of $K$ and $m \sin i$ presented here are likely to be slightly larger than their true values.
\item The values for argument and time of periastron ($T_P$ and $\omega$) become unconstrained at low eccentricities, and are poorly reproduced by this catalog. We release them so that our model curves can be reproduced, but should be taken with a grain of salt.
\end{itemize}
Finally we stress that the systems presented here are \textit{candidates}, and that the orbital parameters presented here may only be characteristic of the true values of the system. In particular, the low-mass and low-visit candidates are the most in need of additional observation.

\bibliographystyle{apj}

\LongTables
\begin{deluxetable*}{cccc}
\tabletypesize{\scriptsize}
\tablecaption{Data model for the 1st extension of the \texttt{goldOrbit-dr12.fits} file}
\tablewidth{0.98\textwidth}
\tablecolumns{4} 
\tablehead{
\colhead{Field Name} & \colhead{Data Type} & \colhead{Units} & \colhead{Description}  
}
\startdata
APOGEE\_ID & char[18] &\nodata & TMASS-STYLE object name\\
LOCATION\_ID & int16  & \nodata & APOGEE field location ID number\\
FIELD & char[16] &\nodata & APOGEE field name\\
NVISITS & int16 & \nodata & Number of RV measurements used in the fit \\
SNR & float32 & \nodata & median S/N per pixel in combined frame (at apStar sampling) \\
J & float32 & mag & 2MASS J mag \\
J\_ERR & float32 & mag &uncertainty in 2MASS J mag \\
H & float32 &  mag & 2MASS H mag \\
H\_ERR & float32 & mag & uncertainty in 2MASS H mag\\
K & float32  & mag & 2MASS Ks mag \\
K\_ERR & float32 & mag & uncertainty in 2MASS Ks mag \\
AK & float32 & mag & K-band extinction adopted \\
AK\_SRC & char[17] & \nodata & Method used to get targeting extinction\\
RA & float64 & degrees & Right ascension (J2000) \\
DEC & float64 & degrees & Declination (J2000) \\
GLON & float64 & degrees & Galactic longitude \\
GLAT & float64 & degrees & Galactic latitude \\
PMRA & float32 & mas/yr & One proper motion measurement \\
PMDEC & float32 & mas/yr & One proper motion measurement \\
PM\_SRC & char[20] & \nodata & Catalog used for PM \\
EXTRATARG & int32 & \nodata & bitmask that identifies main survey targets and other classes\tablenotemark{a} \\
APOGEE\_TARGET1 & int32 & \nodata & bitwise OR of first APOGEE target flag of all visits\tablenotemark{a} \\
APOGEE\_TARGET2 & int32 & \nodata & bitwise OR of second APOGEE target flag of all visits\tablenotemark{a} \\
TARGFLAGS & char[116] & \nodata & target flags in English\\
STARFLAG & int32 & \nodata & Flag for star condition taken from bitwise OR of individual visits\tablenotemark{a} \\
STARFLAGS & char[129] & \nodata & STARFLAG in English \\
ASPCAPFLAG & int32 & \nodata & Flag for ASPCAP analysis\tablenotemark{a} \\
ASPCAPFLAGS & char[114] & \nodata & ASPCAPFLAG in English \\
TEFF & float32 & K & Adopted $T_{\rm eff}$ for the primary star \\
TEFF\_ERR & float32 & K & Adopted $T_{\rm eff}$ uncertainty \\
LOGG & float32 & log (cgs) & Adopted $\log g$ for the primary star \\
LOGG\_ERR & float32 & log (cgs) & Adopted $\log g$ uncertainty \\
FE\_H & float32 & dex & Adopted [Fe/H] for the primary star \\
FE\_H\_ERR & float32 & dex & Adopted [Fe/H] uncertainty \\
SPARAMTYPE  & int16 & \nodata & Source of the stellar parameters adopted\tablenotemark{b} \\
STARTYPE & char[3] & \nodata &Classification applied to host star (See \S \ref{sec:hostStarSort}) \\
MSTAR	   & float32 & $M_{\odot}$ & Mass of the primary based on the available stellar parameters \\ 
MSTAR\_ERR	   & float32 & $M_{\odot}$ & Uncertainty of the primary mass \\ 
RSTAR	   & float32 & AU & Radius of the primary based on the available stellar parameters \\ 
RSTAR\_ERR	   & float32 & AU & Uncertainty of the primary radius \\ 
DIST	   & float32 & pc & Adopted distance of the primary star\\ 
DIST\_ERR   & float32 & pc & Uncertainty of the distance \\
MSTAR\_SRC  & int16  & \nodata & Source/method of mass/radius/distance estimation\tablenotemark{c} \\
VJITTER   & float32 & m s$^{-1}$ & Estimated intrinsic RV jitter of the star. \\
BASELINE    & float32 & days &  Maximum baseline of RV data included in fit \\
SIGMA\_V  & float64 & m s$^{-1}$ & Median of RV errors, $\sigma_v$\\
SIG\_RVVAR & float64 & \nodata & Significance of the RV variations (See \S \ref{sec:selRVs})\\
JD      & float64[50]& JD & Julian Date of observations included in fit \\
RV 	   & float64[50] & m s$^{-1}$ & Radial velocities of observations included in fit \\
RV\_ERR & float64[50] & m s$^{-1}$ & Error in radial velocities of observations included in fit \\
MODEL & float64[50] & m s$^{-1}$ & RVs of best-fit orbital model. \\
RESID & float64[50] & m s$^{-1}$ & Residuals of best-fit orbital model. \\
PERIOD 	   & float64 & days & Best-fit orbital period, $P$, of the system \\
PERIOD\_ERR 	   & float64 & days  & Uncertainty in $P$ \\
SEMIAMP    & float64 & m s$^{-1}$  & Best-fit RV semiamplitude, $K$, of the system \\
SEMIAMP\_ERR  & float64 & m s$^{-1}$ & Uncertainty in $K$\\
ECC 	   & float64 & \nodata & Best-fit eccentricity,$e$, of the system \\
ECC\_ERR  & float64 & \nodata & Uncertainty in $e$\\
OMEGA & float64 & degrees & Argument of periastron, $\omega$\\
T0 & float64 & JD & Epoch of Transit \\
TPERI   & float64 & JD   & Epoch of periastron \\
V0  & float64 & m s$^{-1}$ & Intercept of the global trend applied to the RVs \\
SLOPE      & float64 & m s$^{-1}$ day$^{-1}$ & Slope of the global trend applied to the RVs \\
NITER & int16 & \nodata & Number of iterations used to converge on a period (see \S \ref{sec:KepOrbits})\\
CHI2 & float64 & \nodata & $\chi^2$ (not reduced) of the fit\\
DOF 	   & float64 &\nodata  & Degrees of freedom of the fit \\
FIT\_RMS   & float64 & m s$^{-1}$ & Root-Mean-Square of the residuals of the fit\\
PUI    & float64 & \nodata & Phase Uniformity Index (see \S\ref{sec:fitCrit})\\
VUI     & float64 & \nodata & Velocity Uniformity Index (see \S\ref{sec:fitCrit})\\
MASSFN     & float64 & $M_{\odot}$ & Mass function of the system \\
MSINI 	   & float64 & $M_{\odot}$ & Estimated $m \sin i$ of the companion \\
SEMIMAJ	   & float64 & AU & Estimated orbital semimajor axis of the companion 
\enddata
\tablenotetext{a}{See \url{http://www.sdss.org/dr12/algorithms/bitmasks/} for APOGEE bitmask definitions}
\tablenotetext{b}{0=calibrated ASPCAP, 1=uncalibrated ASPCAP, 2=RV mini-grid (see \S \ref{sec:spDer})}
\tablenotetext{c}{0=\cite{Torres2009} relation, 1=APOKASC, 2=RC, 3=spectrophotometric, 4=TRILEGAL, 5=young cluster distance (See \S \ref{sec:spDer})}
\label{tab:DataModel}
\end{deluxetable*}

\end{document}